\documentclass[11pt]{article}  
\usepackage{amsfonts}

\usepackage{geometry}

\usepackage{amsfonts} 
\usepackage{graphicx}
\usepackage{color}
\usepackage{rotating}
\usepackage{amsmath}
\usepackage{amssymb}
\usepackage{amsthm}
\usepackage{float}
 \usepackage{url}
\usepackage{subfigure}

\usepackage{natbib} 

\usepackage{enumerate}
\usepackage{enumitem}

\usepackage{dsfont}
 
 \usepackage{algorithm}
\usepackage{algorithmic}
\usepackage{array,arydshln}

\usepackage{soul} %% PARA TACHAR CON \st
\usepackage{hyperref} %PARA QUE AL PINCHAR VAYAS A LA ECUACION

 \usepackage{xcolor}
\hypersetup{
    colorlinks,
    linkcolor= {blue!90!black}, %{red!80!black},
    citecolor={blue!60!black},
    urlcolor={blue!80!black}
}

\newcommand\bh {\mathbf h}

\newcommand\bv {\mathbf v}

\newcommand\bx {\mathbf x}

\newcommand\itN {{\mathcal{N}}}

\newcommand\itU {{\mathcal{U}}}

\newcommand\bbe {\mbox{\boldmath $\beta$}}

\newcommand\bbech {\mbox{\scriptsize${\bbe}$}}

\newcommand\bla {\mbox{\boldmath $\lambda$}}

\newcommand\walfa {\widehat{\alpha}}

\newcommand\wbeta {\widehat{\beta}}
\newcommand\wbbe {\widehat{\bbe}}

\newcommand\wlam {\widehat{\lambda}}
\newcommand\wblam {\widehat{\bla}}

\newcommand\wmu {\widehat{\mu}}

\newcommand\wsigma {\widehat{\sigma}}

\def\real{\mathbb{R}}

\newcommand{\esp}{\mathbb{E}}
\newcommand{\prob}{\mathbb{P}}

\newcommand{\var}{\mbox{\sc Var}}

\newcommand{\trasp}{^{\mbox{\footnotesize \sc t}}}

\def\median{\mathop{\mbox{median}}}

\def\mad{\mathop{\mbox{\sc mad}}}

\def\argmin{\mathop{\mbox{argmin}}}

\newcommand\noi{\noindent}

\def\ep{\vspace{1cm}}
\def\square{\ifmmode\sqr\else{$\sqr$}\fi}
\def\sqr{\vcenter{
         \hrule height.1mm
         \hbox{\vrule width.1mm height2.2mm\kern2.18mm
\vrule width.1mm}
         \hrule height.1mm}}

\newcommand{\tuk}{\mbox{\scriptsize \sc t}}

\newcommand{\ini}{\mbox{\footnotesize \sc ini}}

\newcommand\mm {\mbox{\scriptsize \sc mm}}
\newcommand\wmm {\mbox{\scriptsize \sc wmm}}

\newcommand\nuevo {\mbox{\scriptsize \sc n}} 
\newcommand\mmnew {\mbox{\scriptsize \sc mm}_{\nuevo}}
\newcommand\wmmnew {\mbox{\scriptsize \sc wmm}_{\nuevo}}
\newcommand\hmmnew {\mbox{\scriptsize \sc hmm}_{\nuevo}}
\newcommand\hwmmnew {\mbox{\scriptsize \sc hwmm}_{\nuevo}}

\newcommand\hmm {\mbox{\scriptsize \sc hmm}}
\newcommand\hwmm {\mbox{\scriptsize \sc hwmm}}

\newcommand\LS {\mbox{\sc LS}}
\newcommand\HLS {\mbox{\sc HLS}}
\newcommand\MM {\mbox{\sc MM}}
\newcommand\WMM {\mbox{\sc WMM}}
\newcommand\HMM {\mbox{\sc HMM}}
\newcommand\HWMM {\mbox{\sc HWMM}}

\begin{document}

\title{Robust estimation of heteroscedastic regression models: a brief overview and new proposals}

\author{Concei\c{c}\~{a}o  Amado \\ \small CEMAT, Instituto Superior Técnico, Universidade de Lisboa, Portugal \\ Ana M. Bianco \\ \small Universidad de Buenos Aires and CONICET \\ Graciela Boente \\\small Universidad de Buenos Aires and CONICET \\ Isabel M. Rodrigues \\ \small CEMAT, Instituto Superior Técnico, Universidade de Lisboa, Portugal}

\date{}
\maketitle

%%%%%%%%%%%%%%%%%% ABSTRACT%%%%%%%%%%%%%%%%%%%%%%%%%%%%%%%%
\begin{abstract}

We collect robust proposals given in the field of regression models with heteroscedastic errors. Our motivation stems from the fact that the practitioner frequently faces the confluence of two phenomena in the context of data analysis: non--linearity and heteroscedasticity. The impact of heteroscedasticity on the precision of the estimators is well--known, however the conjunction of these two phenomena makes handling outliers more difficult.

An iterative procedure to  estimate the parameters of a heteroscedastic non--linear model is considered. The studied estimators combine weighted $MM-$regression estimators, to control the impact of high leverage points, and a robust method to estimate the parameters of the variance function.
\end{abstract}

\ep

\normalsize
\newpage

\section{Introduction}{\label{sec:intro}}
Non--linear regression models are applied to a great variety of problems in different disciplines, such as chemistry, toxicology and pharmacology. 
Both theoretical and heuristic considerations may suggest an adequate functional form  between the response and the predictor variables. 
An assumption beneath these models is that some available prior knowledge  enables them to pose a specific functional relationship. Sometimes, the non--linear relation can be transformed to obtain a linear model, however, there are situations in which the transformed model does not retain the main characteristics of the original problem, see \citet{Seber:Wild:2003}. Additionally, the parameters of the linear model are not so easy to interpret as they are in the original non--linear one. The interpretation and meanfulness of the estimates are the main reasons for the wide application of non-linear regression models.

In practice, in many situations non--linearity is not an isolated effect and heteroscedasticity is also encountered. 
Regarding the identification of outliers, the combination of these two problems makes this task harder.
It is well known that in linear regression analysis heteroscedasticity always poses a challenge to robust estimators: it may mask outliers or either anomalous data may vitiate the diagnosis of heteroscedasticity.  Unfortunately, non--linear models are not an exception to this statement, making the problem even more complicated.

It is worth noting that, even when there is a lack of homogeneity in the variance of the responses, the estimators of the regression parameters computed assuming  homoscedasticity  are still consistent, but they lose efficiency. Furthermore, this is often reflected in inaccurate confidence intervals. Even in linear regression, $M-$estimators do not protect against the unduly influence of high leverage points, that is why more sophisticated methods were proposed.

The existence of heteroscedasticity warns the practitioner that the variance may depend on the independent variables. Thus, if the variance is not constant,  it is common to pose a model such as
\begin{equation}
y= g(\bx,\bbe_0)+ \sigma(\bx) \; \epsilon\, , \label{eq:modelo0}
\end{equation}
where $y$ is the dependent variable, which is related to the covariates $\bx \in {\cal{X}} \subseteq \real^{k}$ through a known function $g$ that depends on the unknown parameters $\bbe_0 \in \Theta \subseteq \real^p$, $\sigma(\bx)$ is a positive real function and $\epsilon$ is a random error independent of the covariates. In the classical setting, it is assumed that $\esp(\epsilon)=0$ and $\var(\epsilon)=1$.

Different approaches were given to handle heteroscedasticity, either modelling the variance $\sigma(\bx)$ nonparametrically or parametrically. In this article, we focus on the latter approach, in which we admitted that the variance function has a given parametric form, that is $\sigma(\bx)= \sigma_0 \upsilon(\bx,\bla_0,\bbe_0)$, where $\upsilon:\real^k \times \real ^q\times \real^k \to \real^+$ is a known function up to the parameters $\bbe_0$ and $\bla_0$ which, as $\sigma_0>0$, are unknown.  Hence, we assume the regression model is given by
\begin{equation}
y= g(\bx,\bbe_0)+ \sigma_0 \; \upsilon(\bx,\bla_0,\bbe_0) \; \epsilon\, . \label{eq:modelo}
\end{equation}

Some models for the variance function $\upsilon$ have been considered in the literature. Among them, \citet{Box:Hill:1974} introduced the function
$\upsilon(\bx,\lambda,\bbe) = (1 + |\bx \trasp\bbe|)^{\lambda}$, while \citet{Bickel:1978} considered variance functions such that $\upsilon(\bx,\lambda,\bbe) =  \exp\{\lambda\,|\bx\trasp\bbe|\}$ and $\upsilon(\bx,\lambda,\bbe) = \exp\{\lambda h(\bx)\}$. Note that, in all cases $\lambda \in \real$, furthermore, in the former situation, the variance depends on the regression parameter, while in the latter it does not.

All these models have in common that the ratio $\left [ \partial \upsilon(\bx,\lambda,\bbe)/{\partial\lambda}\;\right]/ \,\upsilon(\bx,\lambda,\bbe)$ does not depend on $\lambda$, which is very useful for estimation purposes.

In this work, the emphasis is placed on robust inference approaches for the regression parameter $\bbe$ in the context of a situation in which the errors are heteroscedastic. Since outliers may influence the least squares estimator either through the residuals or the leverage, protection against anomalous data should be accomplished by bounding the effect of large residuals and high leverage.
To face this problem,  in Section \ref{sec:lineal}, we start by providing a brief and non--exhaustive review of existing robust estimation proposals in heteroscedastic linear regression models, i.e., when $g(\bx,\bbe) = \bx\trasp\bbe$.  We follow our discussion in Section \ref{sec:nolineal} by revisiting the extensions given when $g(\bx,\bbe)$ is a non--linear function in $\bbe$ and $\sigma(\bx)$. Section \ref{sec:propuesta} introduces two robust stepwise procedures for all the parameters involved in the heteroscedastic non--linear regression model. Our proposals utilize weighted $MM-$procedures to estimate the regression parameter. These estimators constrain large residuals by employing a score function and control high leverage points by incorporating a weight function. Additionally, our method includes a procedure to estimate the parameters of the variance function.
An appealing feature of introducing weights  is that they control the influence of high leverage points on the covariance matrix. These weights, together with   reliable estimators of the variance function,  are of special interest in procedures where the asymptotic distribution is involved, such as in testing and confidence interval problems. Section \ref{sec:montecarlo} summarizes the results of a numerical experiment conducted to evaluate the stability of the given proposals in presence of different type of outliers. Final comments are given in Section \ref{sec:comentarios}.

\section{Robust estimation in heteroscedastic models: An overview}{\label{sec:proposal}}

Let us assume that $(y_i,\bx_i\trasp)$, $1 \le i \le n$ are observations satisfying the regression model \eqref{eq:modelo0}.
In the standard framework, it is usual to assume that: i) the errors are independent and independent of the covariates, ii) $\sigma(\bx) \equiv \sigma_0$ and  iii) the errors have a normal distribution. Condition ii) is usually known as \textsl{homoscedasticity} and if, in addition, iii) holds, the least squares approach coincides with the maximum likelihood one and is the standard procedure. It is well known that ordinary least squares theory under  
heteroscedasticity leads to consistent but inefficient estimators and inconsistent covariance matrix estimators, which impact inference problems such as testing and confidence intervals. Thus, when ii) and/or iii) do not hold, alternative classical procedures appear in the literature, which may be classified into three groups:
\begin{enumerate}
	\item[\textbf{A}.] methods based on data transformation.
	
	\item[\textbf{B}.] methods based on iterative weighted least squares.
	
	\item[\textbf{C}.] methods for repeated measurements.
	
\end{enumerate}
Methods in class \textbf{A} transform the data in such a way that the new regression model has symmetric errors with constant variance. The well--known Box--Cox transformations are usually considered. However, two drawbacks of this procedure become more or less evident: the interpretation of the coefficients in the new model is not directly transferable to the untransformed one, and besides, the rescaled and untransformed models are not necessarily equivalent. \citet{Tsai:Wu:1990} warned about the sensitivity to outliers of data transformations.
Related to methods in group \textbf{A}, we can mention the approaches based on generalized linear models where the error term is not assumed to be neither symmetric nor homoscedastic. However, we will not follow this approach.

In contrast, methods in  groups  \textbf{B} and  \textbf{C} analyse heterogeneous variance data using weighted least squares  and a preliminary scale estimator. The simplest situation corresponds to \textbf{C} where  repeated measurements are available for each design point, in which case sample variances are usually considered as inverse weights in the weighted least squares estimator, see the discussion in \citet{Nanayakkara:Cressie:1991}.  The iterative weighted least squares algorithm may also be implemented when the variance function is modelled and the unknown parameters are estimated.

However, as it is well--known, all these approaches suffer from their lack of stability in the presence of vertical outliers and/or high--leverage points, making them sensitive to these anomalous observations. 
In the following sections, we describe some robust proposals given for linear and non--linear models to overcome this issue.
Even though there is a vast literature on this topic,  we only offer a partial survey of the state of the art in robust estimation for heteroscedastic linear and non--linear regression models. Both the selection of topics and the references are far from being exhaustive. Many interesting ideas and references are left out for the sake of brevity, and in advance, we apologize for the omissions made. In this sense, the most recent references have been preferred over the older ones when overlapping.

\subsection{Linear models}{\label{sec:lineal}}

Among the first contributions to robust estimators in heteroscedastic linear models, we can mention the weighted $M-$estimators given \citet{Carroll:Ruppert:1982}, while  \citet{Nanayakkara:Cressie:1991} faced the case of  repeated measurements from a robust point of view.

An approach extending the ideas of the least trimmed squares estimators to heteroscedastic linear regression was considered by 
\citet{Hadi:Luceno:1997} and \citet{Vandev:Neykov:1998}, and then it was generalized in \citet{Dimova:Neykov:2004} and \citet{Cheng:2005}. However, as mentioned in \citet{Maronna:Martin:SalibianBarrera:Yohai:2019}  these estimators have low efficiency, while according to the numerical results in \citet{Cheng:2011} the estimators of the variance function parameters have a large bias, see also \citet{Gijbels:Vrinssen:2019}.

In robust estimation, the aim is to have methods that are less sensitive to  vertical outliers, i.e., outliers in the error terms, as well as  to leverage points, i.e., observations that are outliers in the space of the covariates. With these ideas in mind,\citet{Giltinan:Carroll:Ruppert:1986} extended the Krasker--Welsch weighted estimators to heteroscedastic linear models, i.e., when \eqref{eq:modelo} holds with $g(\bx,\bbe)=\bx\trasp\bbe$.
Later on,  \citet{Bianco:Boente:Rienzo:2000} and \citet{Bianco:Boente:2002}  adapted to the presence of high leverage points the proposal given in \citet{Carroll:Ruppert:1982}. Those authors introduced a one--step version of the weighted generalized $M-$estimator that starts from initial high--breakdown point estimators of $\bbe_0$, $\bla_0$ and $\sigma_0$ and then improves the estimate of the linear parameter by performing one--step of the Newton--Raphson procedure. The  breakdown properties of the introduced robust estimators are studied, and through an extensive numerical experiment, the performances of the one--step versions and related weighted GM-estimators are compared. 

A different approach based on the forward search for weighted regression data was considered in 
\citet{Atkinson:Riani:Torti:2016}. This procedure starts with the least median of squares estimator for homoscedastic models and improves the regression estimator's efficiency by fitting the model to subsets of the data with increasing size.
	
Another approach to robust estimation in heteroscedastic linear regression adapts the ideas given by \citet{Rousseeuw:Yohai:1984}. More precisely, on the one hand, starting from an $S-$estimator for homoscedastic linear regression models \citet{Slock:VanAelst:SalibianBarrera:2013} introduced an efficient algorithm for the computation of heteroscedastic  $S-$estimators. On the other hand, to protect against high--leverage points,  \citet{Gijbels:Vrinssen:2019} introduced additional weights to define adaptive $S-$estimators pursuing, at the same time, the aim of simultaneous robust estimation and variable selection.

\subsection{Non--linear models}{\label{sec:nolineal}}

We begin this section by briefly summarizing the existing robust proposals for homoscedastic non--linear regression models. In fact, classical inference methods in this setting are based on the least squares method, which may become very unstable in the presence of outliers. 

In the homoscedastic case, many robust proposals are extensions of procedures developed for linear regression models. In fact, an early article, \citet{Fraiman:1983} proposed a general $M-$estimate of bounded influence. \citet{Stromberg:Ruppert:1992}  addressed the breakdown point concept in the non--linear regression setting and showed that for most non--linear regression functions, the breakdown point of the least squares estimator is $1/n$, where $n$ is the sample size. \citet{Stromberg:1993} introduced an algorithm to compute high breakdown estimators in non--linear regression that only requires a small amount of least squares fits to $p$ points.  \citet{Tabatabai:Argyros:1993} made an extension of $L-$estimators for both the estimation and hypothesis testing problems.
\citet{Mukherjee:1996} discussed a class of robust  minimum distance estimators. \citet{Sakata:White:2001} extended the use of $S-$estimators to non--linear models with dependent observations, while 
\citet{Cizek:2006} studied the asymptotics of the least trimmed squares estimator in non--linear models also under dependency, see also \citet{Cizek:2008}.  More recently, \citet{Fasano:2009} derived the asymptotic theory of $MM-$ and $\tau-$estimators under regular conditions, but assuming  the existence of second moments of the covariates $\bx$. Later on, \citet{Fasano:Maronna:Sued:Yohai:2012} studied the weak continuity, Fisher--consistency and differentiability of estimating functionals that correspond to high breakdown estimates in the context of linear and non--linear regression when the parametric space is compact. More recently, \citet{Bianco:Spano:2019} addressed robust estimation and inference with missing responses under milder assumptions. 

% On the other hand, Markatou and Manos (1996) and Liu \textsl{et al.} (2005) considered the problem of robust hypothesis testing. 

\citet{Davidian:Carroll:1987} focused on robust aspects in the estimation of the variance function under a  parametric  regression model, possibly non--linear.
Later on,  robust proposals under heteroscedasticity and non--linearity were considered  in \citet{ Welsh:Carroll:Ruppert:1994}, where maximum and pseudo-maximum likelihood and weighted regression
quantiles are compared and studied. For fixed carriers, \citet{Sanhueza:Sen:2004} adapted $M-$estimators to take into account a non--linear regression function and heteroscedastic errors. They obtained the asymptotic behaviour of their estimators and formulated Wald--type and likelihood ratio $M-$tests. In the context of pharmacologist studies of dose--response, where it is natural to assume a fixed design, \citet{Lim:Sen:Peddada:2010,Lim:Sen:Peddada:2012,Lim:Sen:Peddada:2013} also considered $M-$estimators.  \citet{Lim:Sen:Peddada:2012} also proposed a procedure  based on a preliminary test estimation and studied the asymptotic covariance of their proposal. Moreover, \citet{Sanhueza:Sen:Leiva:2008} dealt with a related estimator adjusted to the case of repeated measurements. 
\citet{Bianco:Spano:2019} described
a possible extension of their robust estimation method to the case of non constant
variance. In Section \ref{sec:propuesta} we modify their proposal in order to better capture the variance function.

\section{Robust stepwise procedures for non--linear models}{\label{sec:propuesta}}
In this section, we will introduce two stepwise procedures to estimate all the parameters under a heteroscedastic non--linear regression model. The first proposal is related to the one given,  for linear models, in \citet{Giltinan:Carroll:Ruppert:1986}, while the second one adapts the ideas considered   for heteroscedastic linear regression models  in  \citet{Maronna:Martin:SalibianBarrera:Yohai:2019}. % to the considered situation.

Let $(y_i, \bx_i\trasp)$, $1\le i \le n,$ be a random sample that satisfies the non--linear model
\eqref{eq:modelo},
where the errors $\epsilon_i$ are independent, independent of covariates $\bx_i$ and identically distributed with a symmetric distribution around $0$. It is worth noticing that this symmetry assumption on the errors distribution is usual in robust regression as a way  to circumvent the stronger assumption of a zero mean of the errors, since no moments conditions are preferred in this context. In order to warranty the identifiability of the model, we assume that
\begin{eqnarray}
\prob\big(g(\bx,\bbe_0) = g(\bx,\bbe) \big) < 1 \, \mbox{ for all } \bbe \neq \bbe_0 \, . \label{eq:identifica}
\end{eqnarray}
Henceforth, we will assume that the variance function is parametrically modelled and satisfies
\begin{equation}
\label{eq:upsilon}
\upsilon(\bx_i,\bla,\bbe)=\exp\{\bla\trasp \bh(\bx_i,\bbe)\}\,, \qquad \bla \in \real^q\,. 
\end{equation}
In the linear case, that is when $g(\bx,\bbe)=\bx\trasp \bbe$, \citet{Giltinan:Carroll:Ruppert:1986} introduced generalized $M-$estimators in the heteroscedastic case as the solution $(\wbbe, \wblam, \wsigma)$ of the system of equations
\begin{align*}
\frac 1n \sum^n_{i=1}\Psi \left(\frac
{y_i-\bx_i\trasp\wbbe}{\wsigma
\upsilon(\bx_i,\wblam,\wbbe)}\right) w_1 \left(\frac{\bx_i}{\upsilon(\bx_i,\wblam,\wbbe)}\right) \frac {\bx_i}{\upsilon(\bx_i,\wblam,\wbbe)}&= 0 \\
\frac 1n \sum^n_{i=1} \chi \left(\frac {y_i-\bx_i\trasp\wbbe}{\wsigma \upsilon(\bx_i,\wblam,\wbbe)}\right)&=  0 \\
\frac 1n \sum^n_{i=1} \chi \left(\frac {y_i-\bx_i\trasp \wbbe}{\wsigma \upsilon(\bx_i,\wblam,\wbbe)}\right) w_2(\bh(\bx_i,\wbbe))\bh(\bx_i,\wbbe)&= 0\;. 
\end{align*}
Usually, $\chi(u)=\rho_c(u)-b$, where $c>0$ is a user chosen tuning constant and $\min(b,1-b)$ is the breakdown point of the $M-$scale estimator when $\|\rho_c\|_{\infty}=1$. When $b=1/2$ and  $\rho_c=\rho_{\,\tuk,\,c}$  with $\rho_{\,\tuk,\,c}(t) =\min\left(1 - (1-(t/c)^2)^3, 1\right)$   the Tukey's bisquare function,   the choice  $c=1.54764$  leads to Fisher--consistent scale estimators under normality with 50\% breakdown point. On the other hand,  $\Psi$ is a bounded score function, such as the Huber or Tukey's functions. The weight functions $w_1$ and $w_2$ are nonnegative functions that downweight the leverage of the carriers.

Taking into account that linear regression $MM-$estimators are still consistent under heteroscedasticity, to improve their efficiency,  \citet{Maronna:Martin:SalibianBarrera:Yohai:2019} suggested to compute first an initial $MM-$regression estimator, $\wbbe_{\ini}$, as if the linear model were homocedastic, and from it the residuals $r_i(\wbbe_{\ini})=y_i-\bx_i\trasp\wbbe_{\ini}$. Using these residuals, the parameters $\sigma_0$ and $\bla_0$ are estimated by a robust linear fit of  $\log (r_i(\wbbe_{\ini}))$ on $\bh(\bx_i,\wbbe_{\ini})$. In their proposal, the final $MM-$estimator is obtained through a robust $MM-$linear regression estimator for an homoscedastic regression model based on the transformed variables $y_i^{\star}= y_i/\upsilon(\bx_i,\wblam,\wbbe_{\ini})$ and  $\bx_i^{\star}= \bx_i/\upsilon(\bx_i,\wblam,\wbbe_{\ini})$, where $\wblam$ stands for the estimator of $\bla_0$.

In order to adapt the previous proposals to the non--linear setting,  notice that model \eqref{eq:modelo} can be easily transformed in an homoscedastic model as
$$
\frac{y}{\upsilon(\bx,\bla_0,\bbe_0)}= \frac{g(\bx,\bbe_0)}{\upsilon(\bx,\bla_0,\bbe_0)}+ \sigma _0 \; \epsilon \, . 
$$
In other words,  if we define the pseudo--observations and  the pseudo--regression function as
$$y^*= \frac{y}{\upsilon(\bx,\bla_0,\bbe_0)}\qquad \mbox{and}\qquad g^*(\bx,\bbe_0)= \frac{g(\bx,\bbe_0)}{\upsilon(\bx_i,\bla_0,\bbe_0)} \; ,$$
the transformed model reduces to the homoscedastic non--linear model given by
\begin{equation}
y^*= g^*(\bx,\bbe_0)+ \sigma_0  \; \epsilon \, . 
\label{eq:modelohomo}
\end{equation}
Note that in \eqref{eq:modelohomo}, for clarity, we do not make explicit the dependency on $\bla_0$ of the pseudo--observation nor the pseudo--regression function.
Equation  \eqref{eq:modelohomo}  suggests that given preliminary estimators of $\bbe_0$ and  $\bla_0$, after  plugging  these initial estimators in the   function $\upsilon(\bx,\bla,\bbe)$, a final robust estimator of the regression parameter $\bbe_0$ can be obtained using any robust estimation method valid for a homoscedastic non--linear model.  

To simplify the notation, when possible, we denote
$$r_i(\bbe, \bla)= \frac{y_i-g(\bx_i,\bbe)}{ \upsilon(\bx_i,\bla,\bbe)} \,.$$
Focusing on the \citet{Giltinan:Carroll:Ruppert:1986} proposal and taking into account that the weighted $MM-$estimators defined in \citet{Bianco:Spano:2019} for the homoscedastic case are still consistent for non--homogeneous variance errors, the following stepwise procedure can be implemented.

\begin{itemize}
\item[\textbf{Step 1}.] Compute an initial robust estimator of $\bbe_0$, $\wbbe_{\ini}$, as if the model were homoscedastic. $MM-$estimators and their weighted versions are possible candidates, which correspond to our choice in the numerical study described in Section \ref{sec:montecarlo} and will be denoted henceforth by \MM ~ and \WMM, respectively.

\item[\textbf{Step 2}.] Obtain the estimators, $(\wsigma, \wblam)$, of the nuisance parameters $(\sigma_0, \bla_0)$, as the solution of the following system  
\begin{align*}
\frac 1n \sum^n_{i=1} \chi \left(\frac {r_i(\wbbe_{\ini}, \wblam)}{\wsigma }\right)&=  0 \\
\frac 1n \sum^n_{i=1} \chi \left(\frac{r_i(\wbbe_{\ini}, \wblam)}{\wsigma }\right) 
w_2(\bh(\bx_i,\wbbe_{\ini}))\bh(\bx_i,\wbbe_{\ini})&= 0\;, 
\end{align*}
where $\chi(u)=\rho_0(u)-b$ with $\rho_0$ a $\rho-$function as defined in \citet{Maronna:Martin:SalibianBarrera:Yohai:2019}. The resulting estimators of   $\sigma_0$ and $\bla_0$ will be labelled as $ \wsigma_{\mm}$ and $ \wblam_{\mm}$ when $MM-$estimators are used in \textbf{Step 1} and as  $ \wsigma_{\wmm}$ and $ \wblam_{\wmm}$ when weighted $MM-$estimators are considered.

\item[\textbf{Step 3}.] Calculate the pseudo--variables and the pseudo--regression function
$$y_i^*= \frac{y_i}{\upsilon(\bx_i,\wblam,\wbbe_{\ini})} \qquad \mbox{and} \qquad 
g^*(\bx,\bbe)= \frac{g(\bx,\bbe)}{\upsilon(\bx_i,\wblam,\wbbe_{\ini})} \; ,$$
to compute a robust estimator for the non--linear model \eqref{eq:modelohomo}.  

\item[] For instance, one may consider again the weighted $MM-$ estimators introduced in \citet{Bianco:Spano:2019}, 
where the  proposed estimator  $\wbbe$ is defined as
\begin{eqnarray}
\wbbe &= & \argmin_{\bbech}\frac 1 n\sum_{i=1}^n  \rho_1\left(\frac{y_i-g(\bx_i,\bbe)}{\wsigma \upsilon(\bx_i,\wblam,\wbbe_{\ini})}\right)  w(\bx_i)\;, \label{eq:betaest}
\end{eqnarray} 
where,  $\wsigma$ and $\wblam$ are the estimators obtained in   \textbf{Step 2}, $\rho_1$ is a $\rho-$function   such that $\rho_1\le \rho_0$ and $w$ is a weight function that bounds the effect of high leverage points. These estimators will be denoted as \HWMM ~  for a general weight function, while they will be indicated as \HMM ~ if $w(\bx)\equiv 1$ and $w_2$ in \textbf{Step 2}  equals $w_2(\bh(\bx,\bbe))\equiv 1$.

\item[\textbf{Step 4}.] To improve the performance of the variance parameter estimators computed in \textbf{Step 2} and provide more stable estimators of the variance function, we go further on. Denote  $\wbbe$ the estimator obtained in \textbf{Step 3} and define $z_i= \log(\,|y_i- g(\bx_i,\wbbe)\,\,|\,)$ and $\bv_i=\bh(\bx_i,\wbbe)$.
\item[] Compute a linear regression estimators $ \walfa$ and $ \wblam$ for the parameters $\alpha_0$ and $\bla_0$ in the  pseudo--regression model $z_i= \alpha_0+ \bla_0\trasp \bv_i + u_i$. From now on,   $ \wblam_{\hmm}$ stands for the estimators of $\bla_0$ when the \HMM ~ estimators of $\bbe$ defined in \textbf{Step 3} are used and  $ \wblam_{\hwmm}$ when the weighted counterparts,  \HWMM,  are considered.
\end{itemize}

As shown in the numerical results reported in Section \ref{sec:montecarlo}, the estimators $\wblam_{\mm}$  and $ \wblam_{\wmm}$ computed in \textbf{Step 2} may be biased even under the central Gaussian model. Providing reliable estimators of the variance function plays an important role when computing the asymptotic standard deviations of the estimators defined through \textbf{Step 3}. This is the motivation for including  a fourth step in the stepwise procedure described above.
The method described in \textbf{Step 4} is motivated by the fact that since $y- g(\bx,\bbe_0) = \sigma_0 \; \exp\{\bla_0\trasp \bh(\bx,\bbe)\}\; \epsilon$, after taking logarithm we get $z= \log(\,|y- g(\bx,\bbe_0)\,\,|\,)= \log(\sigma_0)+ \bla_0\trasp \bh(\bx,\bbe_0)+ \log(|\epsilon|)$, meaning that we are in the presence of a linear regression model with covariates $\bv=\bh(\bx,\bbe_0)$ and asymmetric errors $u=\log(|\epsilon|)$.  As it is well known, see Section 4.9.2 in \citet{Maronna:Martin:SalibianBarrera:Yohai:2019}, $MM-$regression estimators provide consistent estimators of the slope (but not necessarily  of the intercept) even in the presence of asymmetric errors and for that reason, the given estimators of $\bla_0$ are appropriate.  

It is worth mentioning that an important difference between our first proposal and the methodology  suggested in \citet{Bianco:Spano:2019} lies in \textbf{Step 2} where we simultaneously estimate $\sigma_0$ and $\bla_0$.

Our second approach is inspired by the proposal given for linear regression models in Section 5.12.2.1 of \citet{Maronna:Martin:SalibianBarrera:Yohai:2019}. In contrast to their proposal, we suggest to take into account that $MM-$regression estimators of the intercept in the transformed model (via the logarithm function) may be biased, even under normal errors. Hence, a modification is needed to provide a consistent scale estimator. Therefore, our second proposal corresponds to the following iterative process:

\begin{itemize}
\item[\textbf{Step N1}.] This step corresponds to \textbf{Step  1} above, that is, we first compute an initial robust estimator of $\bbe_0$, $\wbbe_{\ini}$, as if the model were homoscedastic, i.e., $\upsilon(\bx,\bla,\bbe) \equiv 1$. The resulting estimators will be denoted, as above, as  \MM ~ or \WMM, when the
 $MM-$ or weighted $MM-$estimators defined in  \citet{Bianco:Spano:2019} are computed, respectively.

\item[\textbf{Step N2}.] Define $z_i= \log(\,|y_i- g(\bx_i,\wbbe_{\ini})\,\,|\,)$ and $\bv_i=\bh(\bx_i,\wbbe_{\ini})$.
\begin{itemize}
\item[a)] Compute   linear regression $MM-$estimators $ \walfa$ and $ \wblam$ for the parameters $\alpha_0$ and $\bla_0$ in the  pseudo--regression model $z_i= \alpha_0+ \bla_0\trasp \bv_i + u_i$. 
\item[b)] Compute the estimator $\wsigma$ of $\sigma$ as an $S-$scale of the residuals $r_i(\wbbe_{\ini}, \wblam)$, that is, as the solution of
$$\frac 1n \sum^n_{i=1} \chi \left(\frac{r_i(\wbbe_{\ini}, \wblam)}{\sigma }\right)= 0  \;, $$
where $\chi(u)=\rho_0(u)-b$ as in \textbf{Step 2}. 
\end{itemize}
 
\item[]  
 The resulting estimators of   $\sigma_0$ and $\bla_0$ will be labelled as $ \wsigma_{\mmnew}$ and $ \wblam_{\mmnew}$ when $MM-$estimators are used in \textbf{Step N1} and as  $ \wsigma_{\wmmnew}$ and $ \wblam_{\wmmnew}$ when when the weighted counterparts,  \WMM,  are considered.

\item[\textbf{Step N3}.] Calculate the pseudo--variables  and the pseudo--regression function
$$y_i^*= \frac{y_i}{\upsilon(\bx_i,\wblam,\wbbe_{\ini})}\qquad \mbox{and} \qquad g^*(\bx,\bbe)= \frac{g(\bx,\bbe)}{\upsilon(\bx_i,\wblam,\wbbe_{\ini})} \; ,$$
to compute a robust estimator for the model \eqref{eq:modelohomo} as 
\begin{eqnarray}
\wbbe &= & \argmin_{\bbech}\frac 1 n\sum_{i=1}^n  \rho_1\left(\frac{y_i-g(\bx_i,\bbe)}{\wsigma \upsilon(\bx_i,\wblam,\wbbe_{\ini})}\right)  w(\bx_i)\;, \label{eq:betaestnew}
\end{eqnarray}  
where as above, $\wsigma$ and $\wblam$ are the estimators obtained in   \textbf{Step N2}, $\rho_1$ is a $\rho-$function   such that $\rho_1\le \rho_0$ and $w$ is a weight function that bounds the effect of high leverage points. These estimators will be denoted as \HWMM$_{\nuevo}$ ~  for a general weight function, while they will be indicated as \HMM$_{\nuevo}$   if $w(\bx)\equiv 1$.

\item[\textbf{Step N4}.] This is the counterpart of \textbf{Step 4} and corresponds to apply \textbf{Step N2} to the estimators computed in   \textbf{Step N3}. More precisely, let  $\wbbe$ be the estimator obtained in \textbf{Step N3} and define $z_i= \log(\,|y_i- g(\bx_i,\wbbe)\,\,|\,)$ and $\bv_i=\bh(\bx_i,\wbbe)$.
\begin{itemize}
\item[a)] Compute a linear regression estimators for the parameters $\alpha_0$ and $\bla_0$ in the  pseudo--regression model $z_i= \alpha_0+ \bla_0\trasp \bv_i + u_i$. Denote $ \wblam_{\nuevo}$ the resulting estimator of $\bla_0$.

\item[b)] Compute the estimator $\wsigma_{\nuevo}$ of $\sigma$ as an $S-$scale of the residuals $r_i(\wbbe, \wblam_{\nuevo})$, that is, $\wsigma_{\nuevo}$  satisfies
$$\frac 1n \sum^n_{i=1} \chi \left(\frac
{r_i(\wbbe, \wblam_{\nuevo})}{\wsigma_{\nuevo}}\right)= 0  \;, $$
where $\chi(u)=\rho_0(u)-b$ as in \textbf{Step N2}.  
\end{itemize}
\item[]  
 The resulting estimators of  $\sigma_0$ and $\bla_0$ will be labelled as $ \wsigma_{\hmmnew}$ and $ \wblam_{\hmmnew}$ when  estimator of $\bbe$ denoted \HMM$_{\nuevo}$ is used in \textbf{Step N3} and as  $ \wsigma_{\hwmmnew}$ and $ \wblam_{\hwmmnew}$ when  its weighted counterpart,  \HWMM$_{\nuevo}$,  is considered.

\end{itemize}

\section{Numerical experiments}{\label{sec:montecarlo}}
In this section, we report the results of a numerical study conducted to analyse the stability and sensitivity of the proposal given in Section \ref{sec:propuesta}, when atypical data arise in the sample. For that purpose, we generate synthetic data using a heteroscedastic exponential model given by 
\begin{equation}
	y_i= \beta_{01} \; \exp(\beta_{02} x_i )+ \exp(\lambda_0 (x_i+1)^2) \; \epsilon_i\, , 1 \le i \le 100\, ,
	 \label{eq:modelosimu}
\end{equation}
where $\beta_{01}=5$,  $\beta_{02}=2$, $\lambda_0=1$ and for $1 \le i \le 100\,$ the covariates $x_i \sim \itU(0,1)$ are independent of the errors $\epsilon_i \sim \itN(0,1)$.  Note that, for the considered model, the scale function in \eqref{eq:upsilon} corresponds to 
$\upsilon(x,\lambda,\bbe)=\exp\{\lambda (x+1)^2\}$, that is, $h(x,\bbe)=h(x)=(x+1)^2$ does not depend on the regression parameter and $\sigma_0=1$. For that reason, when computing the weights $w_2$ in \textbf{Step 2}, it is enough to weight the covariates.

Our simulation results are based on $Nrep=1000$ replications and $C_0$ stands for these generated clean samples.

We evaluate the performance of the following estimators: the classical and robust procedures computed as if the model were homoscedastic,  and their weighted counterparts adapted to heteroscedasticity. More precisely, with respect to the classical procedures, we compute the least squares estimator, denoted \LS ~ in all tables and figures, while the weighted least squares one, denoted \HLS, corresponds to a weighted version  using an estimator of the variance function. These estimators were evaluated using the function \texttt{nlreg} available at the library \texttt{nlreg} in \texttt{R}. The robust estimators computed assuming a homoscedastic model as described  in \textbf{Step 1} will be indicated \MM ~ and \WMM. The first ones were obtained using the function \texttt{nlrob} from the library \texttt{robustbase}, and their weighted counterparts were obtained using the code implemented by \citet{Bianco:Spano:2019}. We also compute their heteroscedastic counterparts defined through \textbf{Step 1} to \textbf{Step 3} that will be labelled \HMM ~ and \HWMM, when the weights equal 1 in the former case or when $w(x)$ in \eqref{eq:betaest} and $w_2(h(x))$ in \textbf{Step 2} were computed using the bisquare weight function as
\begin{equation}
w_2(h(x))=w(x)=w_{\tuk,c}\left(\frac{(x-\wmu_x)^2}{\wsigma_x^2}\right)\,,
\label{eq:pesos}
\end{equation}
with $\wmu_x=\median_{1\le i\le n} x_i $,  $\wsigma_x=(4/\sqrt{12})\;\mad_{1\le i\le n}(x_i) $, $c$ the 0.95 quantile of the $\chi_1^2-$distribution and $w_{\tuk,c}(t)=\psi_{\tuk,c}(t)/t$, for $t\ne 0$ and $w_{\tuk,c}(0)=1$ with $\psi_{\tuk,c}(t)=\rho_{\tuk,c}^{\prime}(t)$. The constant $4/\sqrt{12}$ is included to ensure that $\wsigma_x$ is consistent to the standard deviation of a uniform $\itU(0,1)$ distribution. As mentioned in Section \ref{sec:propuesta}, the estimators obtained through  \textbf{Steps N2} to \textbf{N4} will be labelled with a subscript \textsc{n}.

The loss functions $\rho_0$ and $\rho_1$ used  to compute robust estimators  correspond to   $\rho_j= \rho_{\tuk,c_j}$ with   $\rho_{\,\tuk,\,c}(t) =\min\left(1 - (1-(t/c)^2)^3, 1\right)$   the Tukey's bisquare function, $c_0=1.54764$ and $c_1=4.75$.  

To assess the impact of atypical observations  on the estimation of the parameters, we introduce  \textsl{outliers of different kinds}. Our purpose is to explore the stability of different estimation methods in the presence of anomalous data which are potentially harmful. The study is separated into two parts: one focused on vertical outliers and the other on leverage outliers of varying magnitudes. The findings from each part are described in Sections \ref{sec:vertical} and \ref{sec:highlev}, respectively.

\subsection{Vertical Outliers}
\begin{figure}[ht!]
	\begin{center}
		\renewcommand{\arraystretch}{0.1}
				\begin{tabular}{cc}
			$C_0$  & $C_{1}$ \\[-3ex]
			\includegraphics[scale=0.35]{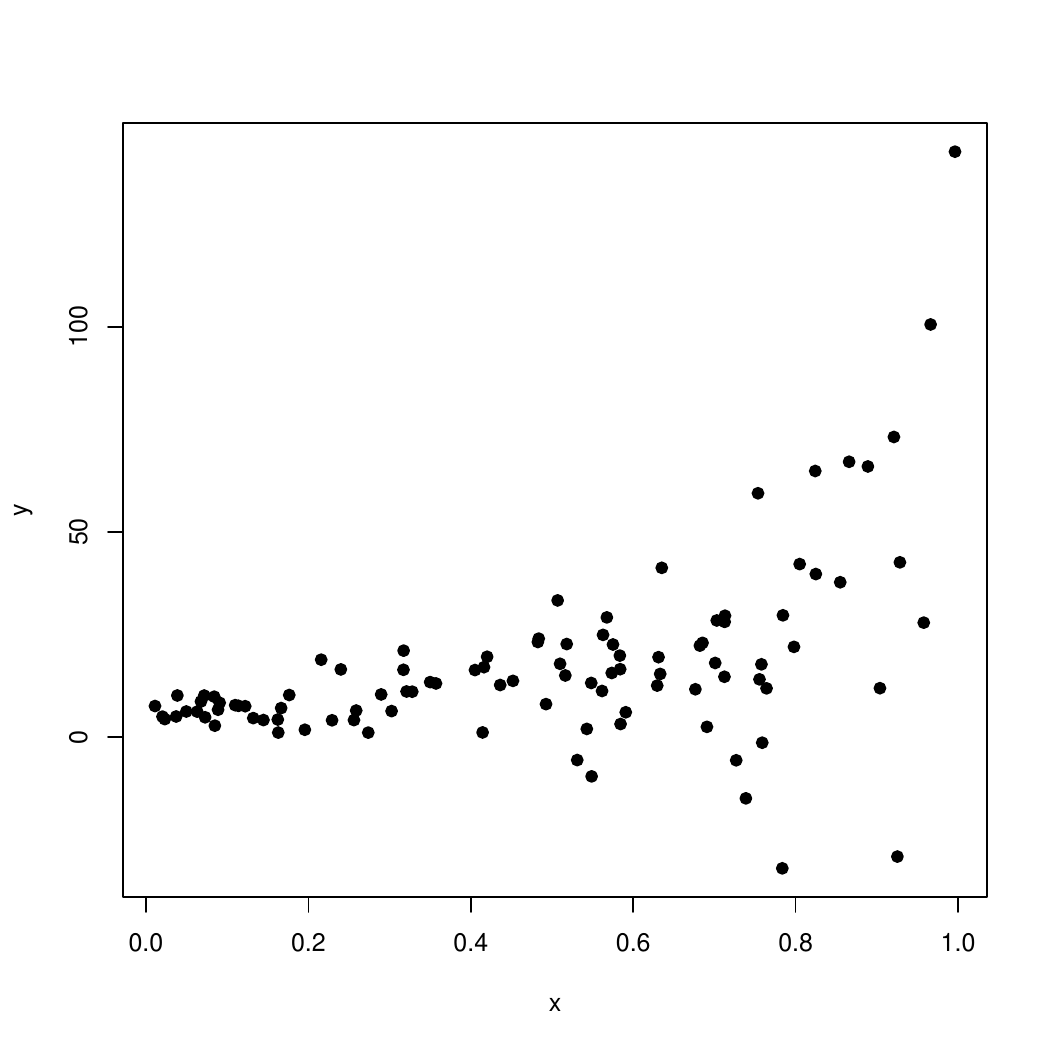} &
			\includegraphics[scale=0.35]{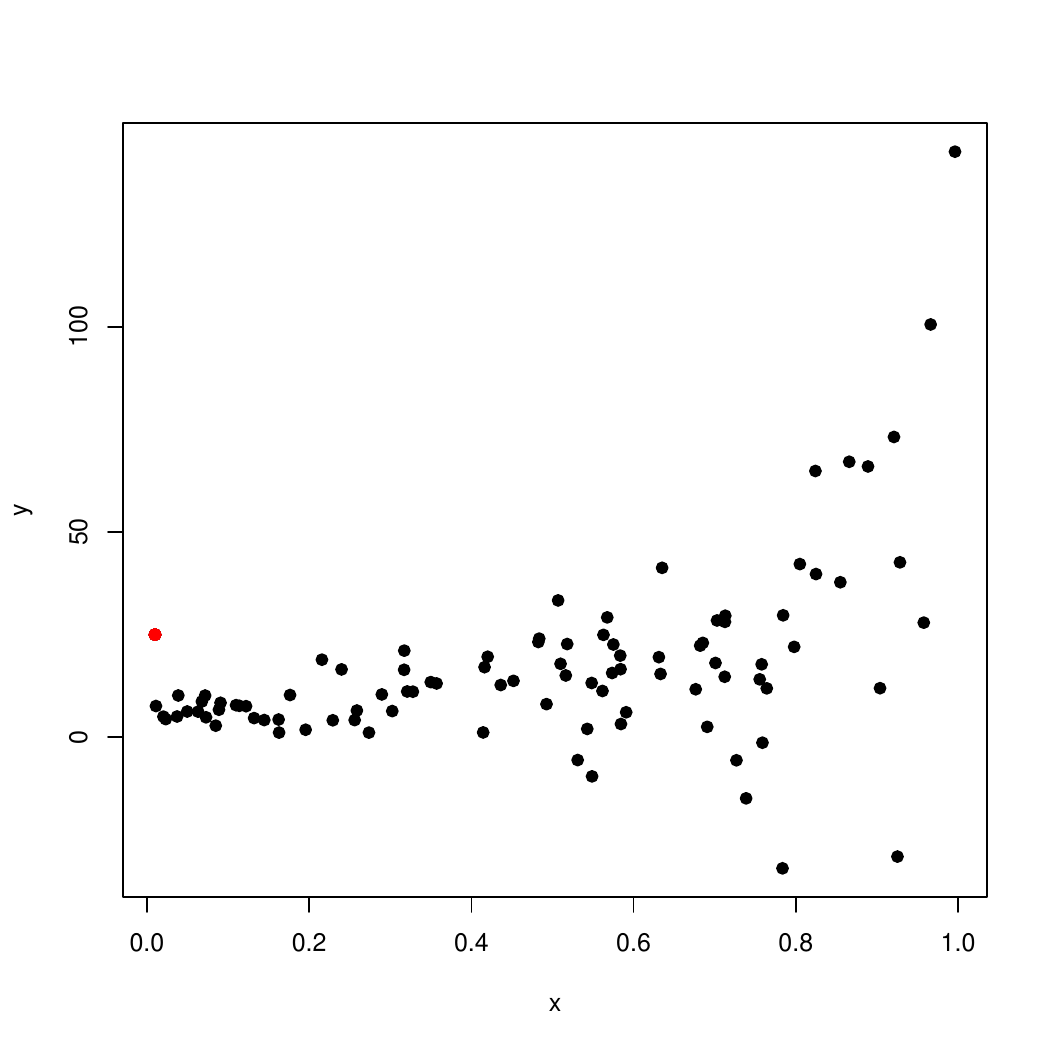} \\
			$C_2$  & $C_{3}$ \\[-3ex]
	 		\includegraphics[scale=0.35]{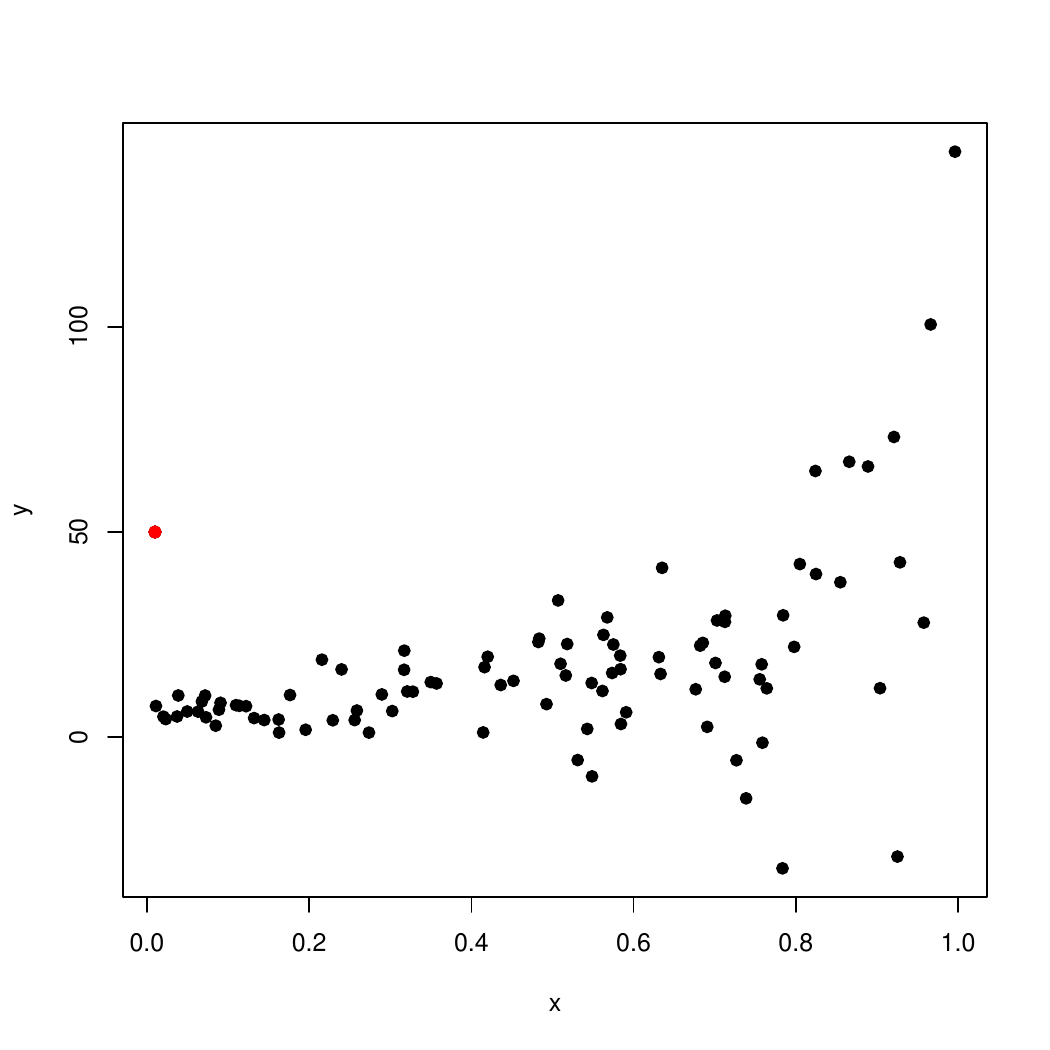} &
	 		\includegraphics[scale=0.35]{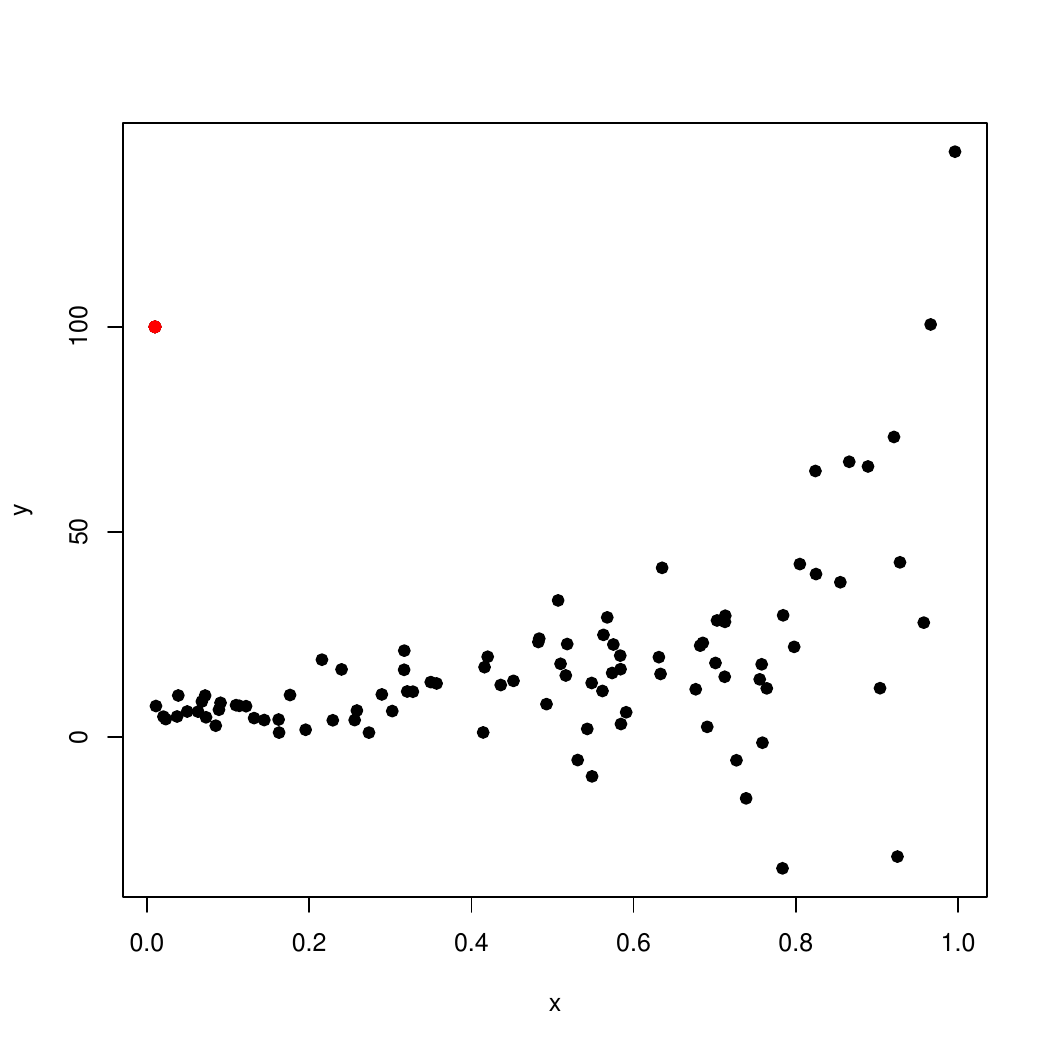} 
		\end{tabular}
		\vskip-0.2in \caption{ \small \label{fig:datos_simulados}  Synthetic data for one of the considered replications: the sample displayed in the upper left panel was generated under scheme $C_0$, the other three panels correspond to contaminations $C_{1}$, $C_{2}$ and $C_{3}$.}
	\end{center} 
\end{figure}

{\label{sec:vertical}} %(\textcolor{magenta}{OJO: corresponde a las salidas 7 conbis2})
In the case of vertical outliers, the last 5\% of observations in the sample were changed to $(x_0,y_0)$, where $x_0=0.01+u$, $u$ is a noise and $y_0$ equals $25$, $50$, or $100$. This created three different contamination settings, which were named $C_1$, $C_2$, and $C_3$. As it is well known, including several repeated points $(x_0,y_0)$  may lead to the numerical instability of the algorithms used
to compute these estimators. For that reason, we have added the  small noise $u$  generated as $N(0,10^{-8})$ to the value $0.01$. 
The panels in Figure \ref{fig:datos_simulados} illustrate the magnitude of the introduced  vertical outliers in one generated sample. 

 Tables \ref{tab:ECM-BIAS-beta01HO} and \ref{tab:ECM-BIAS-beta02HO} show the empirical mean square error (MSE) and  bias for the considered estimators of $\beta_{0j}$, $j=1,2$, for clean and contaminated data.
The reported results  reveal  the gain in efficiency under $C_0$ of  the estimators of the regression parameters   when computed taking into account the heteroscedasticity, that is, when using \HLS, \HMM ~and \HWMM. This effect also becomes evident from the top panels of  Figure \ref{fig:betaH0-C0-C3}. It is worth mentioning that the loss of efficiency of \HMM ~and \HWMM ~ with respect to \HLS ~ is less that 5\%.

\begin{table}[ht!]
\begin{center}
\renewcommand{\arraystretch}{1.2}
    \setlength{\tabcolsep}{4pt} % CAMBIAR POR 2pt SI HACE FALTA ACHICAR 
    \begin{tabular}{|c|c|c|c|c|c|c|c|c|}
	\hline
	& \multicolumn{4}{c|}{MSE} &  \multicolumn{4}{c|}{Bias}\\
	\hline
	& $C_{0}$& $C_{1}$ & $C_{2}$ & $C_{3}$ & $C_{0}$& $C_{1}$ & $C_{2}$ & $C_{3}$ \\
	\hline
\LS & 1.953 & 2.662 & 13.211 & 147364.221  & -0.012 & 1.239 & 3.963 & 9118.835 \\ 
\MM & 1.326 & 1.987 & 1.461 & 1.375 & 0.049 & 1.278 & 0.184 & 0.075 \\ 
\WMM & 1.448 & 2.316 & 1.659 & 1.503 & 0.022 & 1.341 & 0.172 & 0.026\\ 
\hline
\HLS & 0.637 & 3.963 & 15.368 & 84.373 & 0.043 & 3.717 & 7.846 & 25.091\\ 
\HMM & 0.646 & 0.851 & 0.681 & 0.682 & 0.036 & 0.206 & 0.042 & 0.040\\ 
\HWMM & 0.659 & 0.927 & 0.703 & 0.699  & 0.033 & 0.235 & 0.039 & 0.034\\ 	
\hline
\HMM$_{\nuevo}$ & 0.648 & 0.823 & 0.668 & 0.666 & 0.039 & 0.189 & 0.046 & 0.044 \\ 
\HWMM$_{\nuevo}$ & 0.656 & 0.911 & 0.682 & 0.674 & 0.037 & 0.228 & 0.044 & 0.041 \\ \hline
  \end{tabular}
\vskip-0.1in
\caption{ \small \label{tab:ECM-BIAS-beta01HO} MSE and bias of the estimators of $\beta_{01}$. }
\end{center}
\end{table}
 
\begin{table}[ht!]
\begin{center}
\renewcommand{\arraystretch}{1.2}
    \setlength{\tabcolsep}{4pt} % CAMBIAR POR 2pt SI HACE FALTA ACHICAR 
    \begin{tabular}{|c|c|c|c|c|c|c|c|c|}
	\hline
	& \multicolumn{4}{c|}{MSE} &  \multicolumn{4}{c|}{Bias}\\
	\hline
	& $C_{0}$& $C_{1}$ & $C_{2}$ & $C_{3}$ & $C_{0}$& $C_{1}$ & $C_{2}$ & $C_{3}$ \\
	\hline
\LS & 0.978 & 1.014 & 6.713 & 112.143 & 0.113 & -0.173 & -0.808 & -28.635\\ 
\MM & 0.654 & 0.753 & 0.680 & 0.666 & 0.005 & -0.339 & -0.031 & -0.005\\ 
\WMM & 0.703 & 0.836 & 0.750 & 0.731 & 0.029 & -0.342 & -0.005 & 0.031 \\ 
\hline 
\HLS & 0.294 & 1.072 & 10.004 & 26.915 & -0.021 & -0.937 & -2.162 & -6.444\\ 
\HMM & 0.300 & 0.363 & 0.320 & 0.321 & -0.017 & -0.071 & -0.019 & -0.019\\ 
\HWMM & 0.305 & 0.384 & 0.328 & 0.327 & -0.014 & -0.079 & -0.017 & -0.015\\
\hline 
\HMM$_{\nuevo}$ & 0.301 & 0.356 & 0.311 & 0.310 & -0.018 & -0.068 & -0.021 & -0.021 \\ 
\HWMM$_{\nuevo}$ & 0.303 & 0.379 & 0.317 & 0.313 & -0.016 & -0.079 & -0.020 & -0.018 \\ 
\hline  
\end{tabular}
\vskip-0.1in
\caption{ \small \label{tab:ECM-BIAS-beta02HO} MSE  and bias of the estimators of $\beta_{02}$. }
\end{center}
\end{table}

\begin{figure}[ht!]
	\begin{center}
 	 \renewcommand{\arraystretch}{0.4}
 \newcolumntype{M}{>{\centering\arraybackslash}m{\dimexpr.1\linewidth-1\tabcolsep}}
   \newcolumntype{G}{>{\centering\arraybackslash}m{\dimexpr.45\linewidth-1\tabcolsep}}
%\begin{tabular}{MGG}
 
\begin{tabular}{M GG}
			& \small $\wbeta_{01}$  & \small $\wbeta_{02}$ \\[-4ex]
			$C_0$ & 
			\includegraphics[scale=0.33]{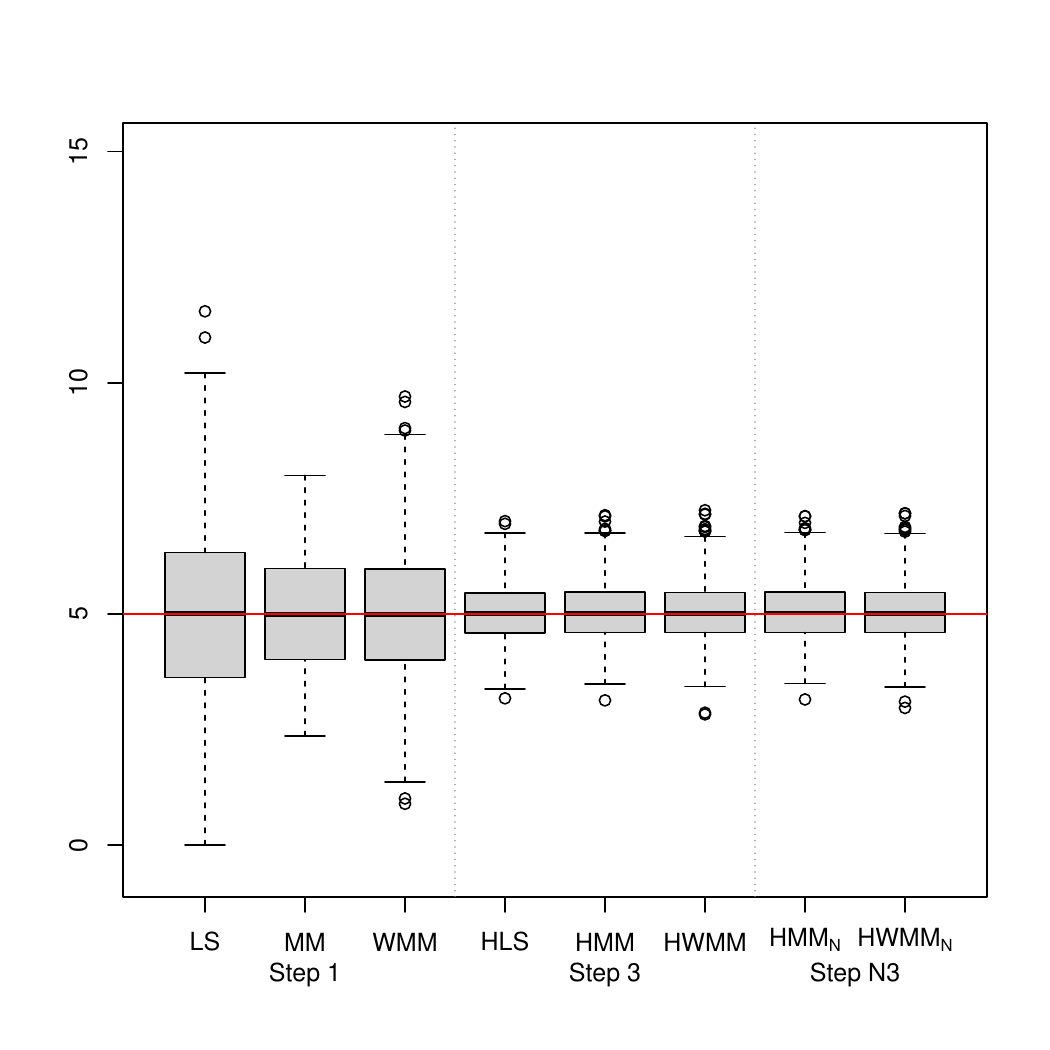} & 
			\includegraphics[scale=0.33]{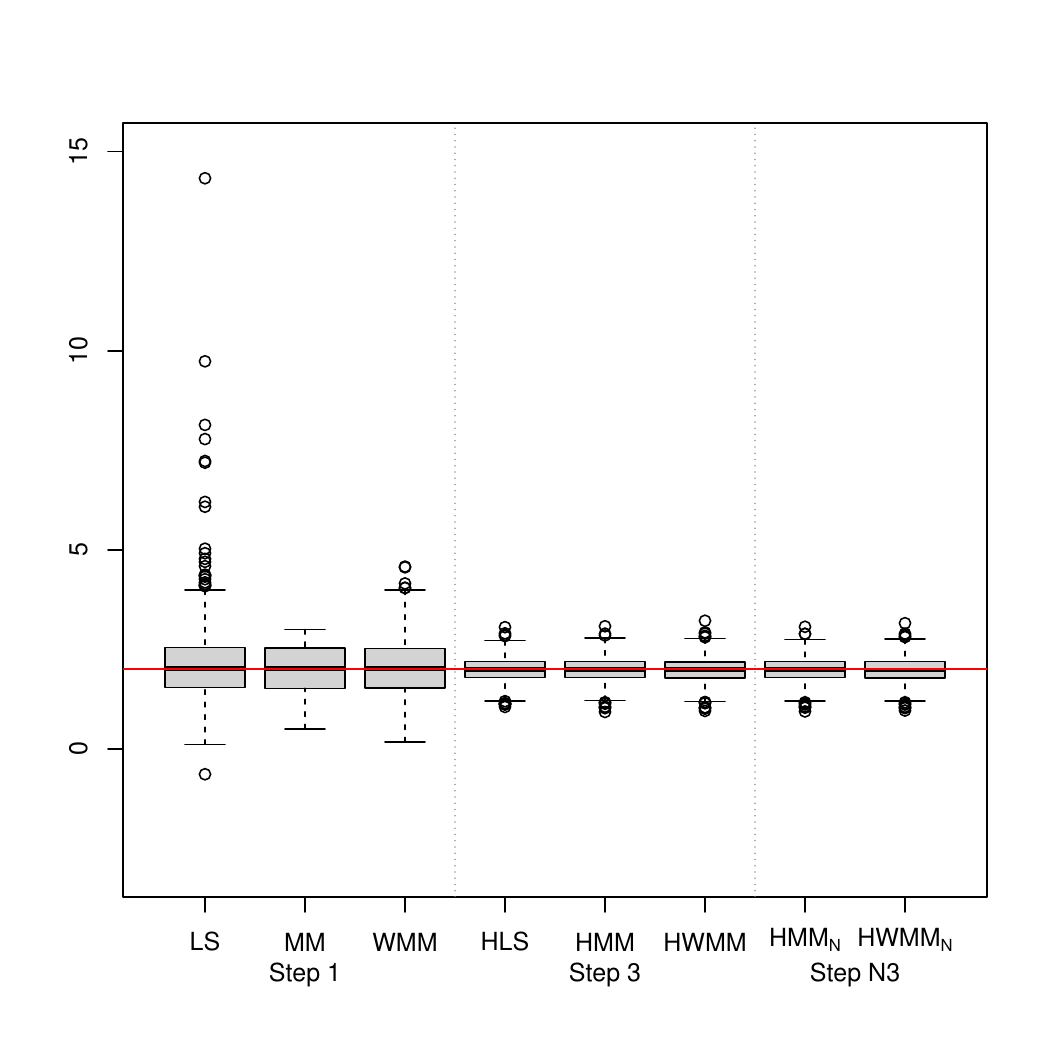} \\[-6ex]
			$C_1$ &  
			\includegraphics[scale=0.33]{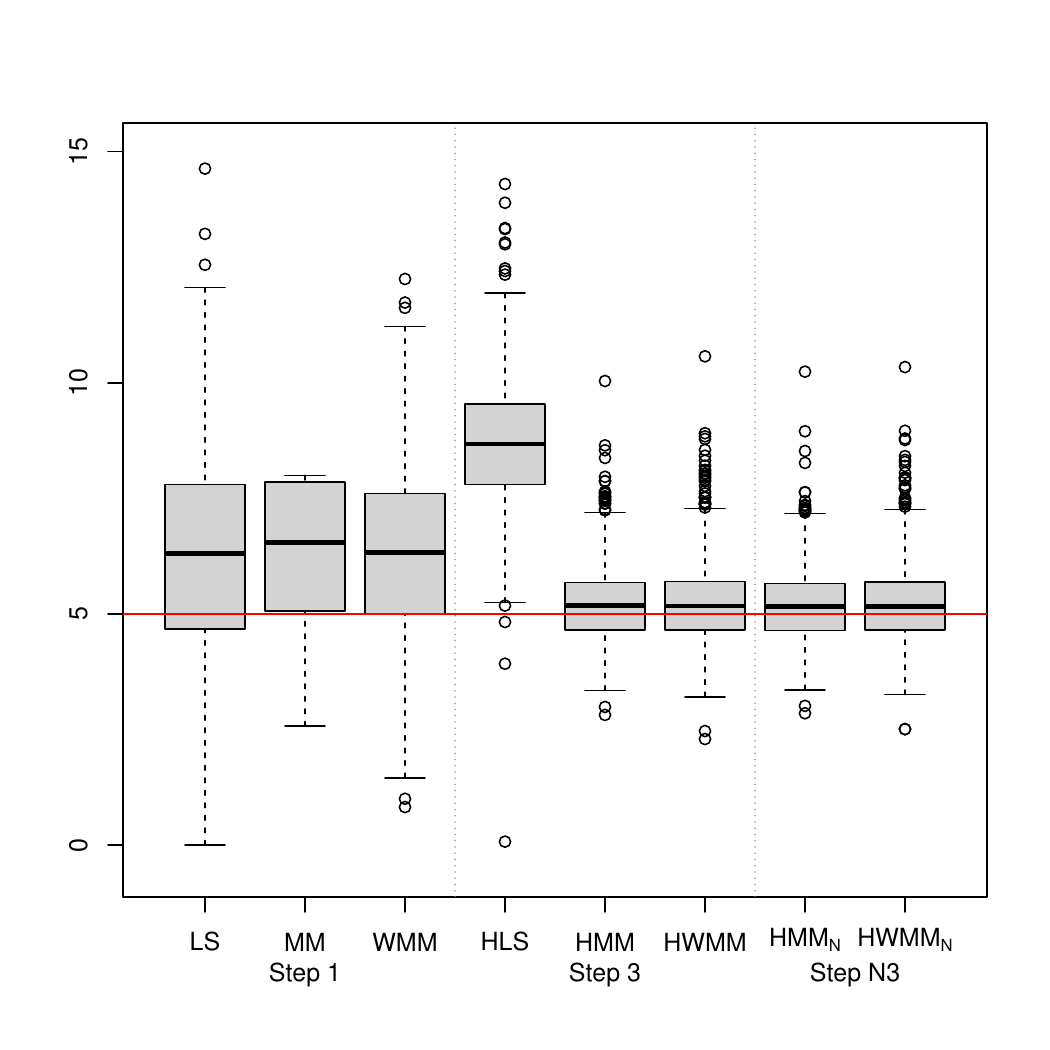} &  
			\includegraphics[scale=0.33]{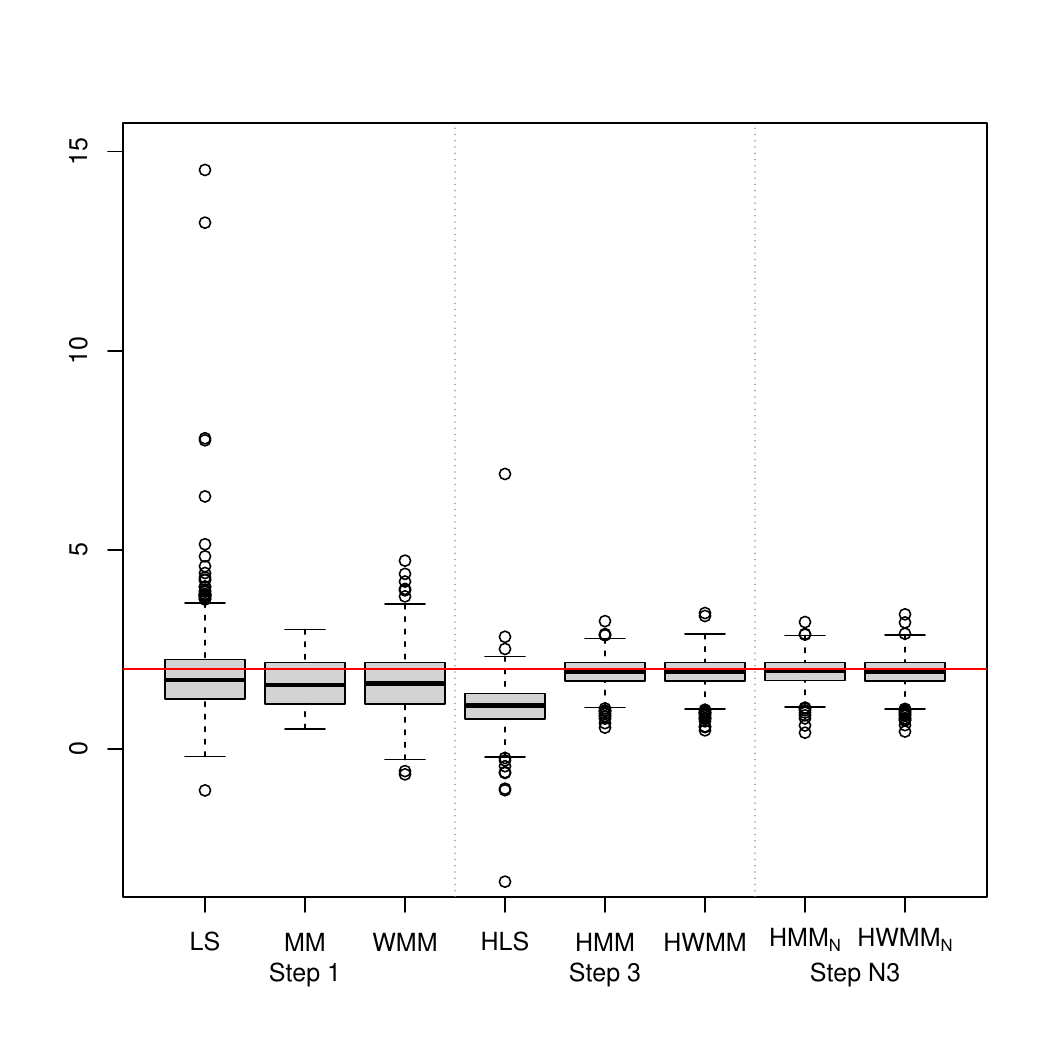} \\[-6ex]
			$C_2$ & 
			\includegraphics[scale=0.33]{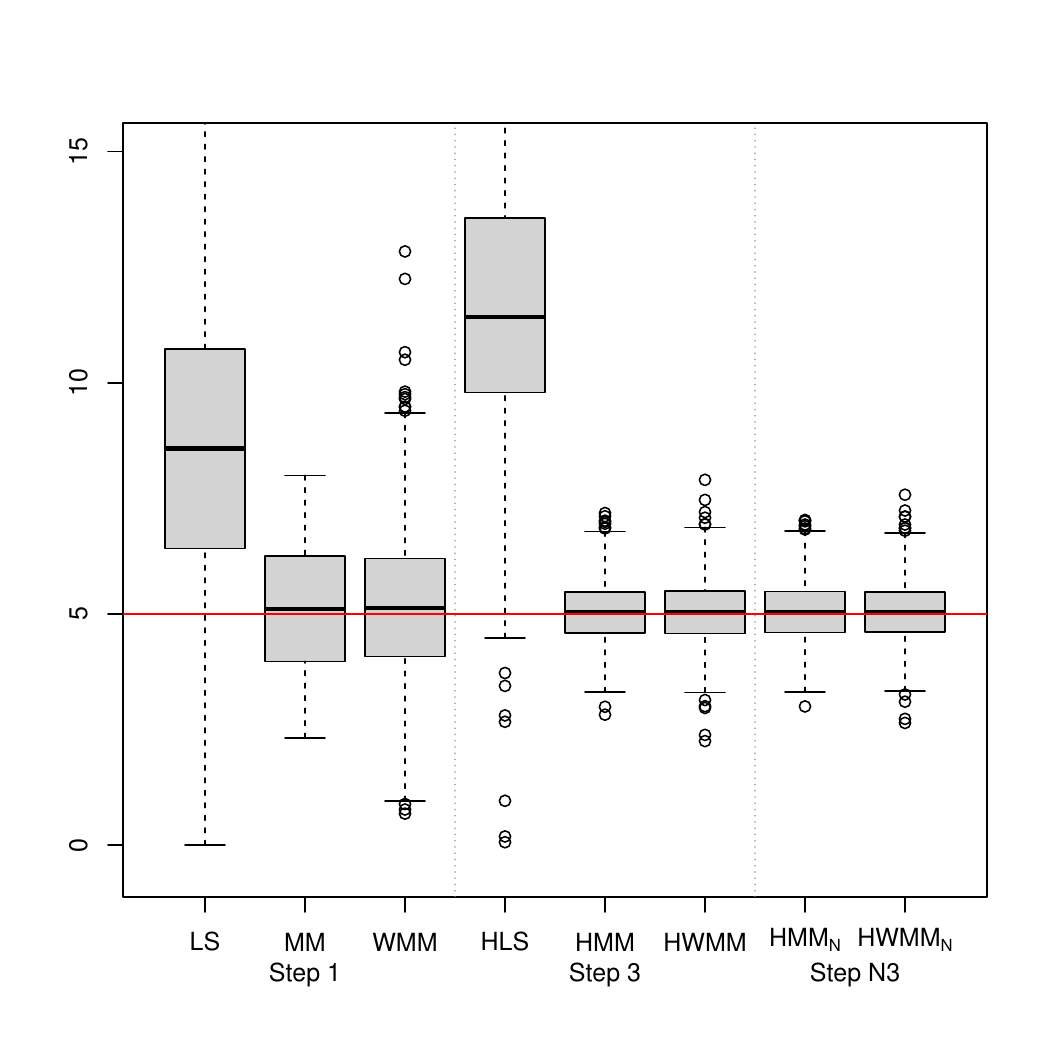} &  
			\includegraphics[scale=0.33]{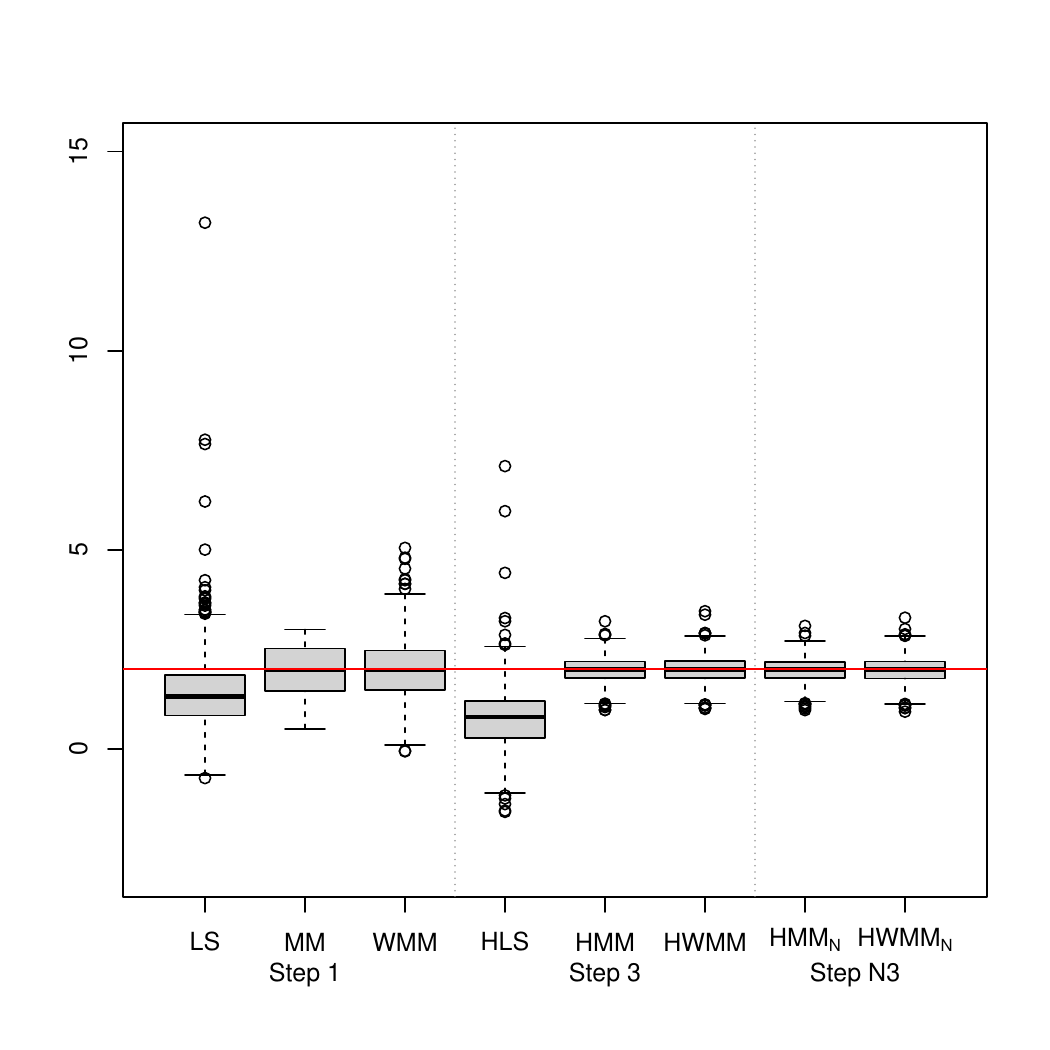} \\[-6ex]
			$C_3$ & 
			 \includegraphics[scale=0.33]{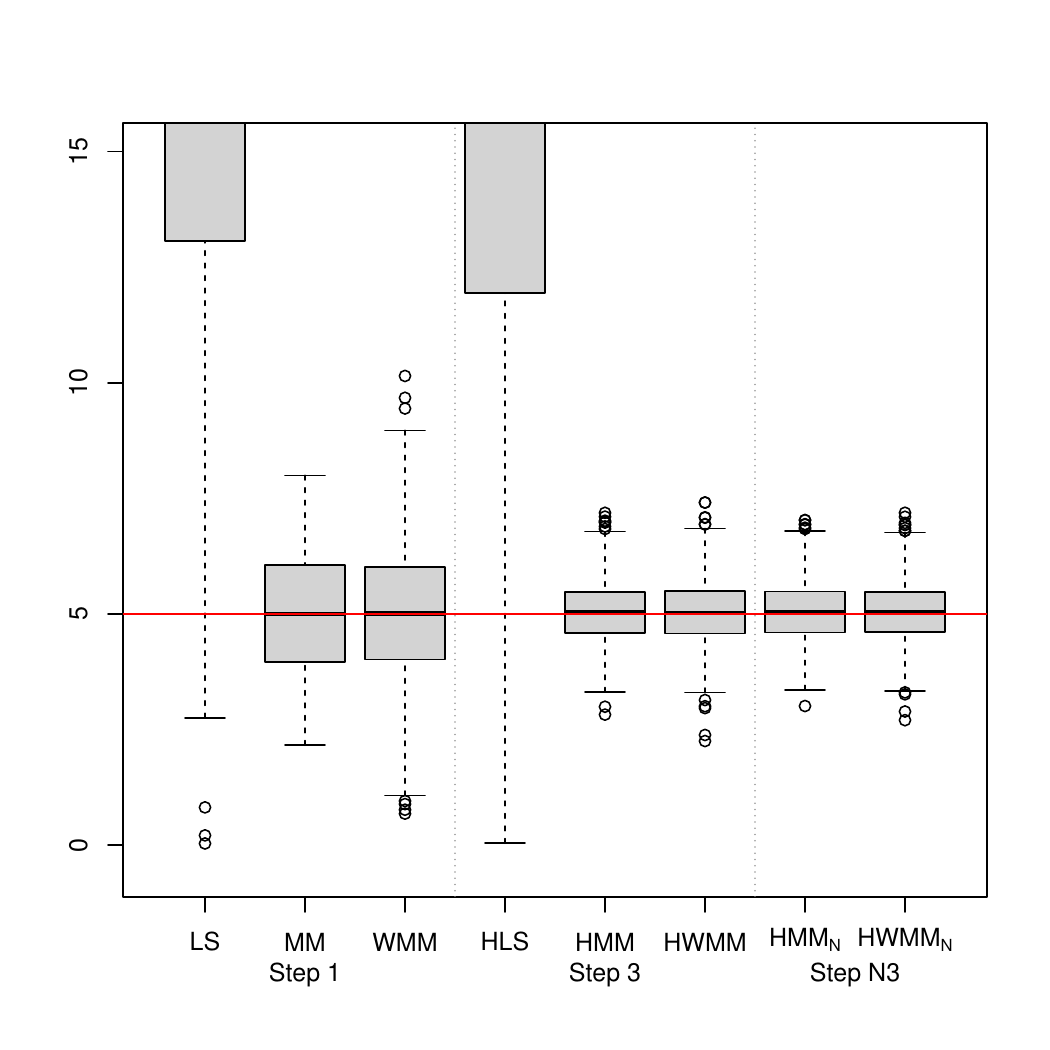} &  
			\includegraphics[scale=0.33]{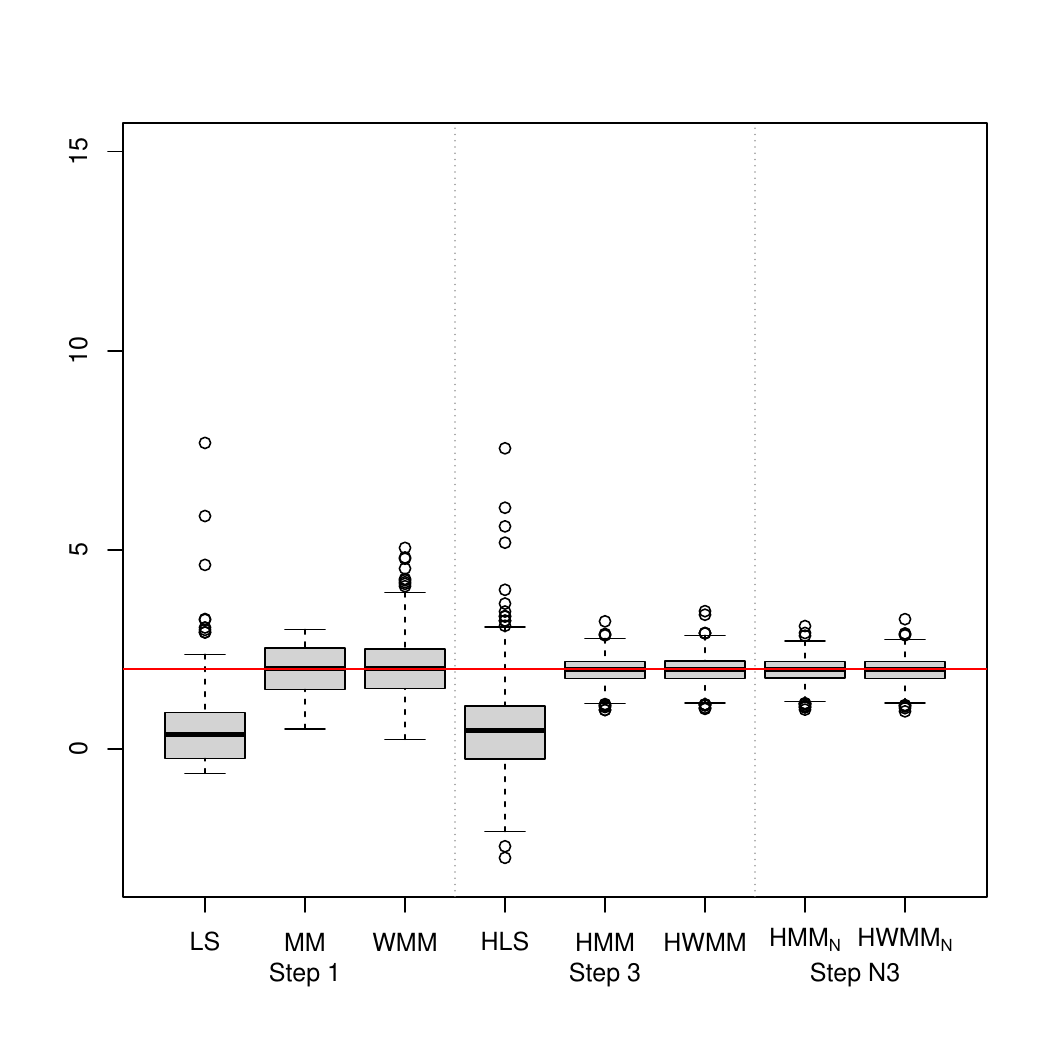} 
		\end{tabular}
		\vskip-0.1in \caption{\small  \label{fig:betaH0-C0-C3-bis} Boxplots of the regression model parameters  estimates, under $C_{0}$ to $C_3$. To facilitate visualization, all graphs are now displayed on the same scale.}
	\end{center} 
\end{figure}

\begin{figure}[ht!]
	\begin{center}
 	 \renewcommand{\arraystretch}{0.4}
 \newcolumntype{M}{>{\centering\arraybackslash}m{\dimexpr.1\linewidth-1\tabcolsep}}
   \newcolumntype{G}{>{\centering\arraybackslash}m{\dimexpr.45\linewidth-1\tabcolsep}}
%\begin{tabular}{MGG}
 
\begin{tabular}{M GG}
			& \small $\wbeta_{01}$  & \small $\wbeta_{02}$ \\[-4ex]
			$C_0$ & 
			\includegraphics[scale=0.33]{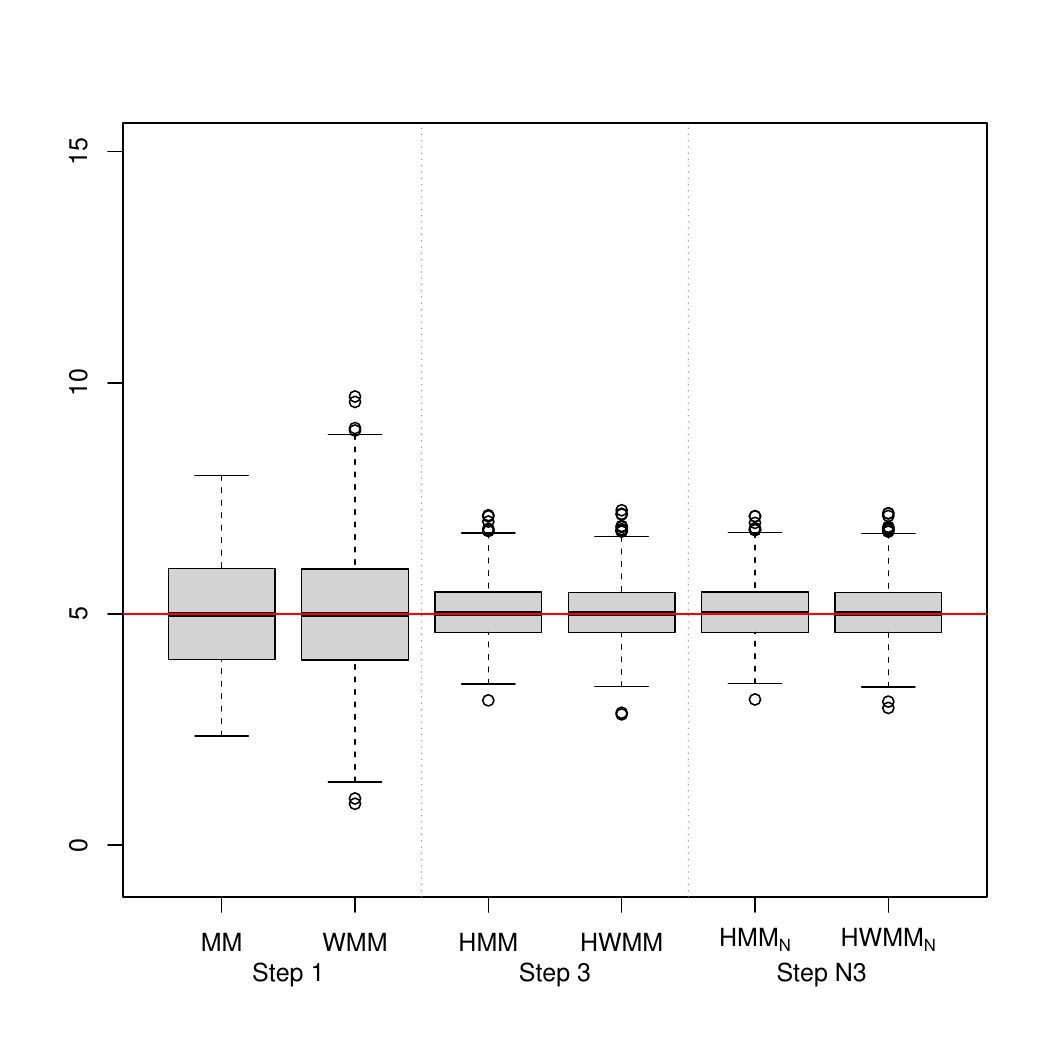} & 
			\includegraphics[scale=0.33]{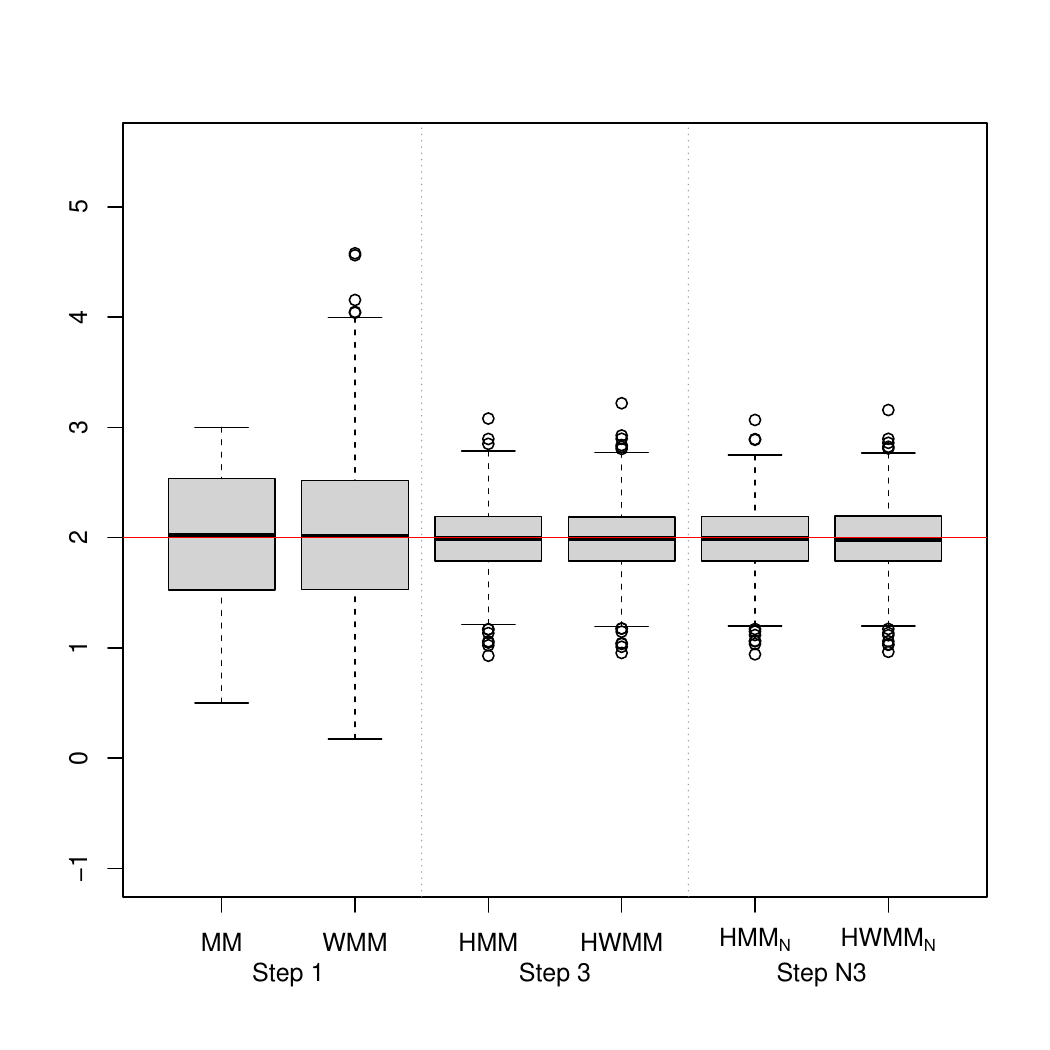} \\[-6ex]
			$C_1$ &  
			\includegraphics[scale=0.33]{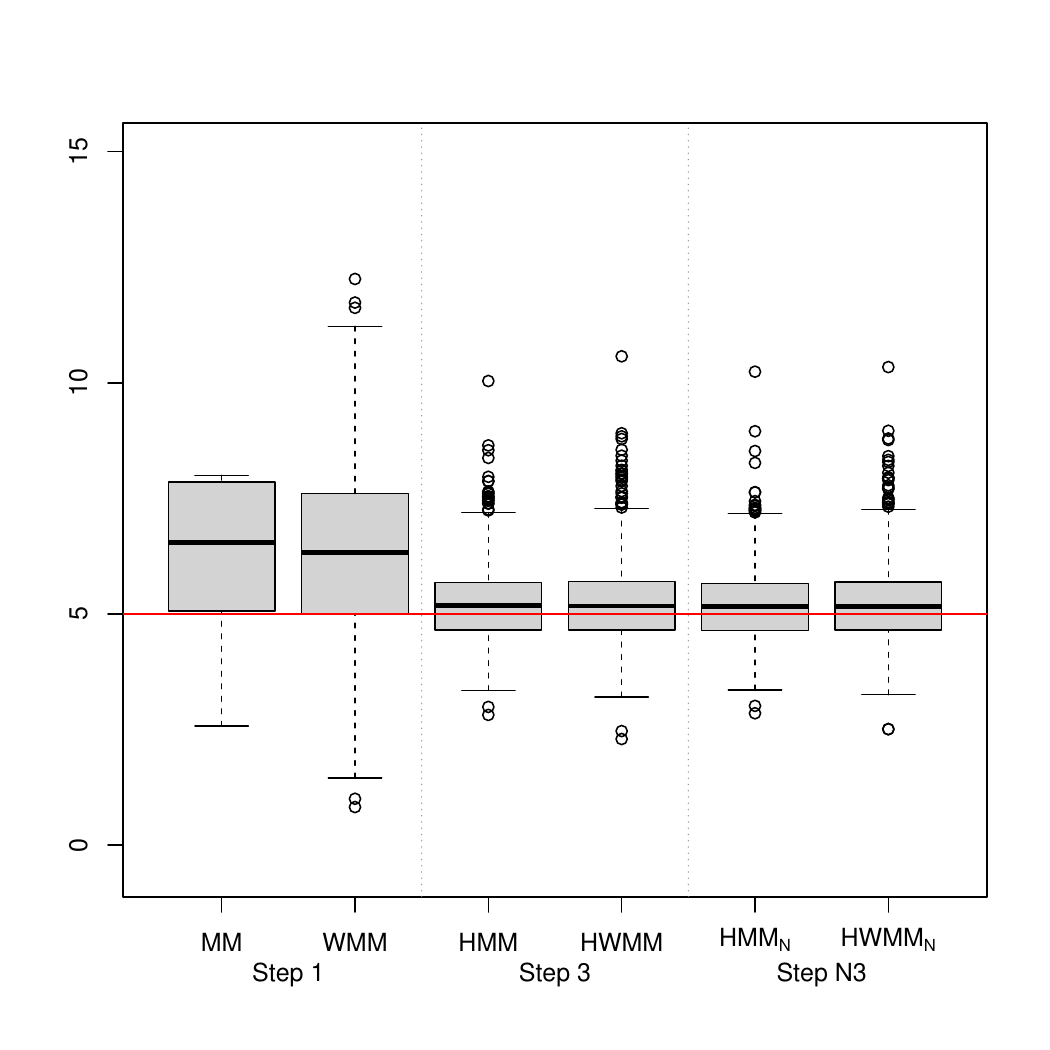} &  
			\includegraphics[scale=0.33]{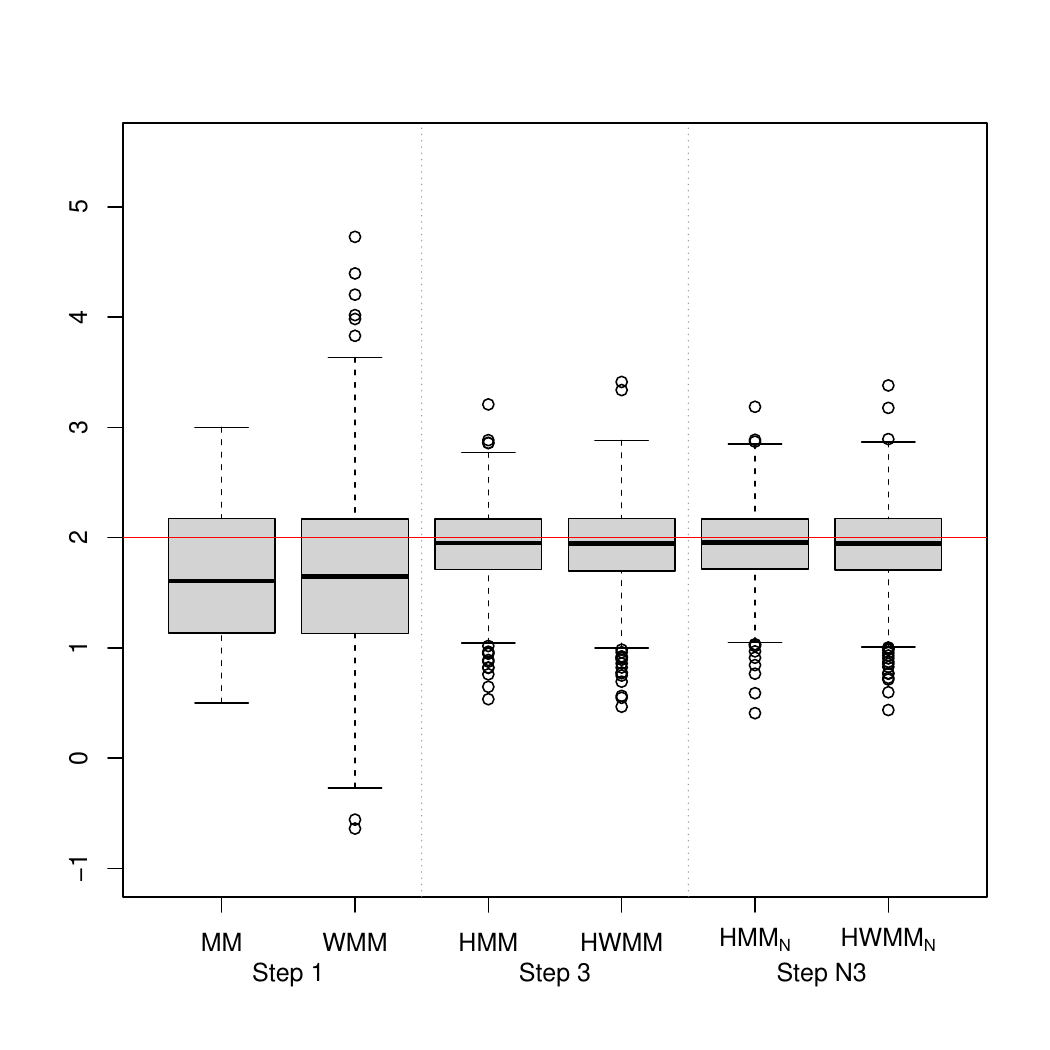} \\[-6ex]
			$C_2$ & 
			\includegraphics[scale=0.33]{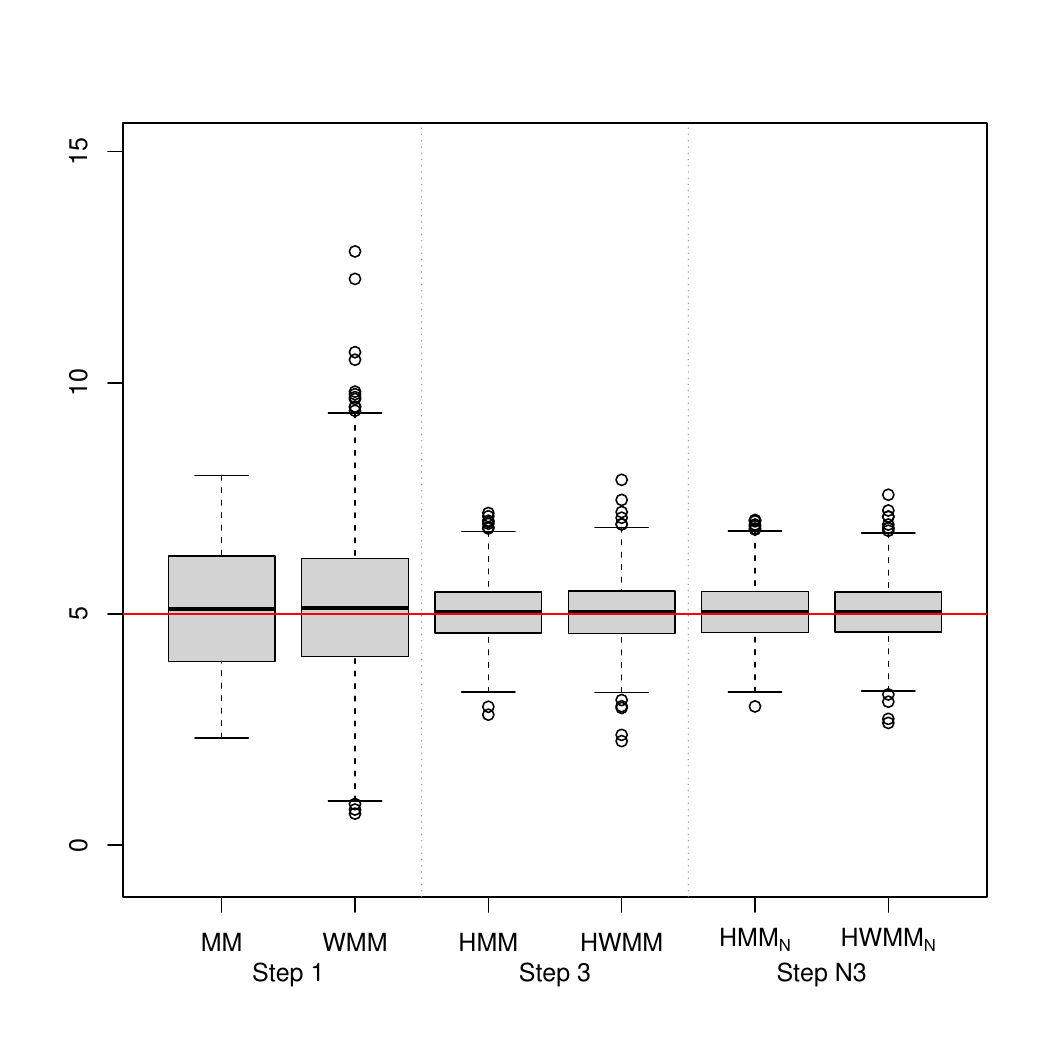} &  
			\includegraphics[scale=0.33]{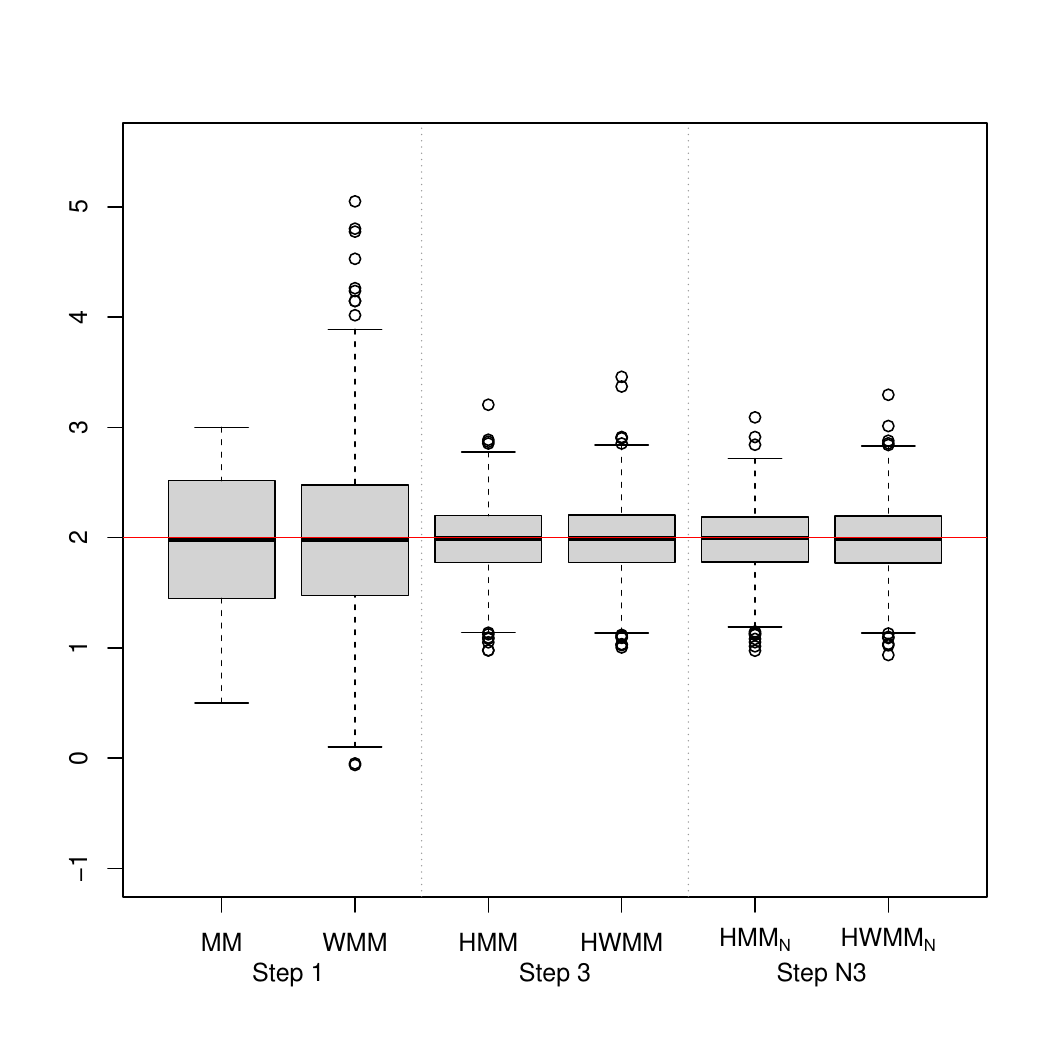} \\[-6ex]
			$C_3$ & 
			 \includegraphics[scale=0.33]{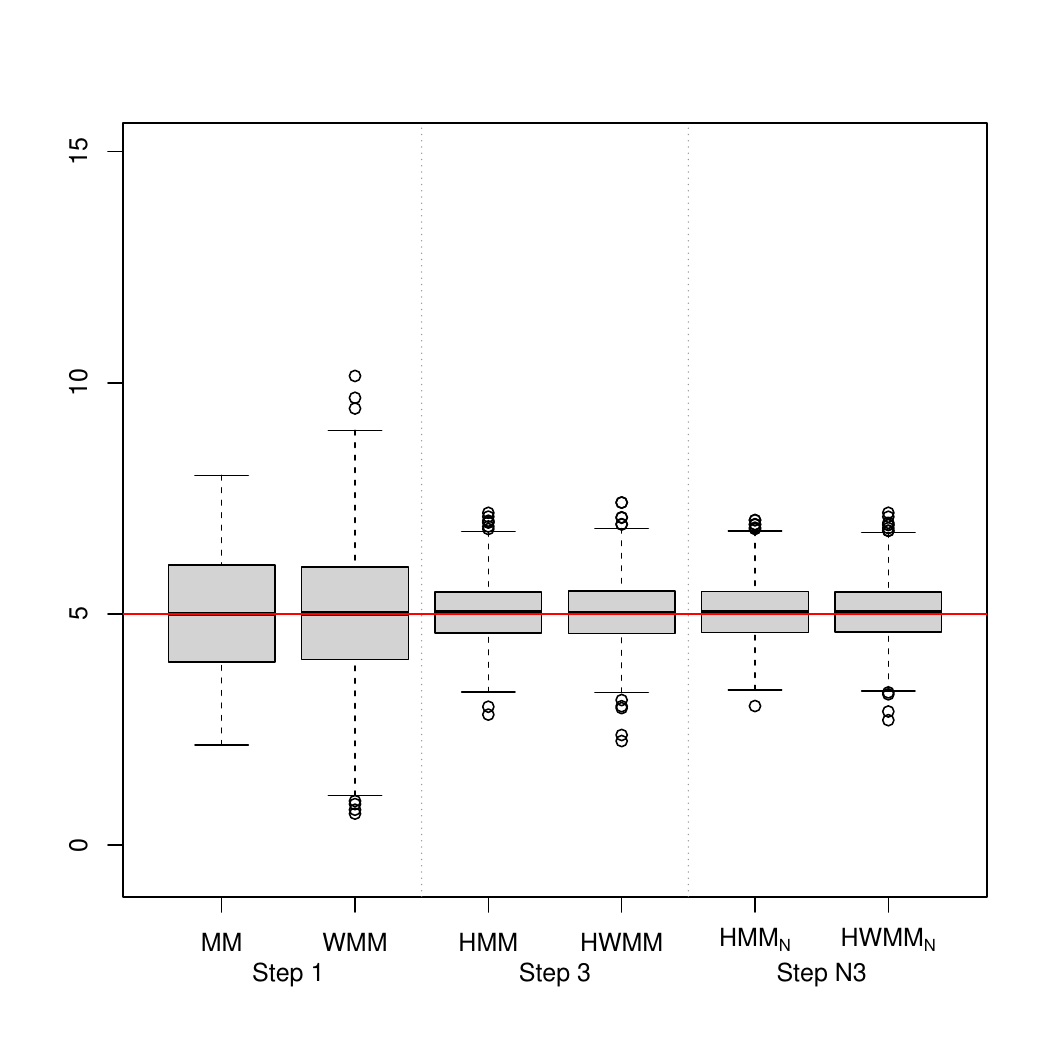} &  
			\includegraphics[scale=0.33]{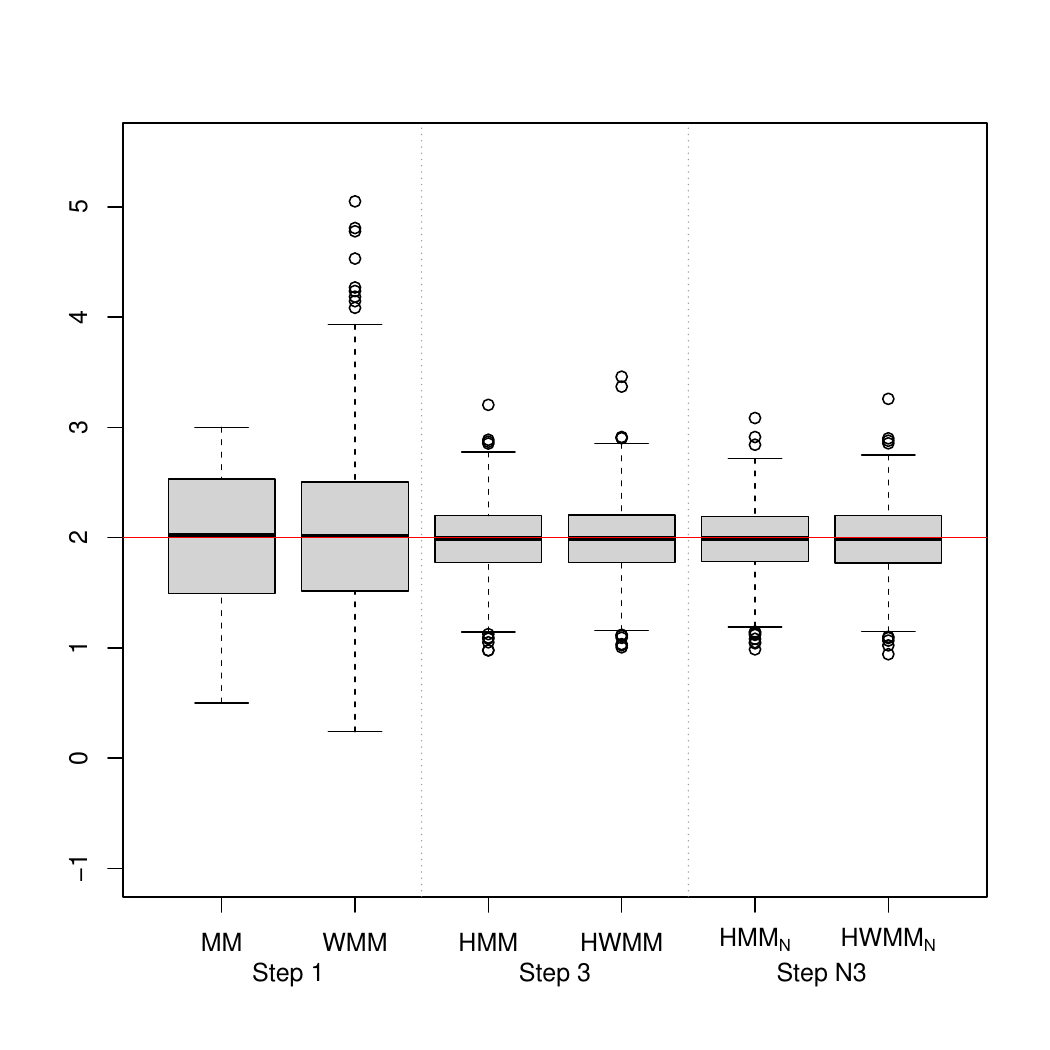} 
		\end{tabular}
		\vskip-0.1in \caption{\small  \label{fig:betaH0-C0-C3-sinLS}  Boxplots of the robust estimates of the regression model parameters, under $C_{0}$ to $C_3$.}
	\end{center} 
\end{figure}

The impact on the estimators of the three outlier magnitudes is very clear from Figures  \ref{fig:betaH0-C0-C3} to \ref{fig:betaH0-C0-C3-sinLS}, where the boxplots of the obtained estimates of  the regression parameters are displayed. The latter figure presents only the values of the robust estimators to facilitate comparisons, since the results under contamination of both \LS ~and \HLS ~enlarge the range of the vertical axis in Figure   \ref{fig:betaH0-C0-C3}. It is worth mentioning that the plots in Figure \ref{fig:betaH0-C0-C3-bis} are always given in the range $[-0.5,15]$ and $[-3,15]$ for the estimators of $\beta_{01}$ and $\beta_{02}$, respectively, to facilitate comparisons even when some boxplots may lie beyond these limits. The effect is devastating on the two versions of the least squares estimator, mainly on $\beta_{01}$ under $C_2$ and  $\beta_{02}$ under $C_3$. As shown in Tables \ref{tab:ECM-BIAS-beta01HO} and  \ref{tab:ECM-BIAS-beta02HO}, the MSE of both least squares estimators increases with the size of the vertical outlier and   explodes under  $C_3$, largely for $\beta_{01}$. On the contrary, the estimates related to the robust estimators \HMM ~ and \HWMM~ show a very stable behaviour in all scenarios. Even when the MSE and the absolute value of the bias of these two estimators increases under $C_1$, which corresponds to mild atypical points, the effect is very moderate and it decreases as the magnitude of the outliers grows. The robust regression estimators   \HMM$_{\nuevo}$  and \HWMM$_{\nuevo}$  show a bias performance similar to that of \HMM ~ and \HWMM, respectively, while  \HWMM$_{\nuevo}$ improves the MSE behaviour under all contamination schemes.

\begin{figure}[ht!]
	\begin{center}
 	 \renewcommand{\arraystretch}{0.4}
 \newcolumntype{M}{>{\centering\arraybackslash}m{\dimexpr.1\linewidth-1\tabcolsep}}
   \newcolumntype{G}{>{\centering\arraybackslash}m{\dimexpr.45\linewidth-1\tabcolsep}}
%\begin{tabular}{MGG}
 
\begin{tabular}{M GG}
			& \small $\wbeta_{01}$  & \small $\wbeta_{02}$ \\[-4ex]
			$C_0$ & 
			\includegraphics[scale=0.33]{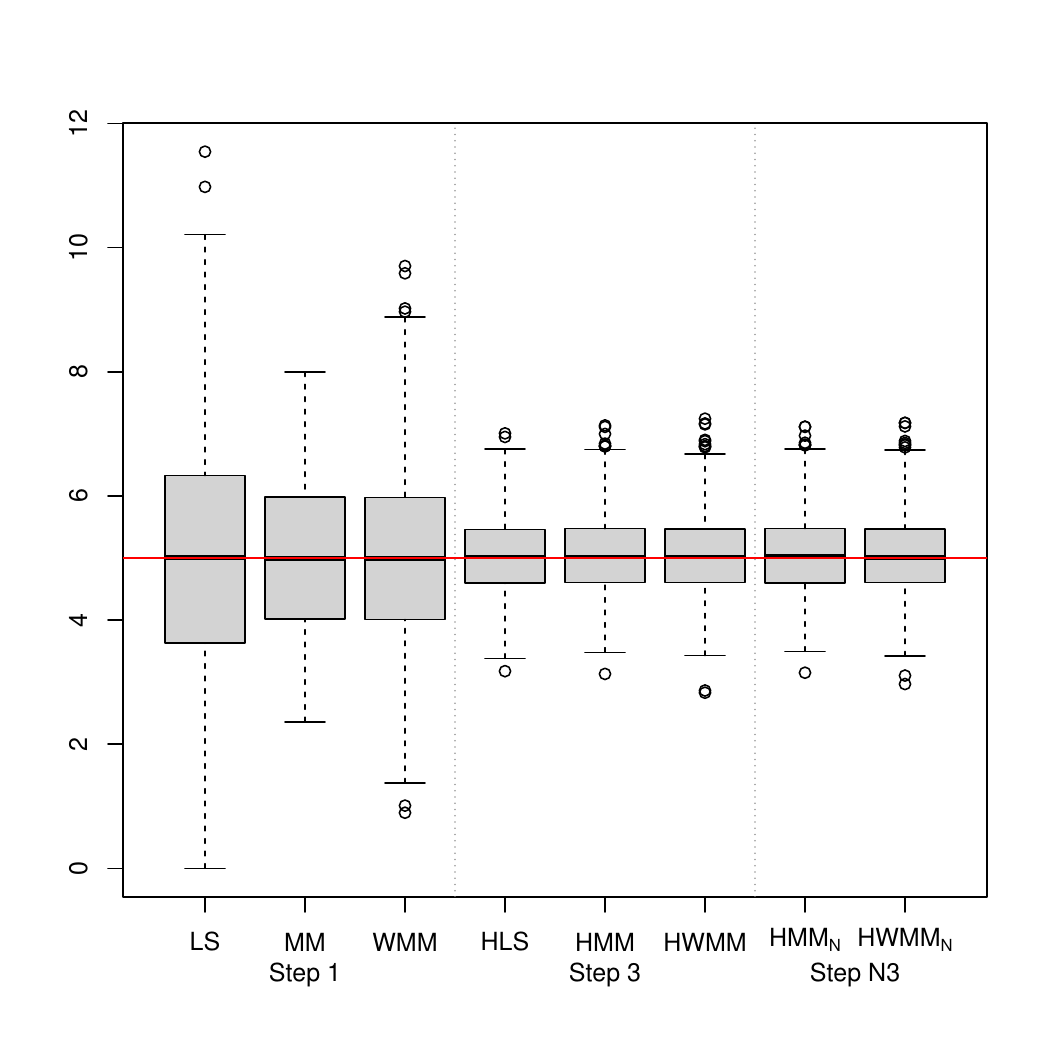} & 
			\includegraphics[scale=0.33]{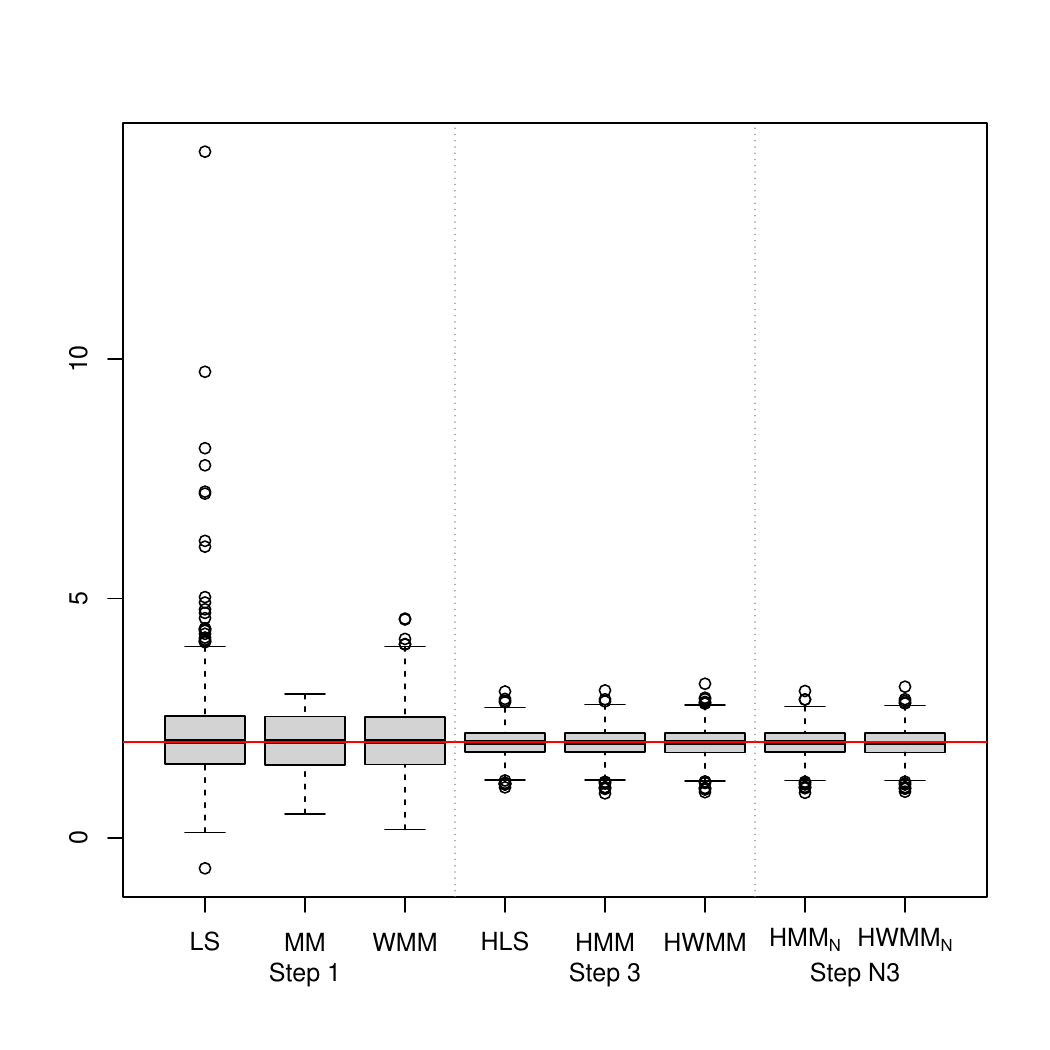} \\[-6ex]
			$C_1$ &  
			\includegraphics[scale=0.33]{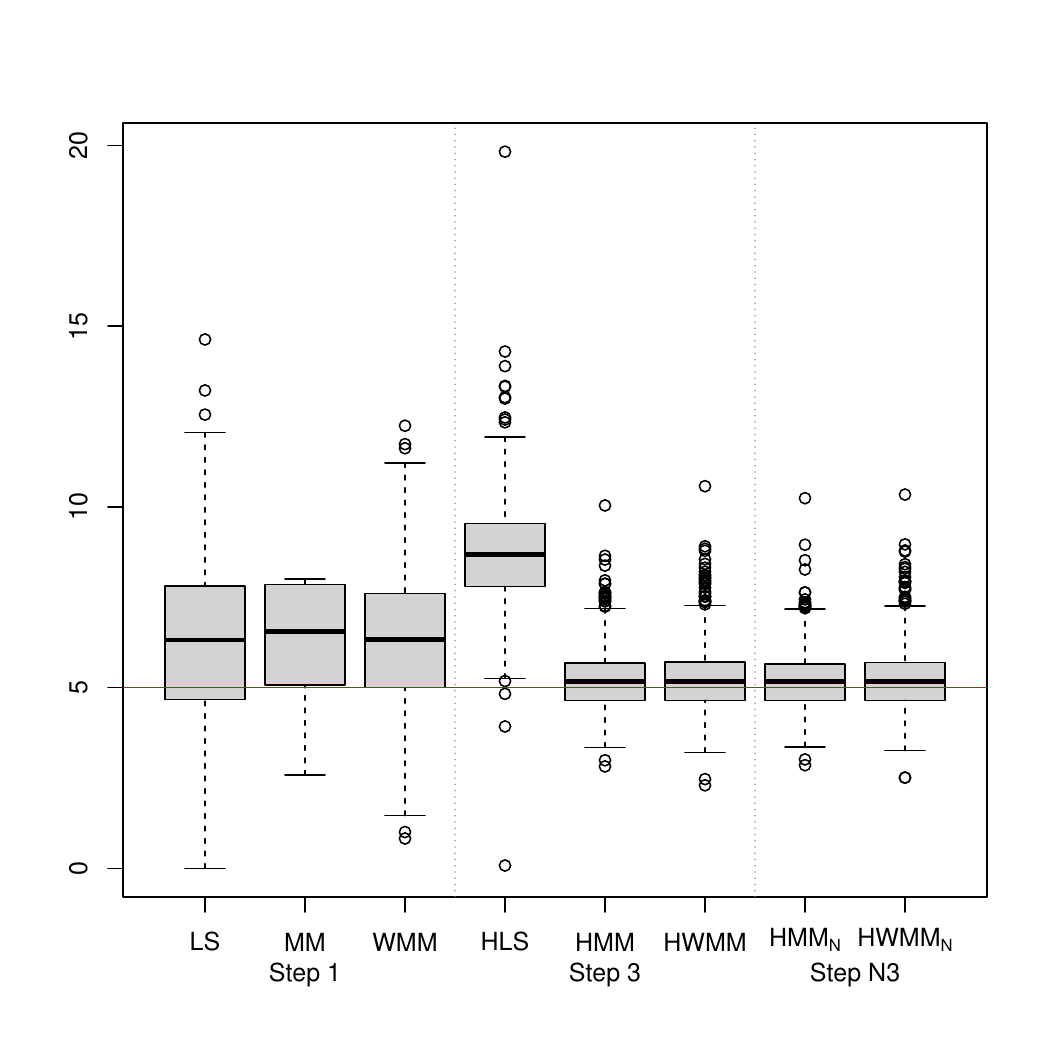} &  
			\includegraphics[scale=0.33]{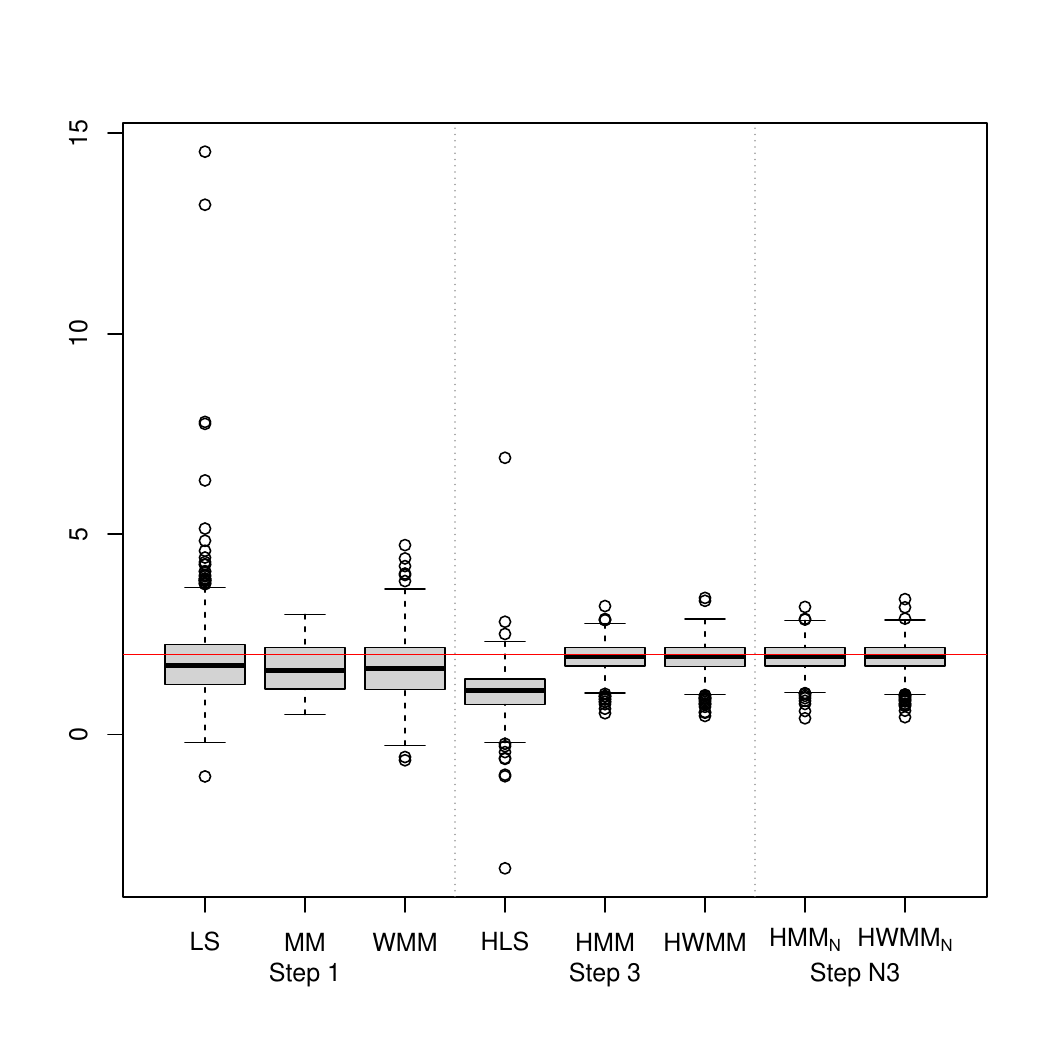} \\[-6ex]
			$C_2$ & 
			\includegraphics[scale=0.33]{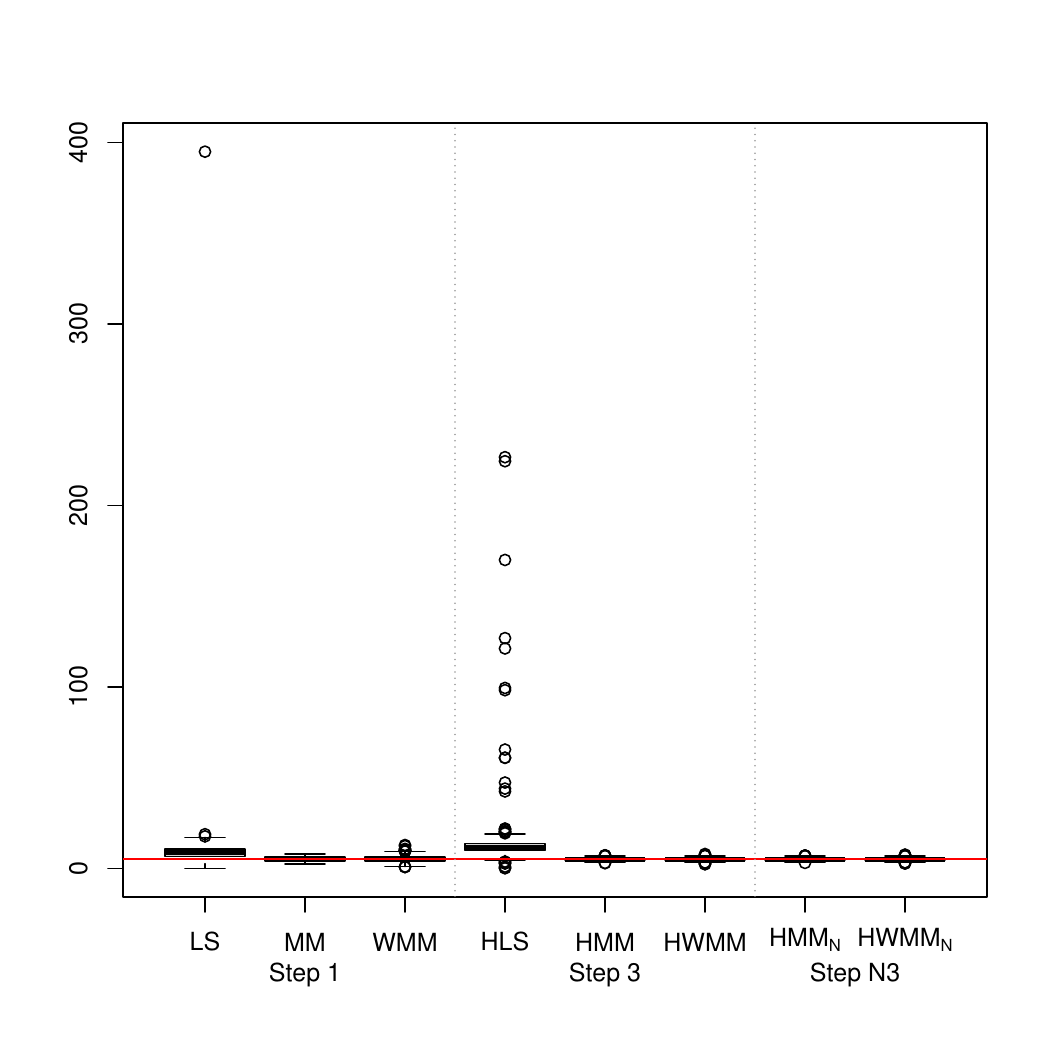} &  
			\includegraphics[scale=0.33]{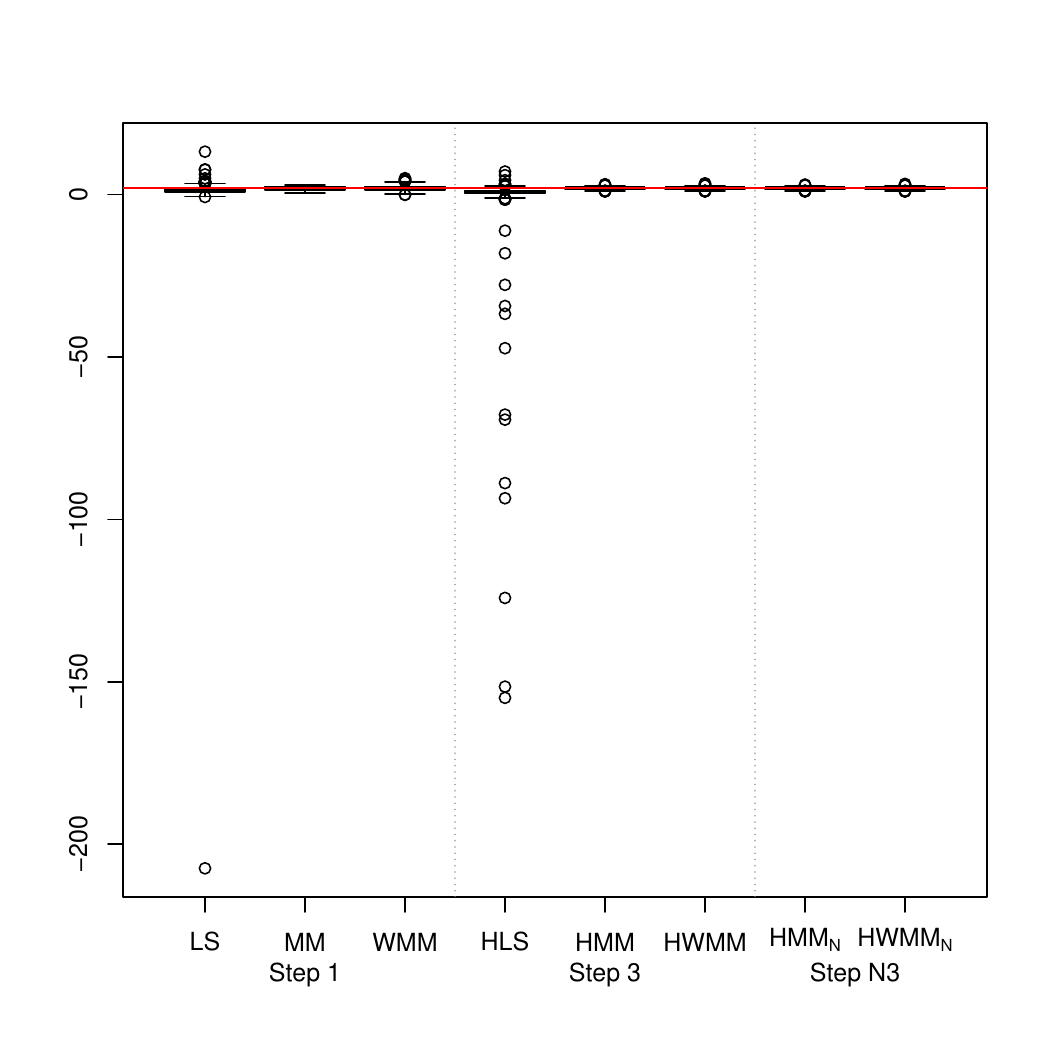} \\[-6ex]
			$C_3$ & 
			 \includegraphics[scale=0.33]{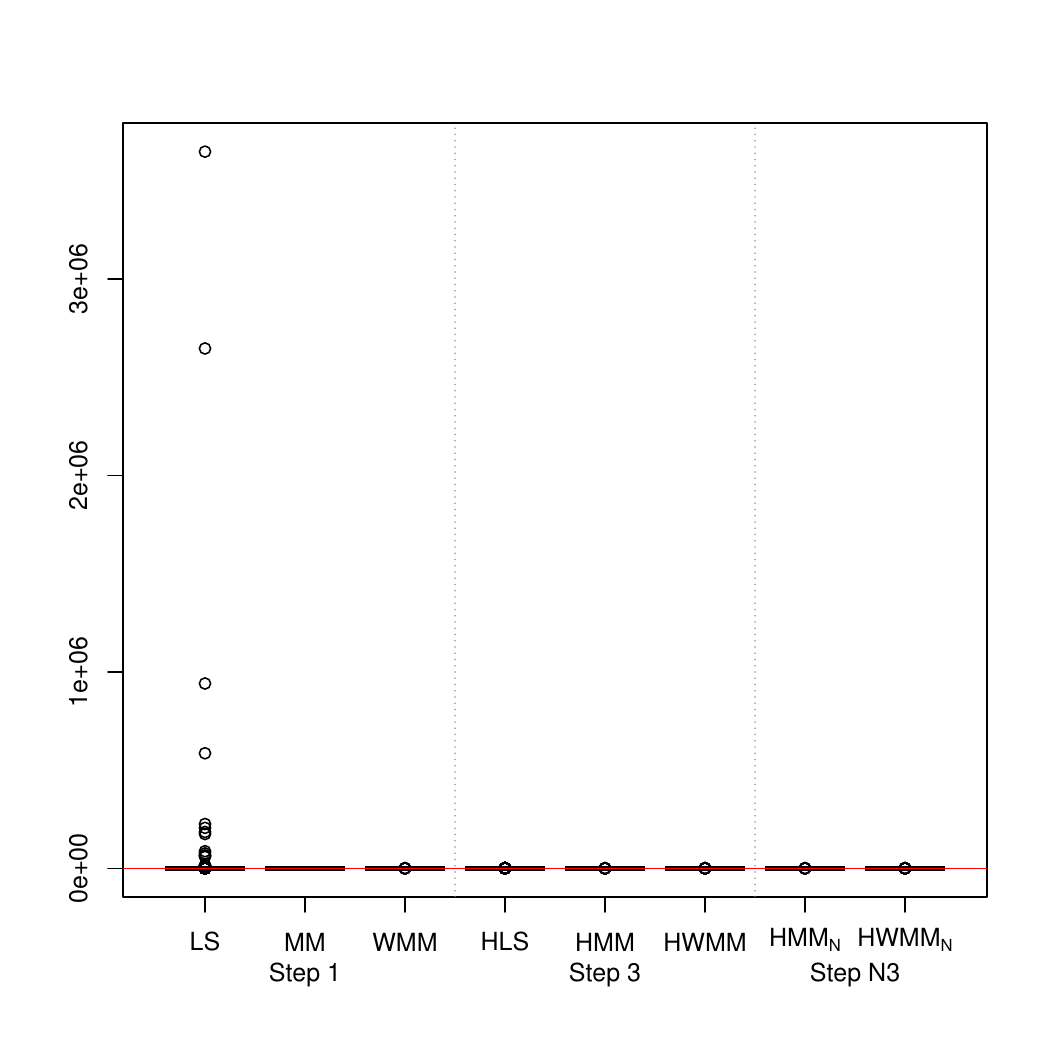} &  
			\includegraphics[scale=0.33]{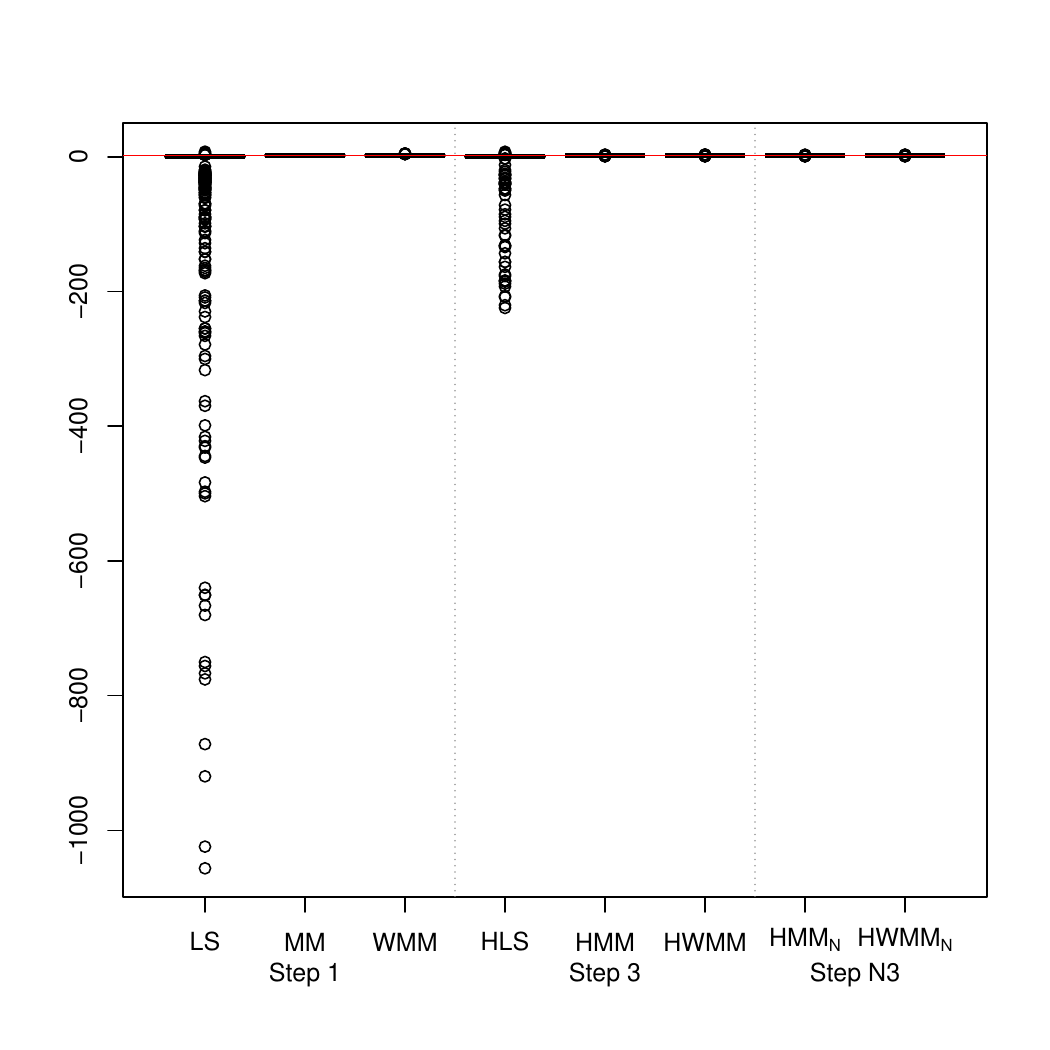} 
		\end{tabular}
		\vskip-0.1in \caption{\small  \label{fig:betaH0-C0-C3}  Boxplots of the regression model parameters  estimates, under $C_{0}$ to $C_3$.}
	\end{center} 
\end{figure}

\begin{figure}[ht!]
	\begin{center}
 	 \renewcommand{\arraystretch}{0.4}
 \newcolumntype{M}{>{\centering\arraybackslash}m{\dimexpr.1\linewidth-1\tabcolsep}}
   \newcolumntype{G}{>{\centering\arraybackslash}m{\dimexpr.45\linewidth-1\tabcolsep}}
%\begin{tabular}{MGG} 
\begin{tabular}{G G}
		$C_0$	& $C_1$   \\[-2ex]
			\includegraphics[scale=0.4]{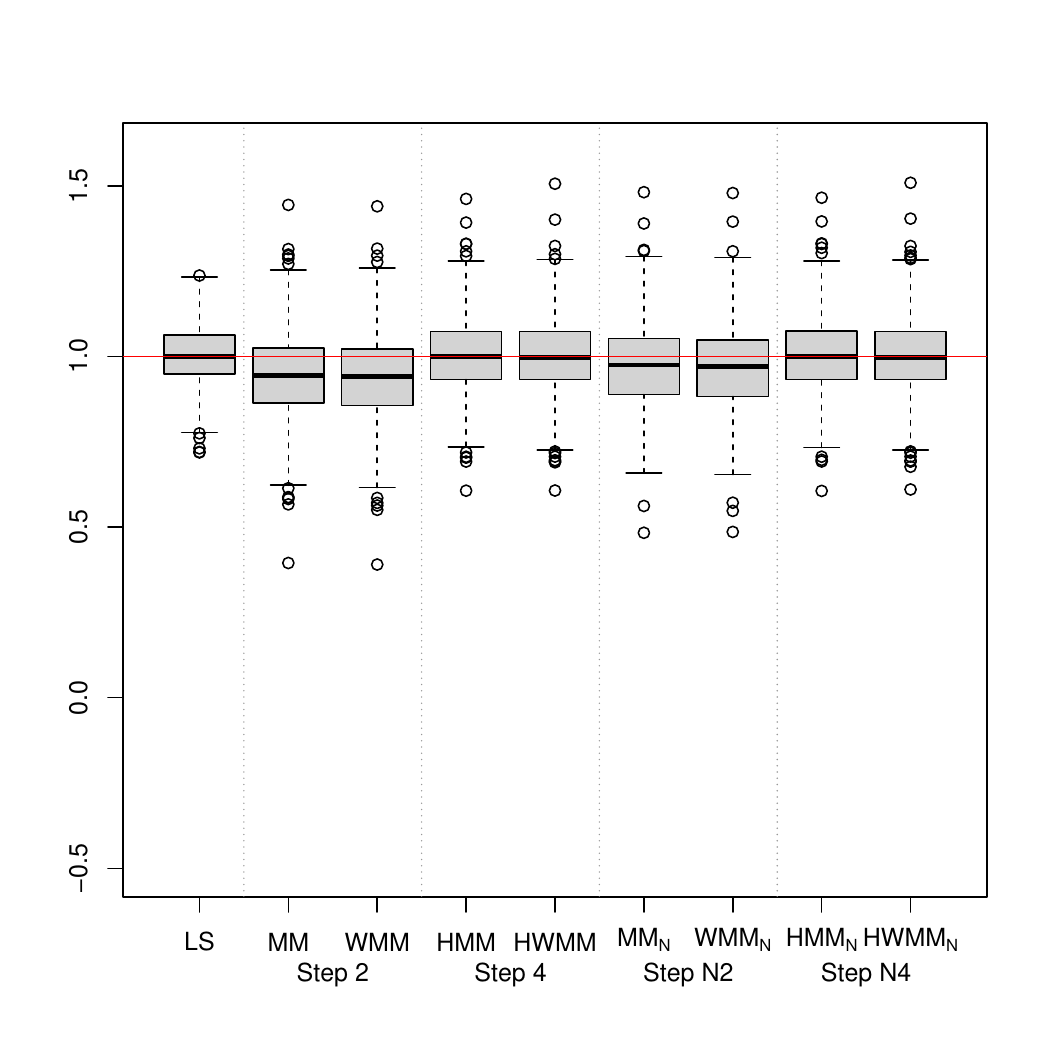}   
			&
			\includegraphics[scale=0.4]{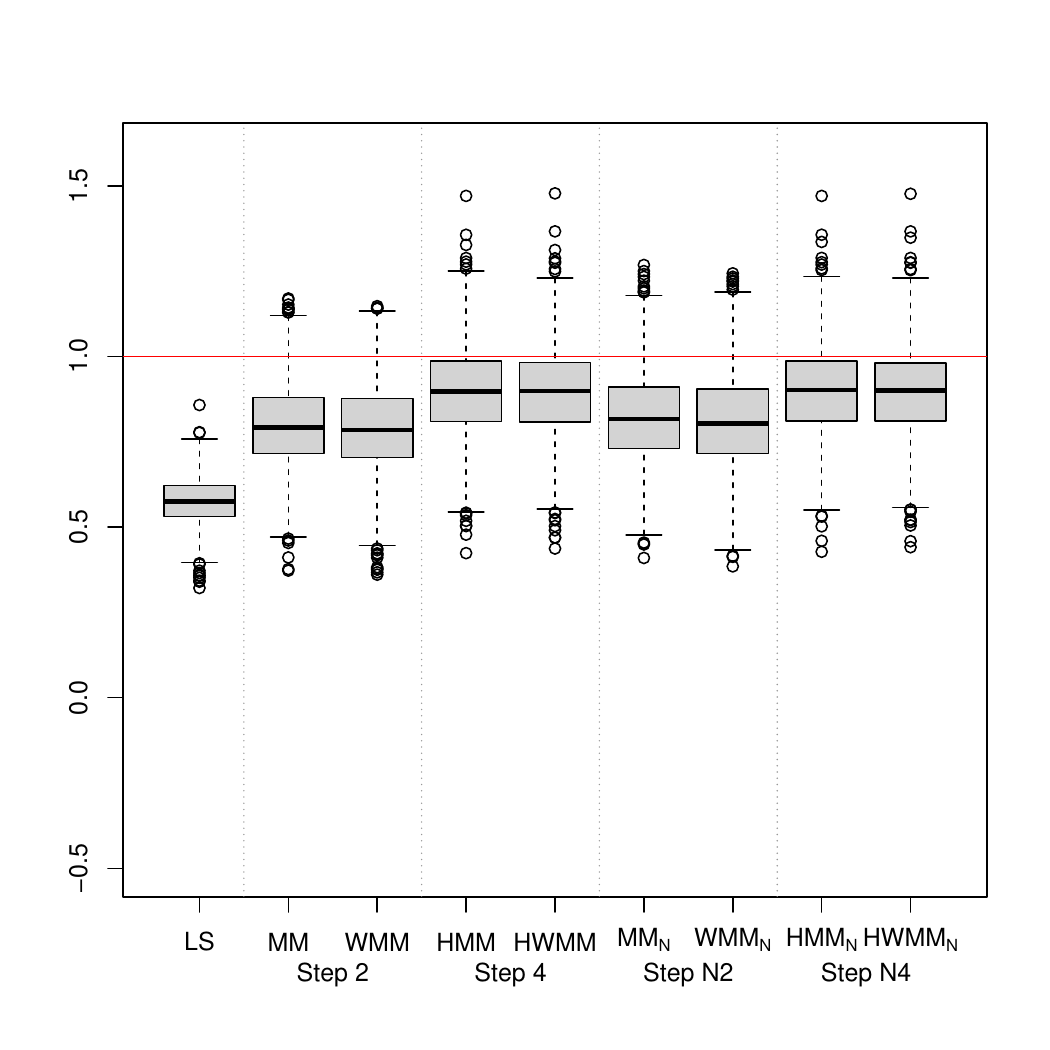}   \\ 
			$C_2$ & $C_3$\\ [-2ex]
			\includegraphics[scale=0.4]{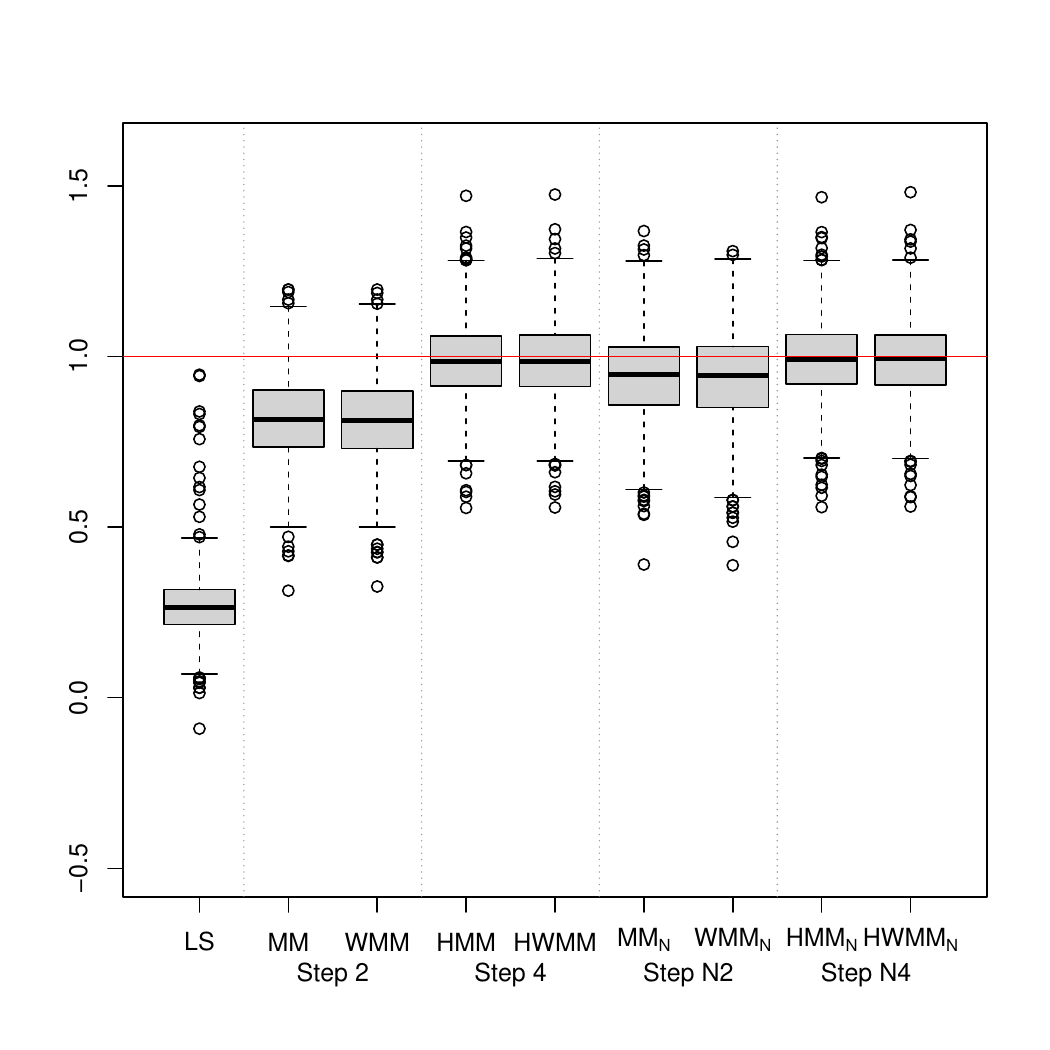}  & 
			 \includegraphics[scale=0.4]{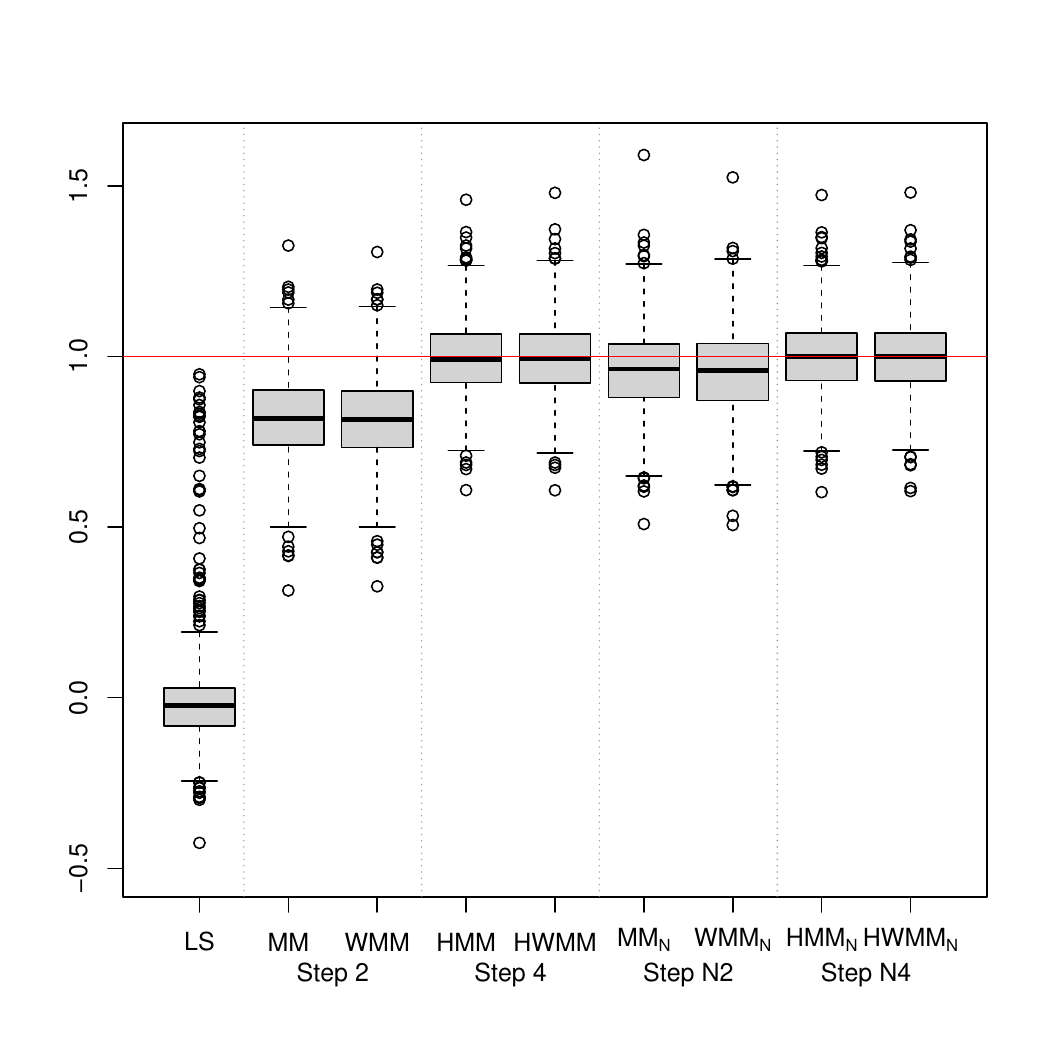}  
		\end{tabular}
		\vskip-0.1in \caption{\small  \label{fig:lambda-bxp} Boxplots of the estimates of $\lambda_0$, under $C_{0}$ to $C_3$.}
	\end{center} 
\end{figure}

Regarding the  estimates of the parameter $\lambda_0$ of the variance function,  Figure \ref{fig:lambda-bxp} presents the boxplots for the robust estimates obtained in \textbf{Step 2}, \textbf{Step 4}, \textbf{Step N2} and \textbf{Step N4}  as well as for the classical estimators, under the different contamination schemes. As indicated above, we labelled  \MM~  and   \WMM ~ the boxplots of the estimators computed in \textbf{Step 2} when the initial estimator $\wbbe_{\ini}$ corresponds to an $MM-$ or a weighted $MM-$estimator assuming an homocedastic model. In contrast, the results for the estimators computed   in \textbf{Step 4} are indicated as \HMM ~ or  \HWMM~ when using the estimators of $\bbe$ defined in \textbf{Step 3} with weights equal to 1 or with the weights   $w(x)$ defined in \eqref{eq:pesos}, respectively.  When using \textbf{Step N2} and \textbf{Step N4}, the estimators are indicated using the subscript \textsc{n}. The classical estimators of $\lambda_0$ are labelled \LS. To facilitate the comparisons all plots have the same vertical axis.

Figure \ref{fig:lambda-bxp} shows that, under $C_0$, all the estimates of $\lambda_0$  have similar behaviour. However, those obtained using \textbf{Step 4} and  \textbf{Step N4} are less biased than those obtained through \textbf{Step 2} or \textbf{Step N2}. The   effect of the introduced  anomalous points on the classical estimator of ${\lambda_0}$ becomes evident in  Figure \ref{fig:lambda-bxp}, since the boxplots of the obtained estimates lie below  the target. The   robust estimators of $\lambda_0$ computed through \textbf{Step 4} or \textbf{Step N4} show their advantage over those computed in   \textbf{Step 2} or \textbf{Step N2}, in particular  under $C_2$ and $C_3$. In addition, it should be stressed that, under these two contaminations, the estimators obtained using  \textbf{Step N2} have better performance than those computed using \textbf{Step 2}.

\begin{figure}[ht!]
	\begin{center}
		 \renewcommand{\arraystretch}{0.4}
 \newcolumntype{M}{>{\centering\arraybackslash}m{\dimexpr.1\linewidth-1\tabcolsep}}
    \newcolumntype{G}{>{\centering\arraybackslash}m{\dimexpr.32\linewidth-1\tabcolsep}}
\begin{tabular}{GGG}
& \multicolumn{2}{c}{\small\textbf{Step 4}}\\
	 	\small	\LS & \small\HMM & \small\HWMM\\[-3ex]	 
	    	\includegraphics[scale=0.3]{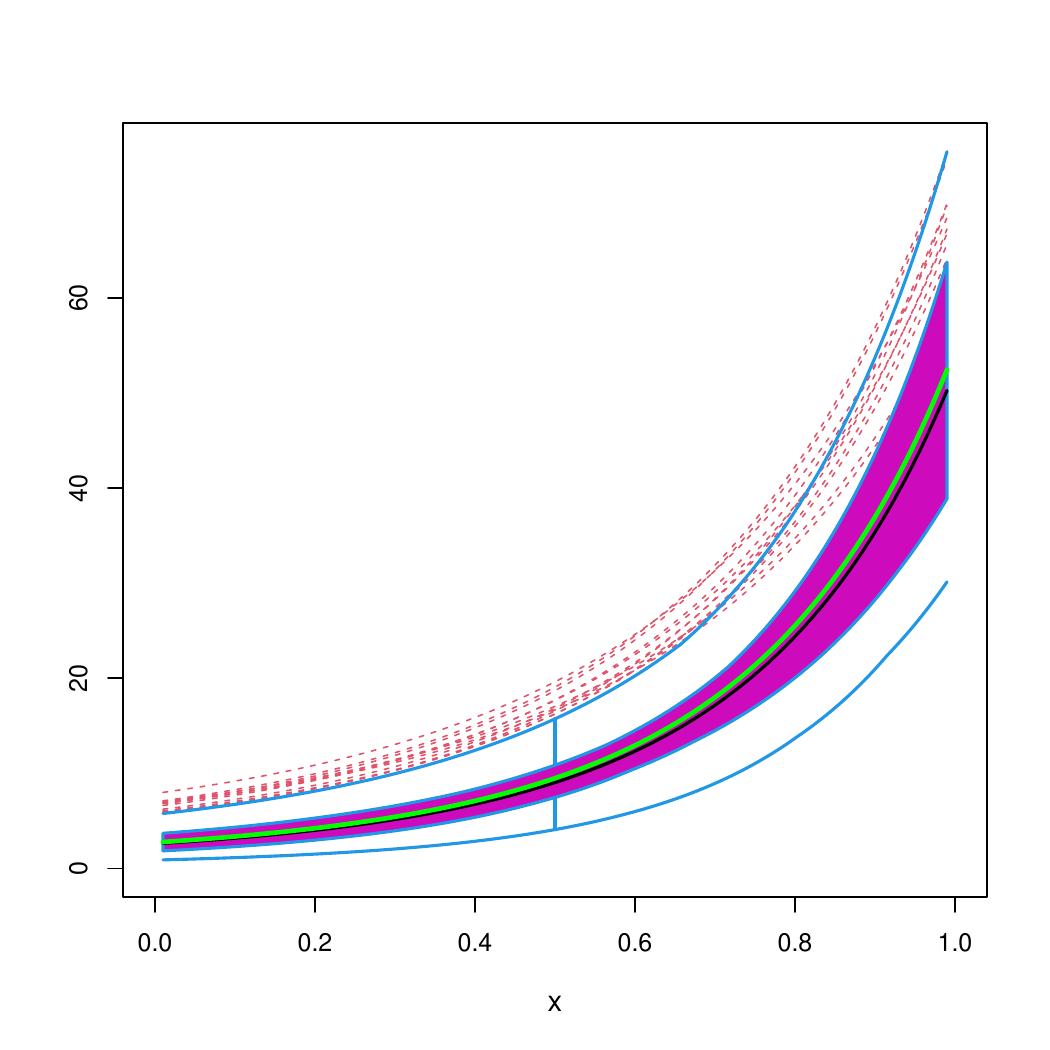} &  
			\includegraphics[scale=0.3]{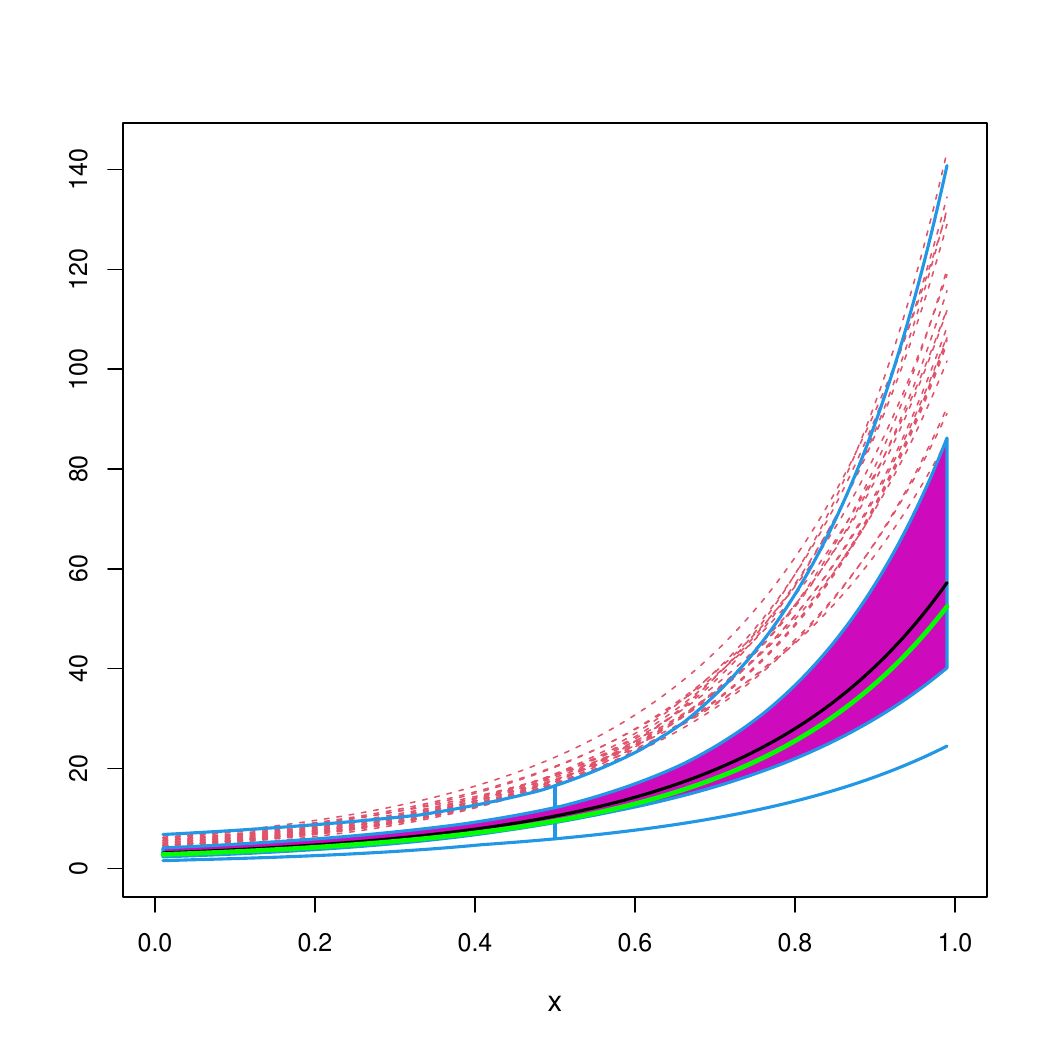}  &  
			\includegraphics[scale=0.3]{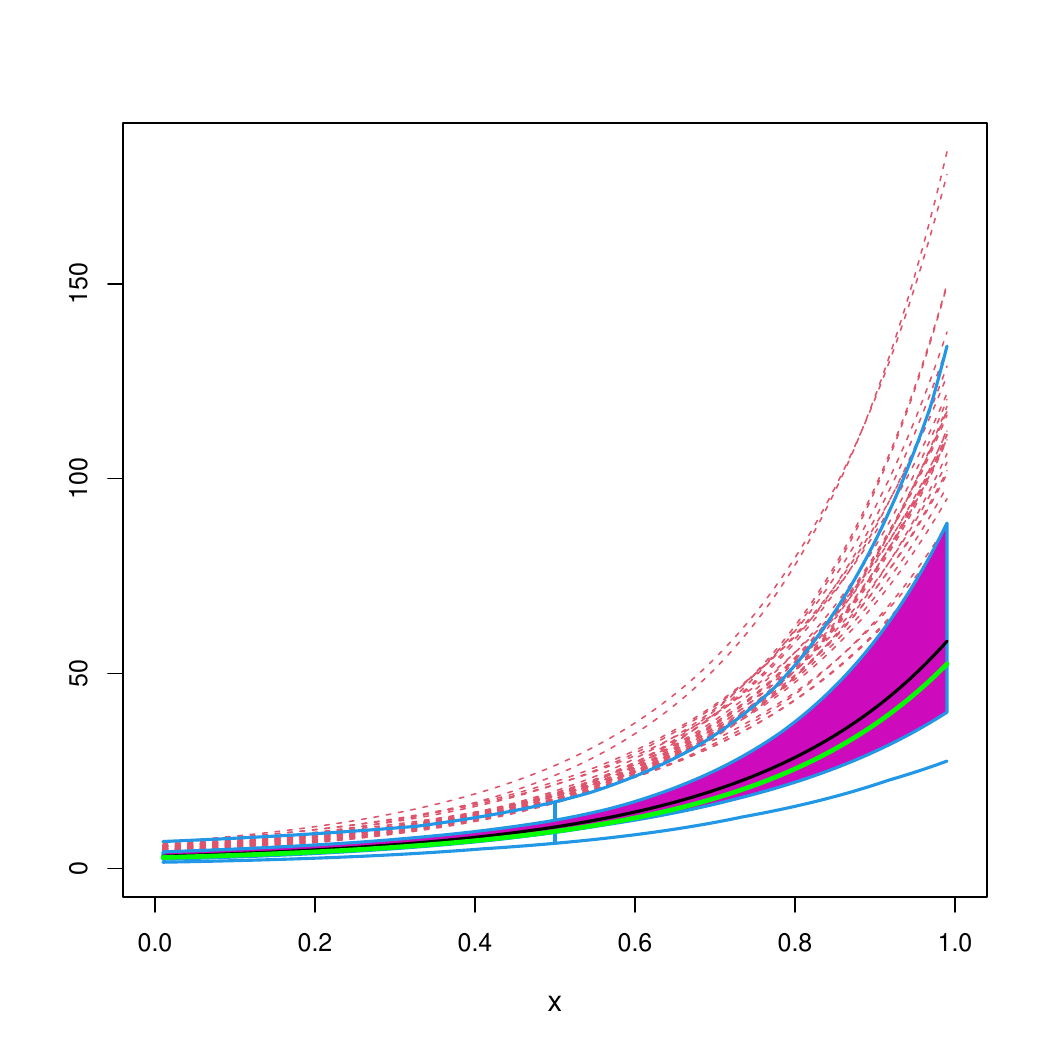}\\
	& \multicolumn{2}{c}{\small\textbf{Step N4}}\\
		 	\small	  & \small\HMM$_{\nuevo}$ & \small\HWMM$_{\nuevo}$\\[-3ex]
	    	  &  
			\includegraphics[scale=0.3]{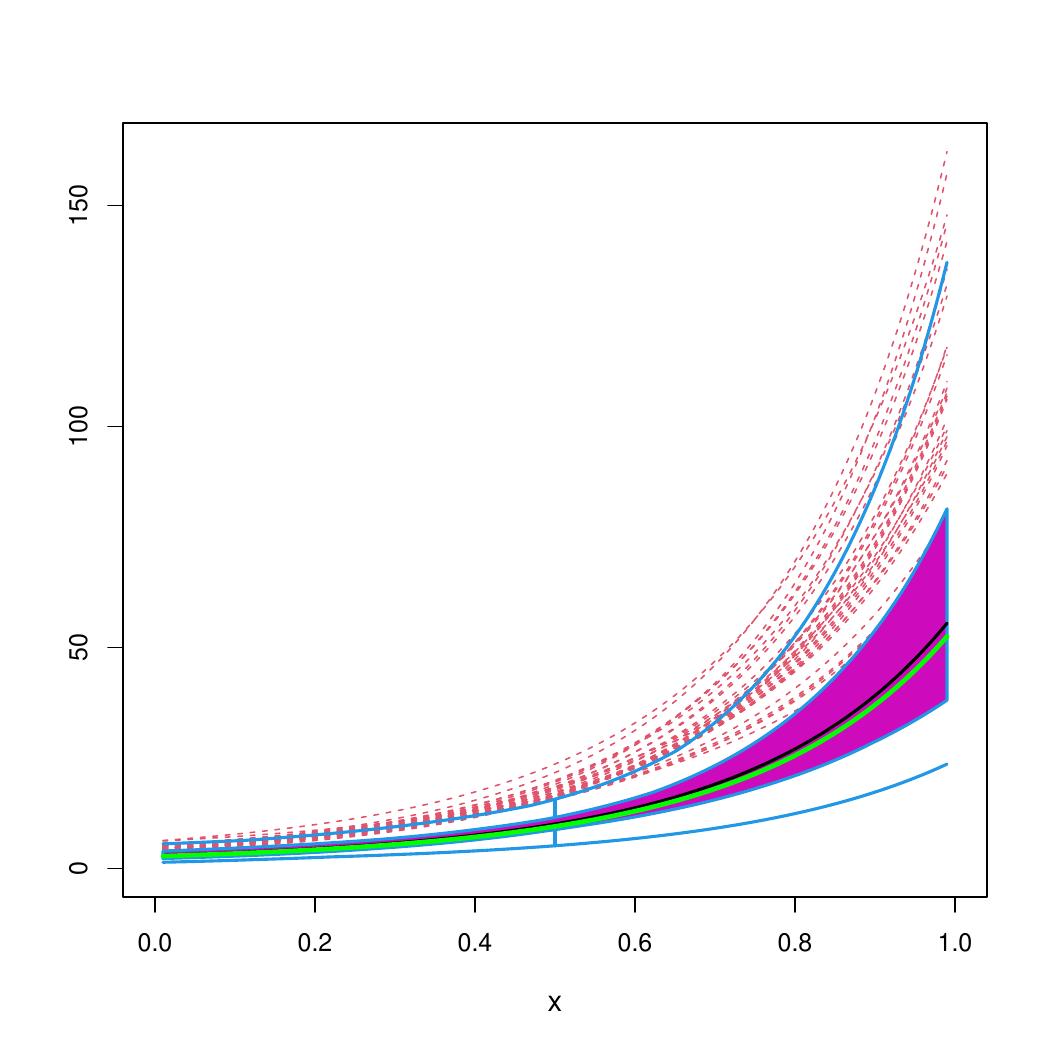}  &  
			\includegraphics[scale=0.3]{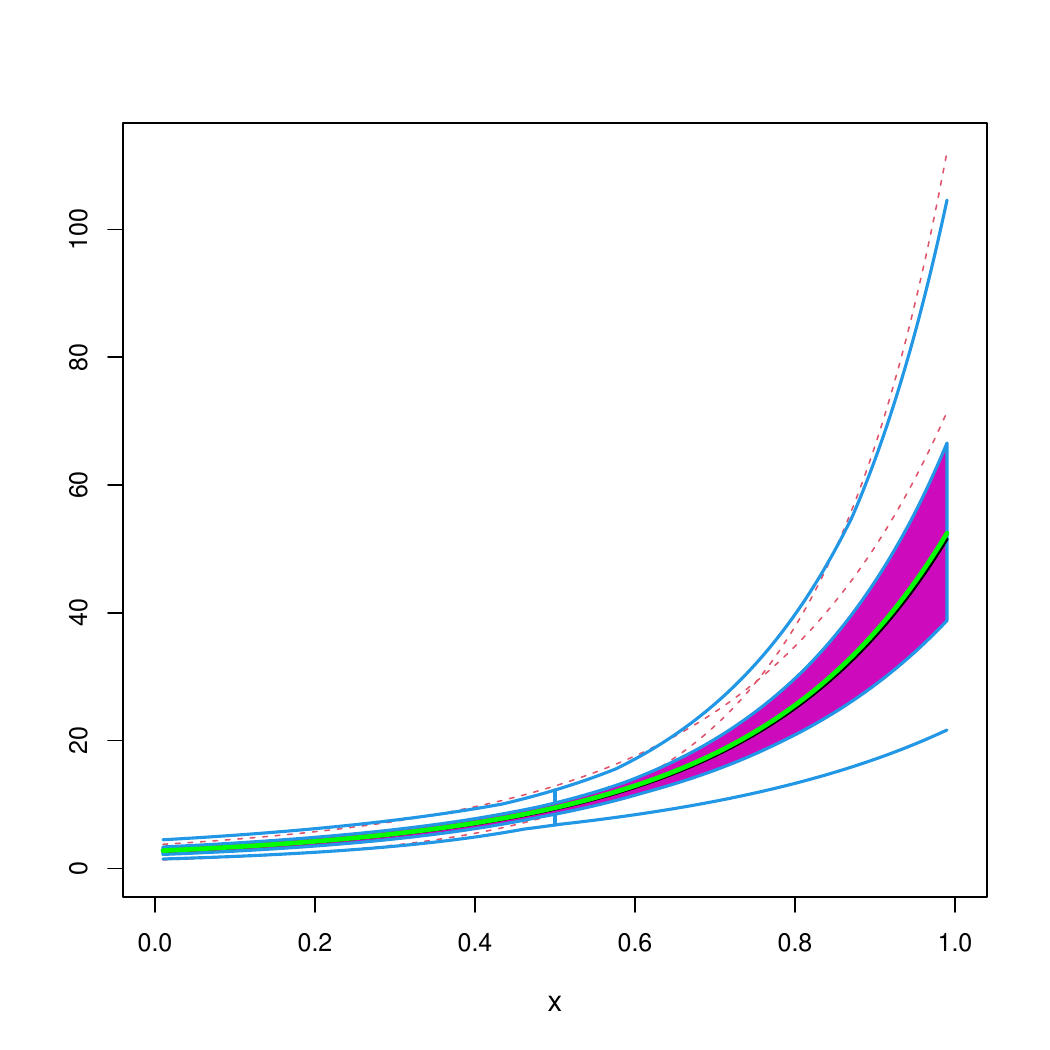}  			
		\end{tabular}
\vskip-0.1in \caption{ \small \label{fig:variance_C0}  Functional boxplots of the estimated variance function under $C_0$, when considering the \LS~ estimators and the robust ones defined through \textbf{Step 4} and \textbf{Step N4}. The green curve corresponds to the true variance function.}
\end{center} 
\end{figure}

\begin{figure}[ht!]
	\begin{center}		
 \newcolumntype{M}{>{\centering\arraybackslash}m{\dimexpr.1\linewidth-1\tabcolsep}}
   \newcolumntype{G}{>{\centering\arraybackslash}m{\dimexpr.32\linewidth-1\tabcolsep}}
\begin{tabular}{GGG}
			\small\LS & \small\HMM & \small\HWMM\\ 	
			\multicolumn{3}{c}{$C_{1}$}\\[-3ex]	
			\includegraphics[scale=0.3]{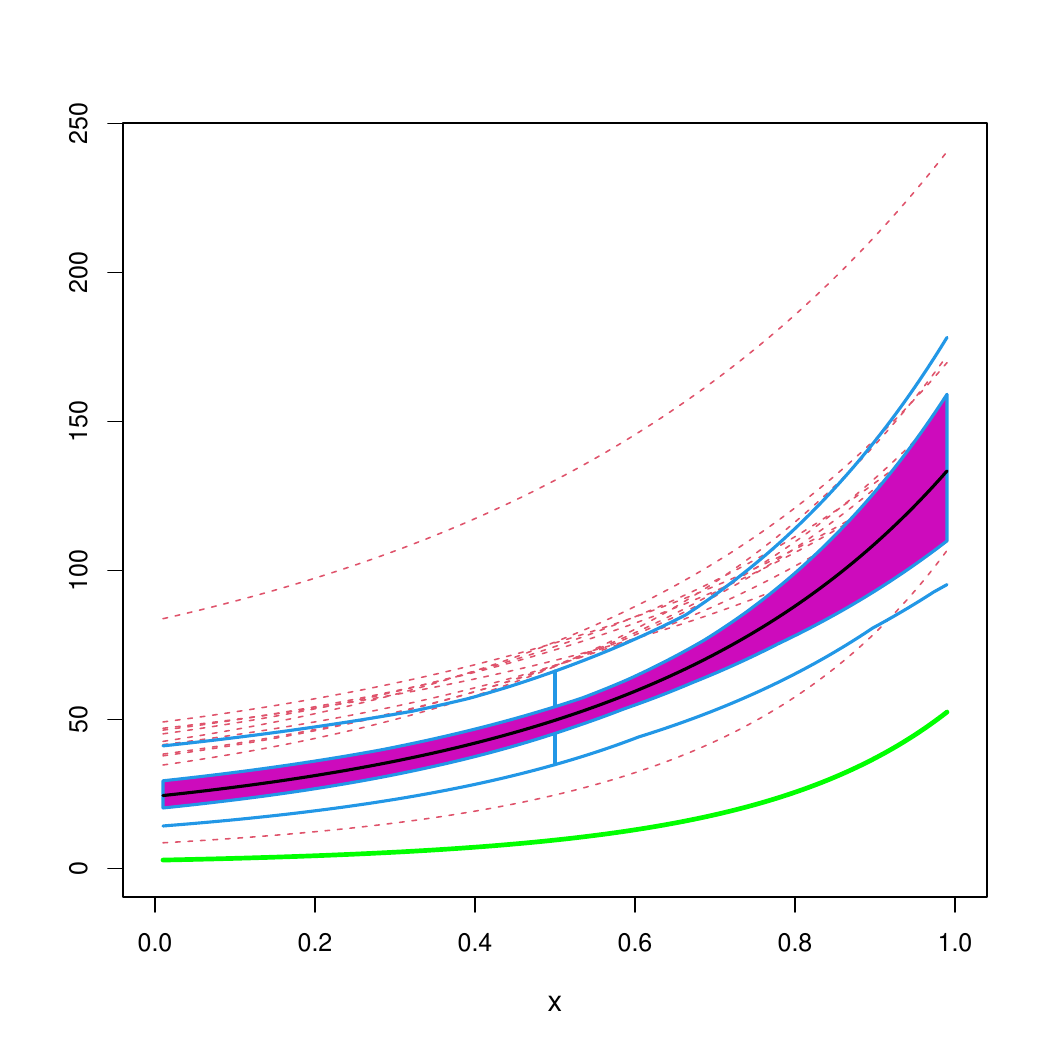} &  
			\includegraphics[scale=0.3]{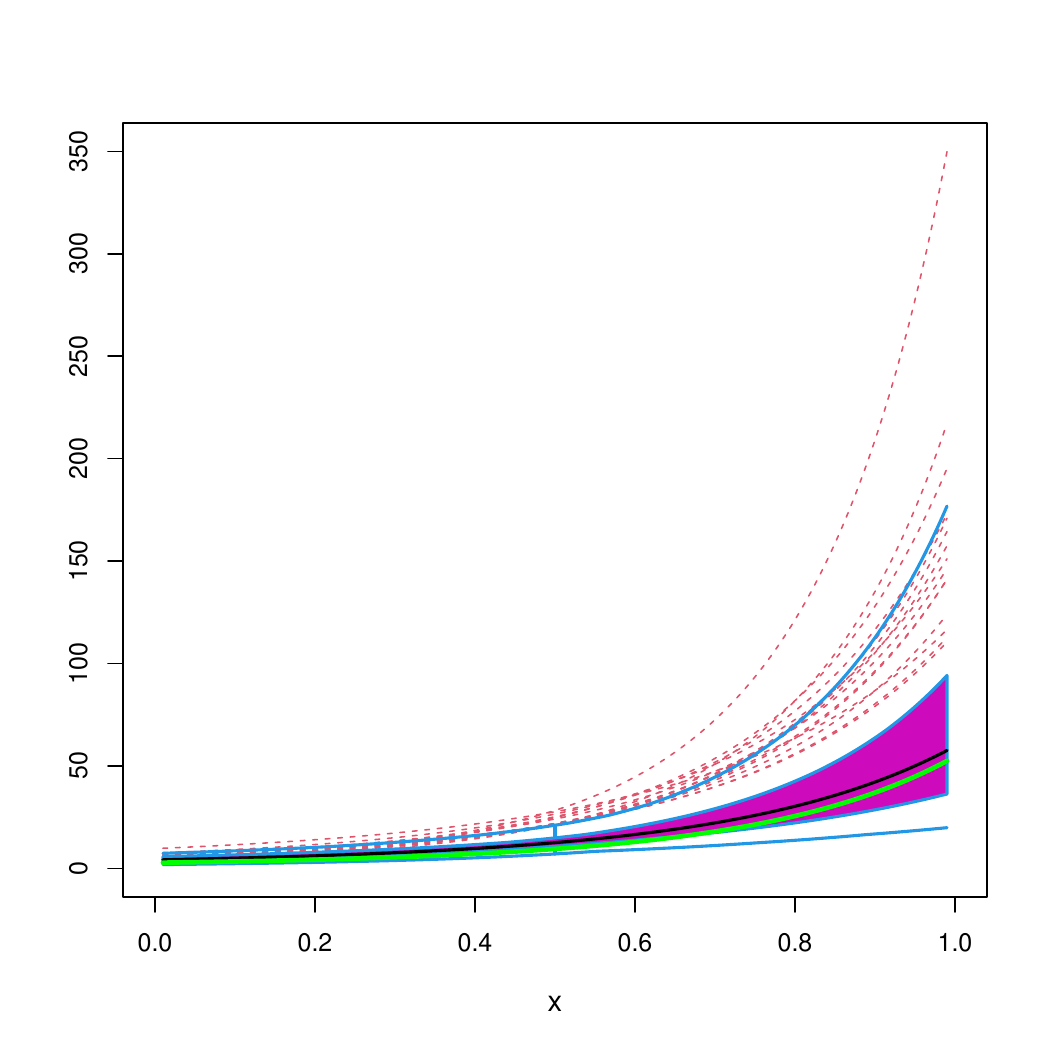}  &  
			\includegraphics[scale=0.3]{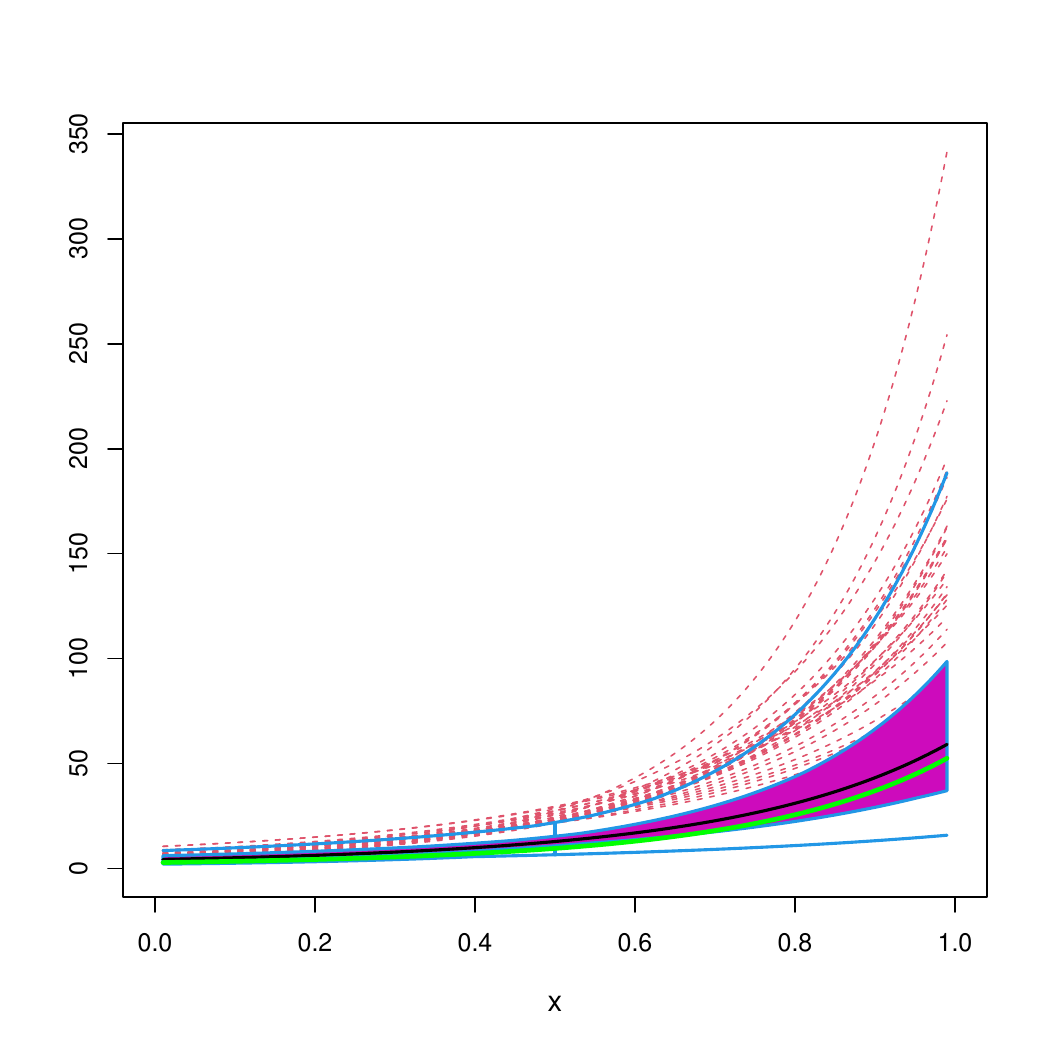}\\ 
			\multicolumn{3}{c}{$C_{2}$}\\[-4ex]				
			\includegraphics[scale=0.3]{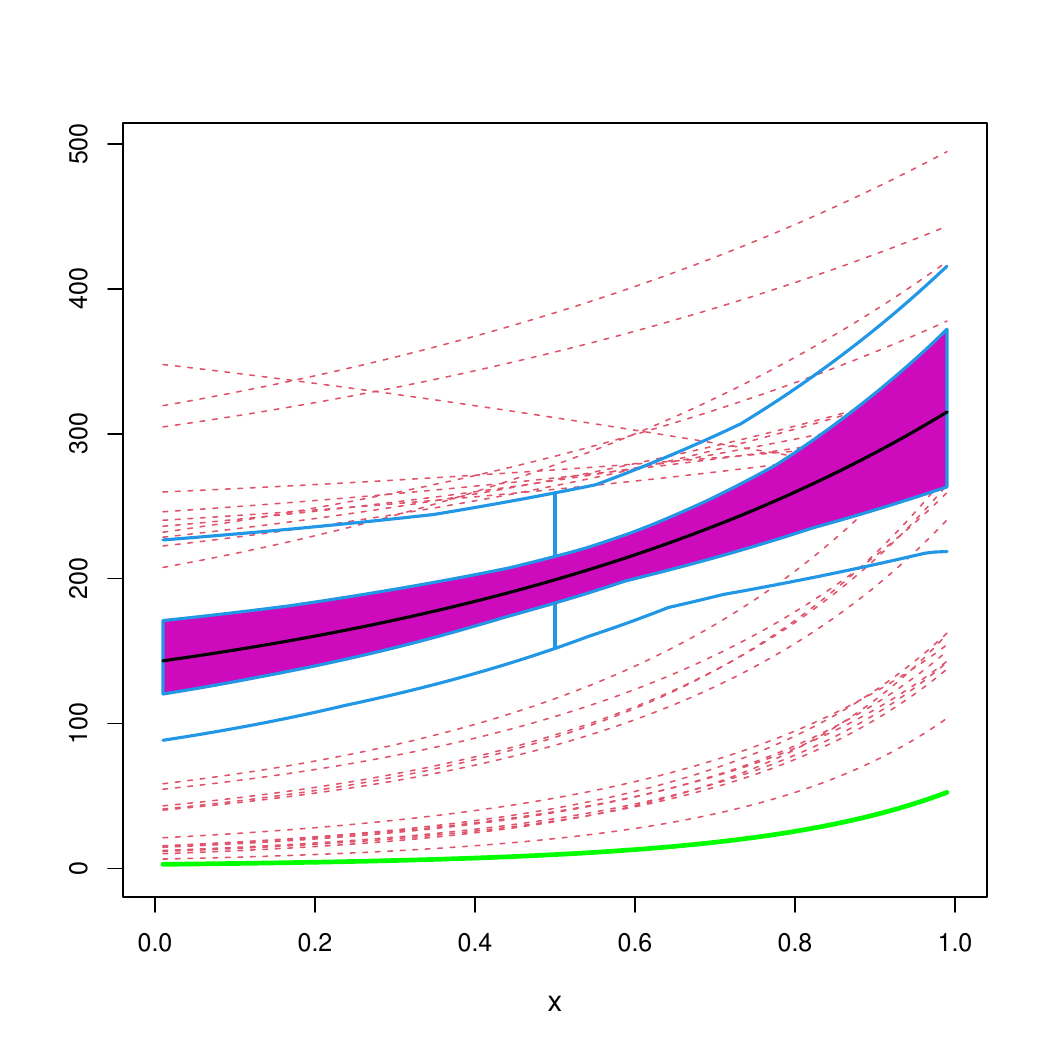} &  
			\includegraphics[scale=0.3]{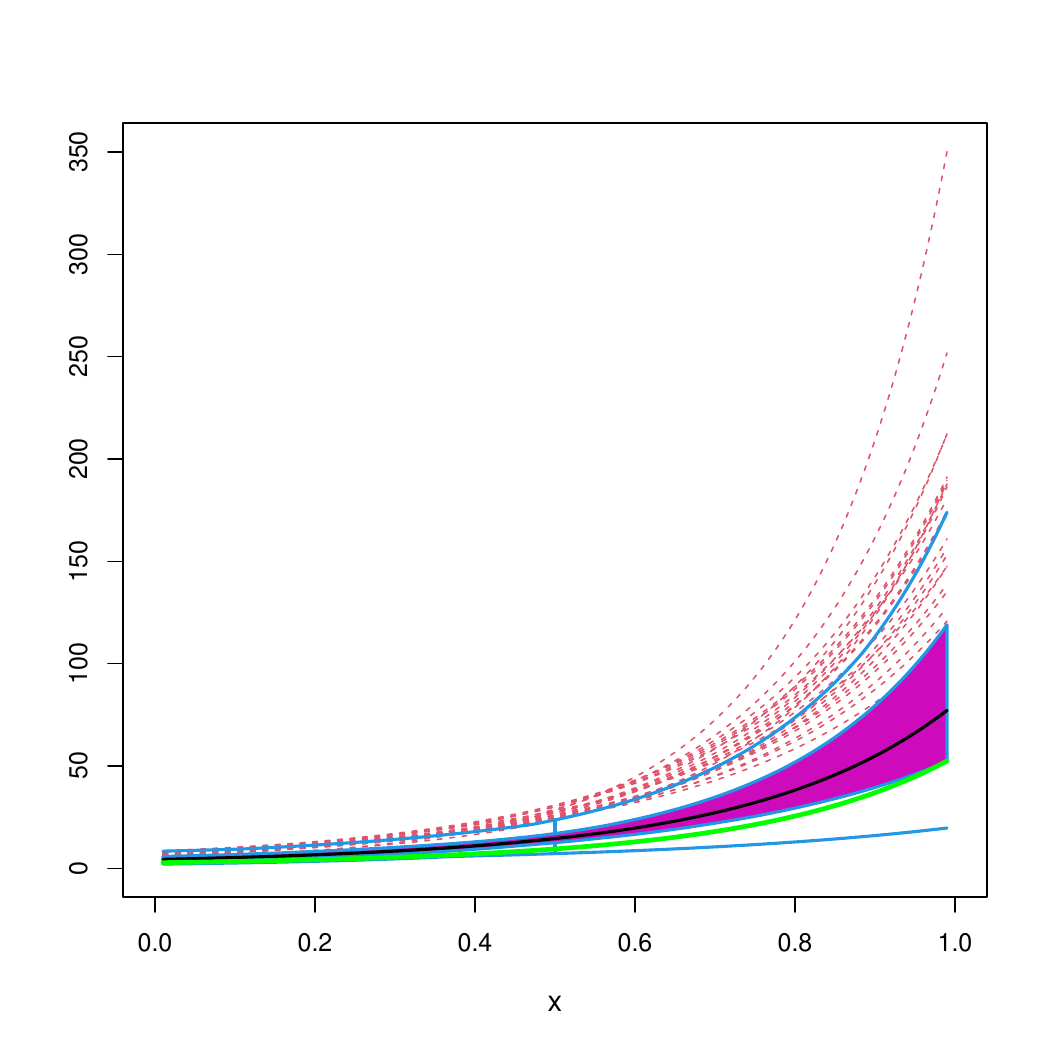}  &  
			\includegraphics[scale=0.3]{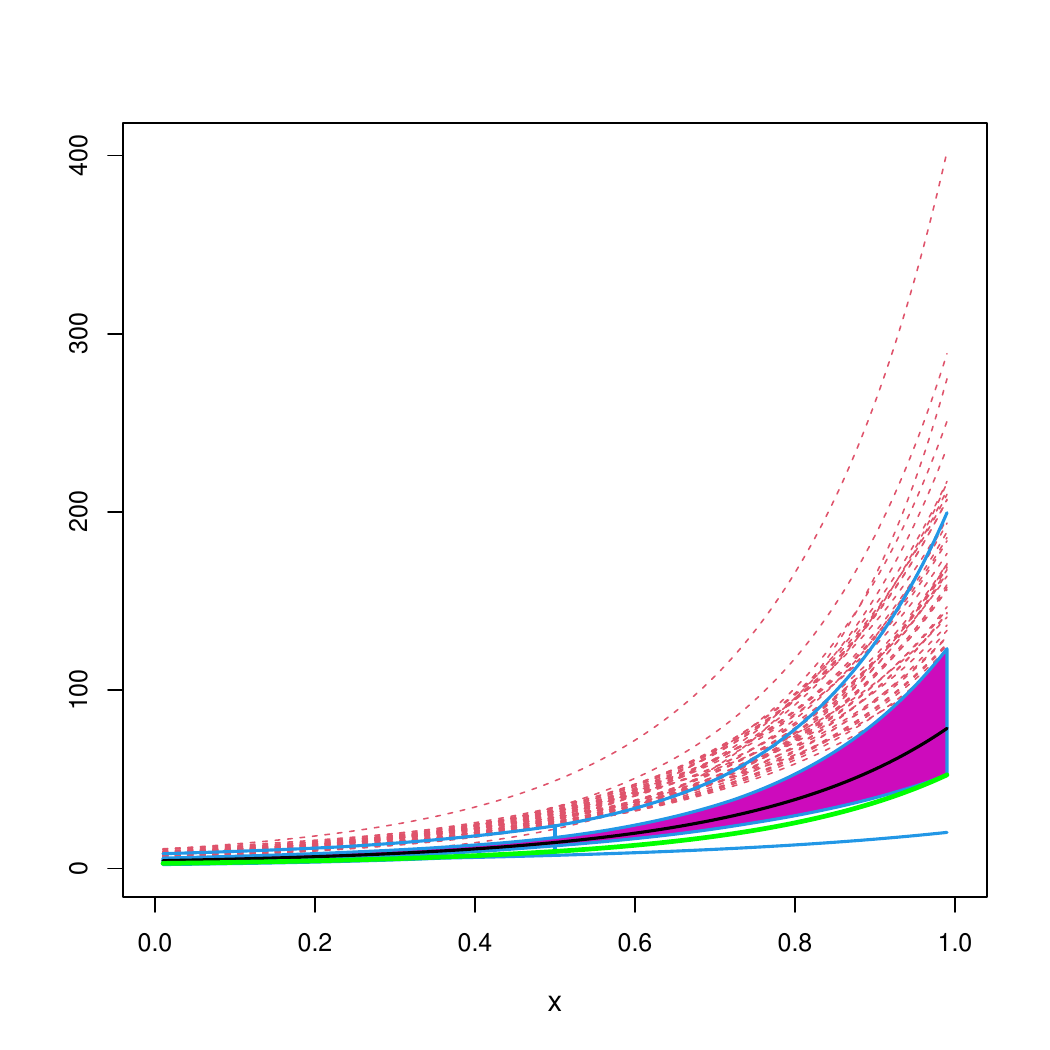}\\ 
			\multicolumn{3}{c}{$C_3$}\\	[-4ex]
			\includegraphics[scale=0.3]{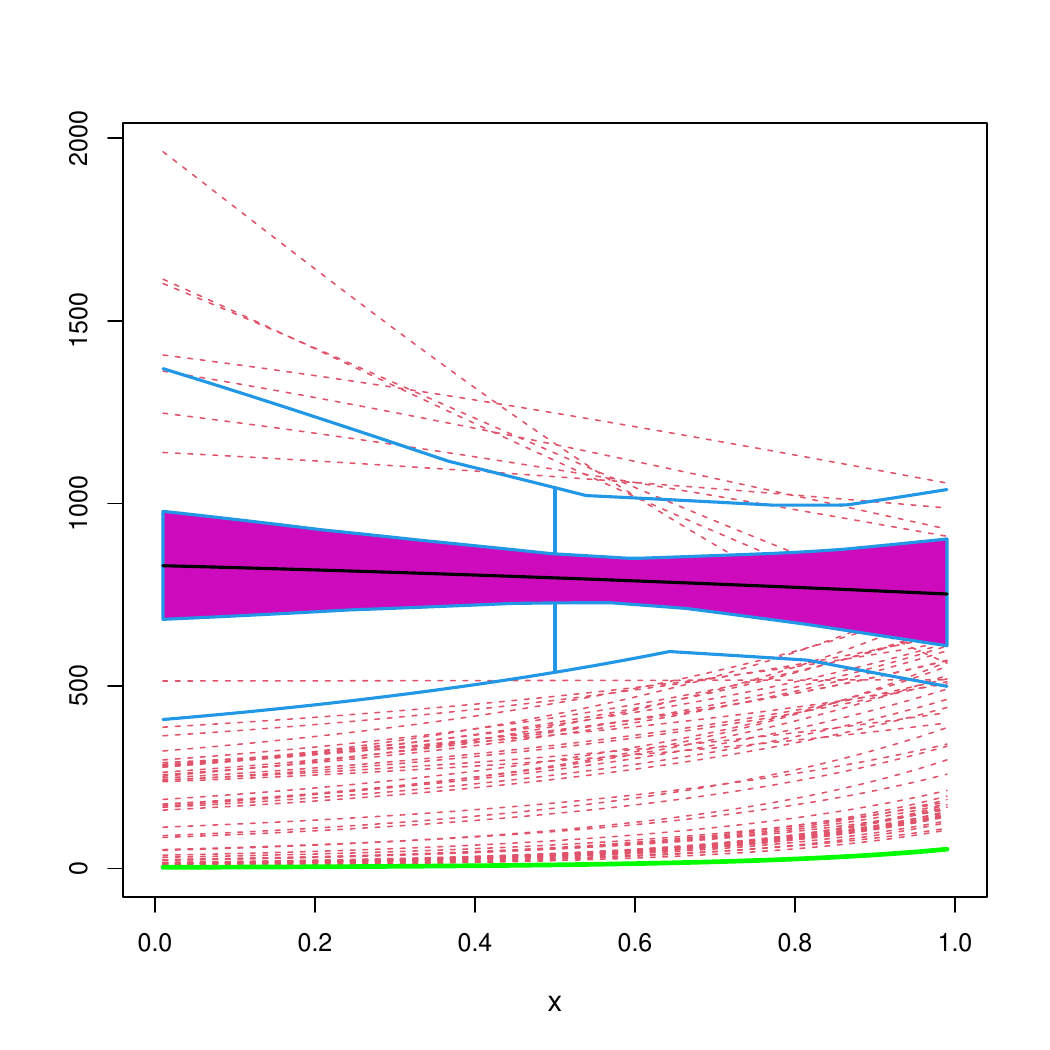} &  
			\includegraphics[scale=0.3]{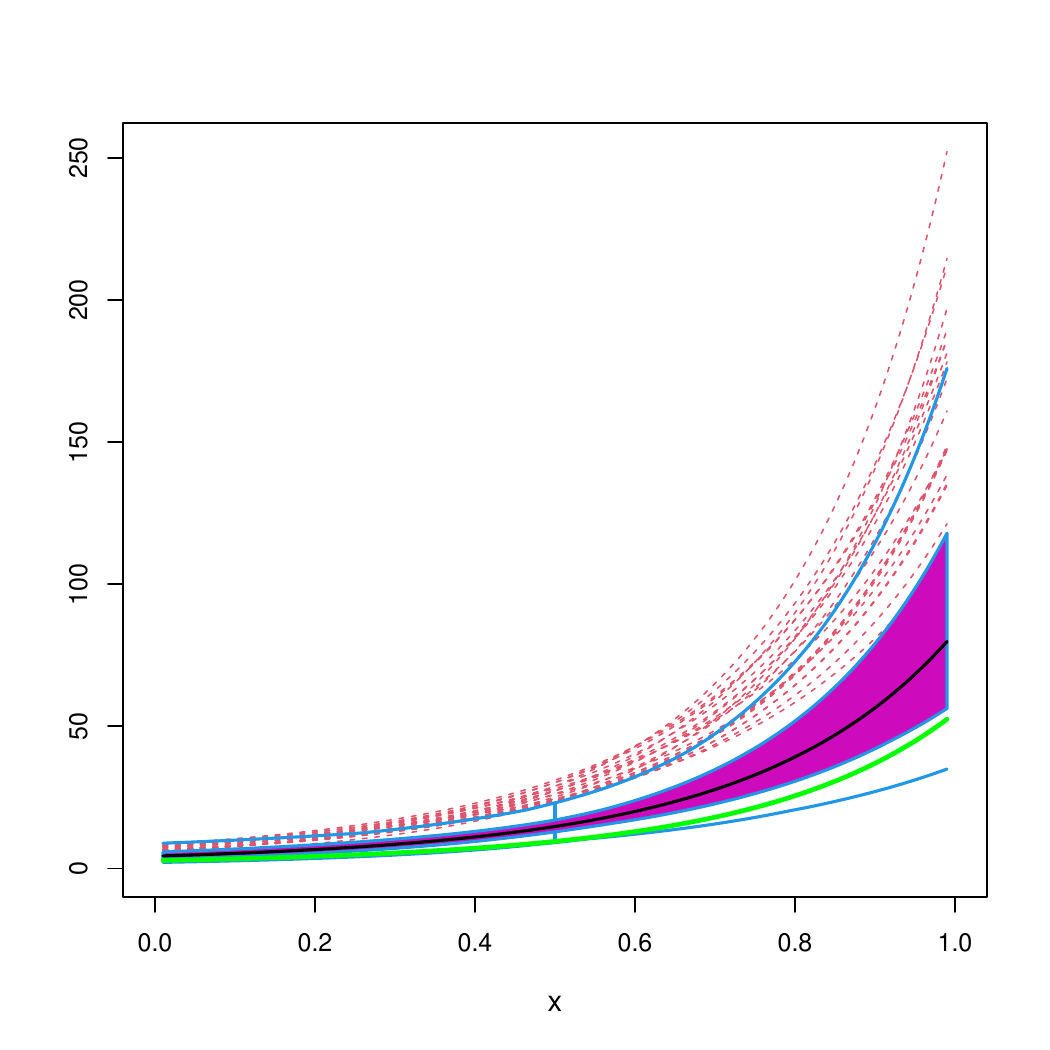}  & 
			 \includegraphics[scale=0.3]{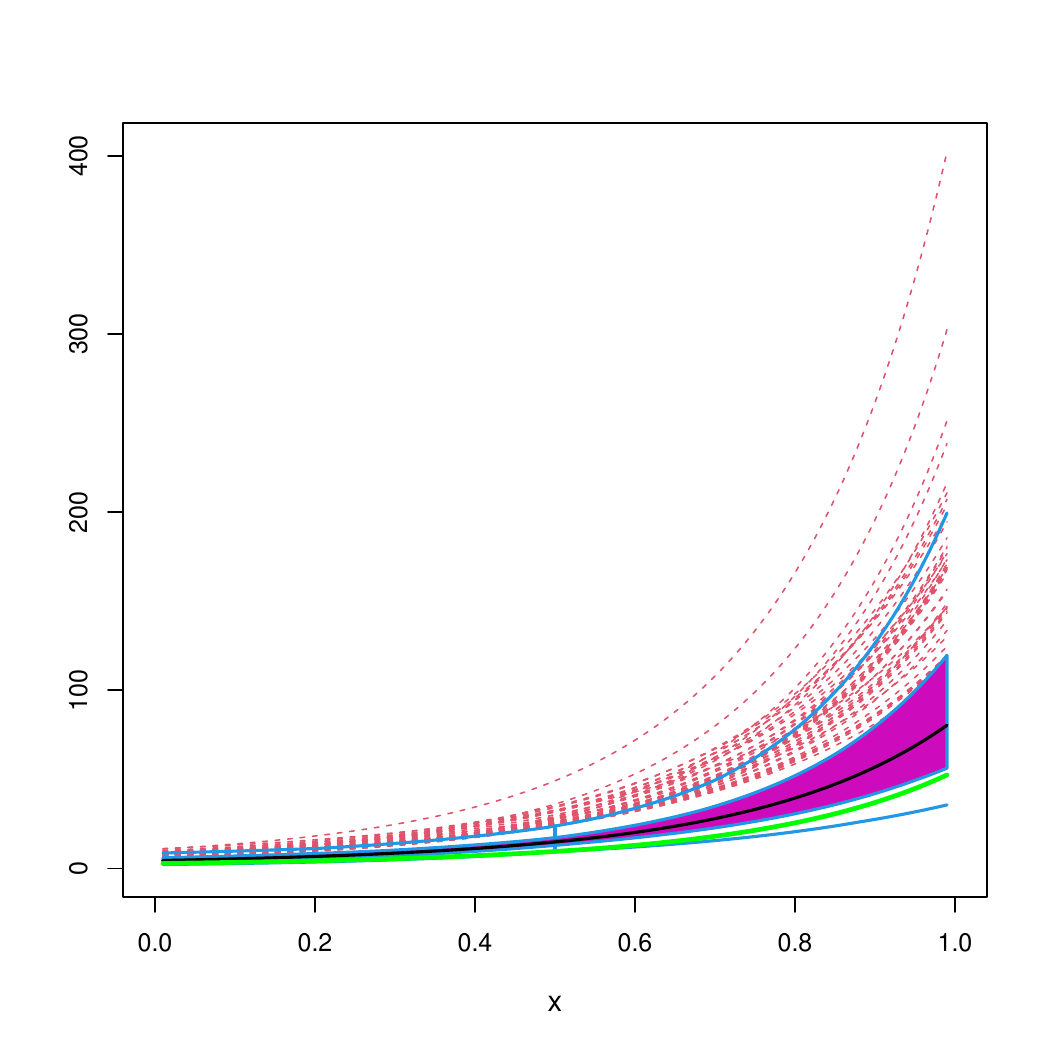} 
			\end{tabular}
		\vskip-0.1in \caption{ \small \label{fig:variance_C123} Functional boxplots of the estimated variance function, when considering the \LS ~ estimators and the robust ones defined through \textbf{Step 4},  for samples with vertical outliers. The green curve corresponds to the true variance function.}
	\end{center} 
\end{figure}

\begin{figure}[ht!]
	\begin{center}		
 \newcolumntype{M}{>{\centering\arraybackslash}m{\dimexpr.1\linewidth-1\tabcolsep}}
   \newcolumntype{G}{>{\centering\arraybackslash}m{\dimexpr.32\linewidth-1\tabcolsep}}
\begin{tabular}{GGG}
			\small\LS & \small\HMM$_{\nuevo}$ & \small\HWMM$_{\nuevo}$\\ 	
			\multicolumn{3}{c}{$C_{1}$}\\[-3ex]	
			\includegraphics[scale=0.3]{homos-C7_1variance_LS_bis.pdf} &  
			\includegraphics[scale=0.3]{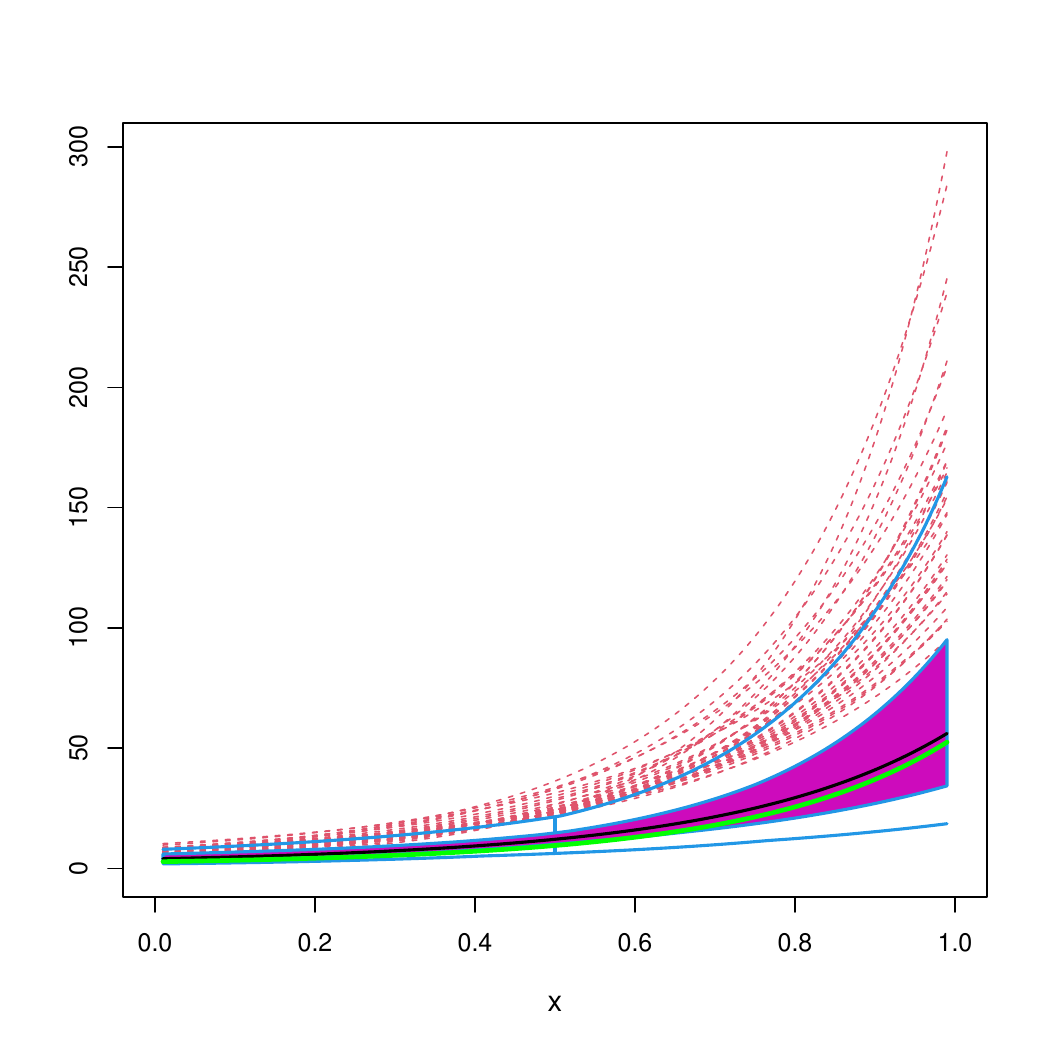}  &  
			\includegraphics[scale=0.3]{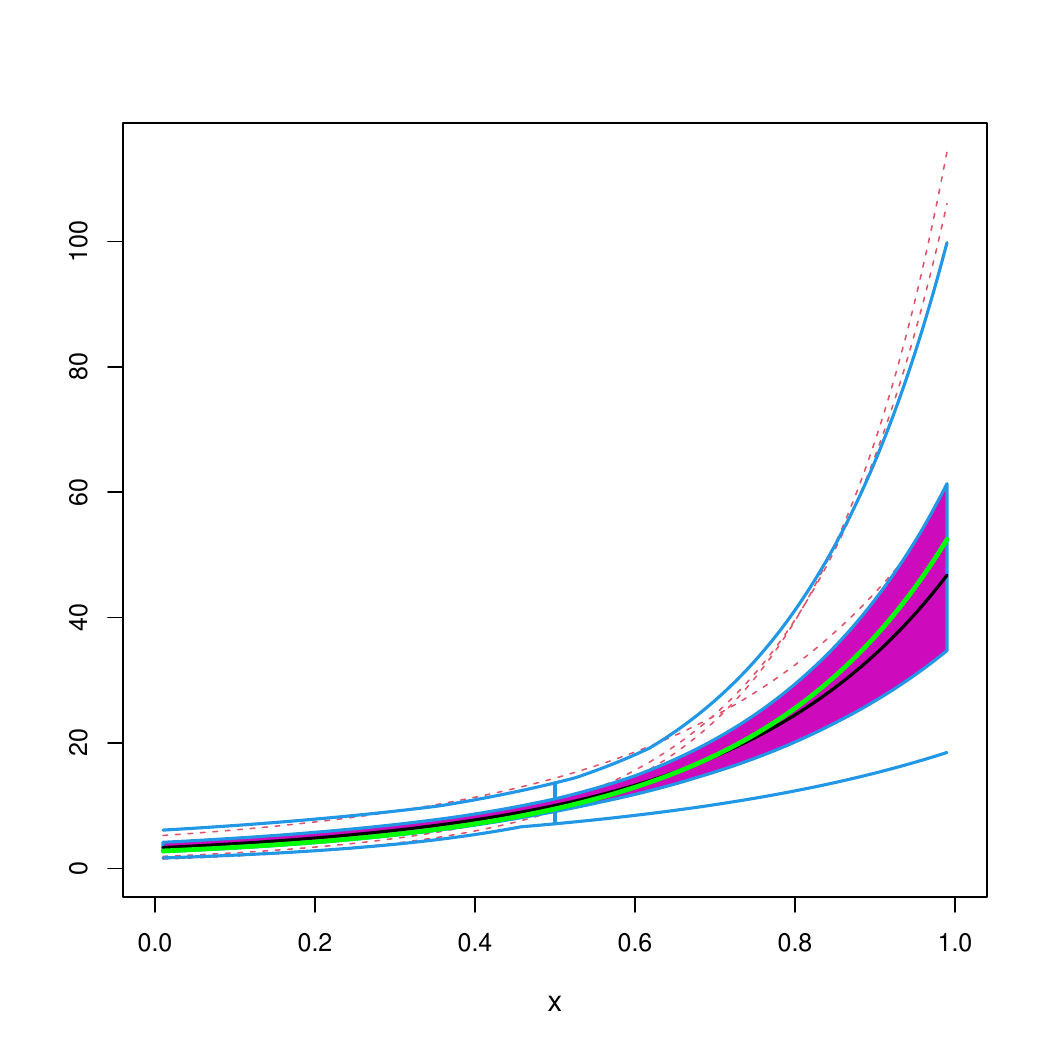}\\ 
			\multicolumn{3}{c}{$C_{2}$}\\[-4ex]				
			\includegraphics[scale=0.3]{homos-C7_2variance_LS_bis.pdf} &  
			\includegraphics[scale=0.3]{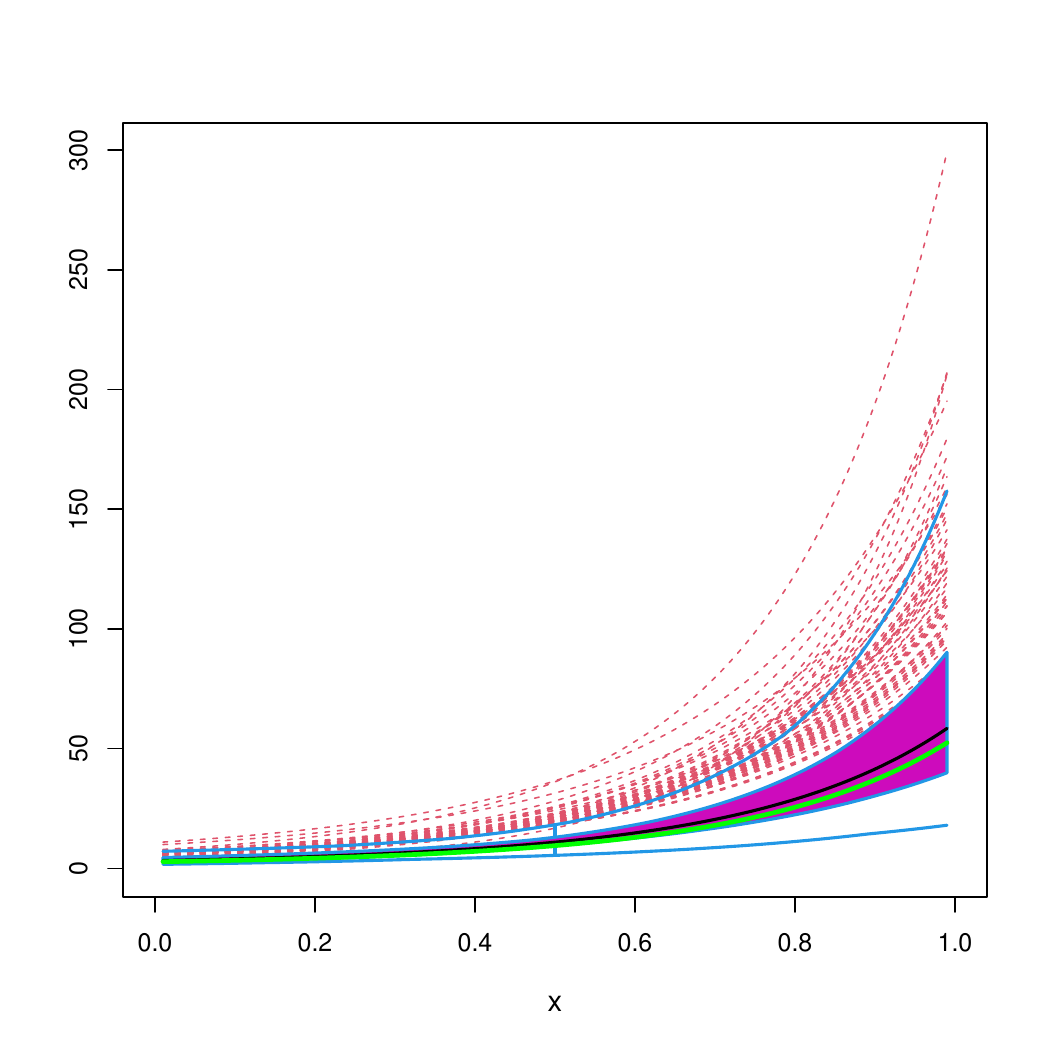}  &  
			\includegraphics[scale=0.3]{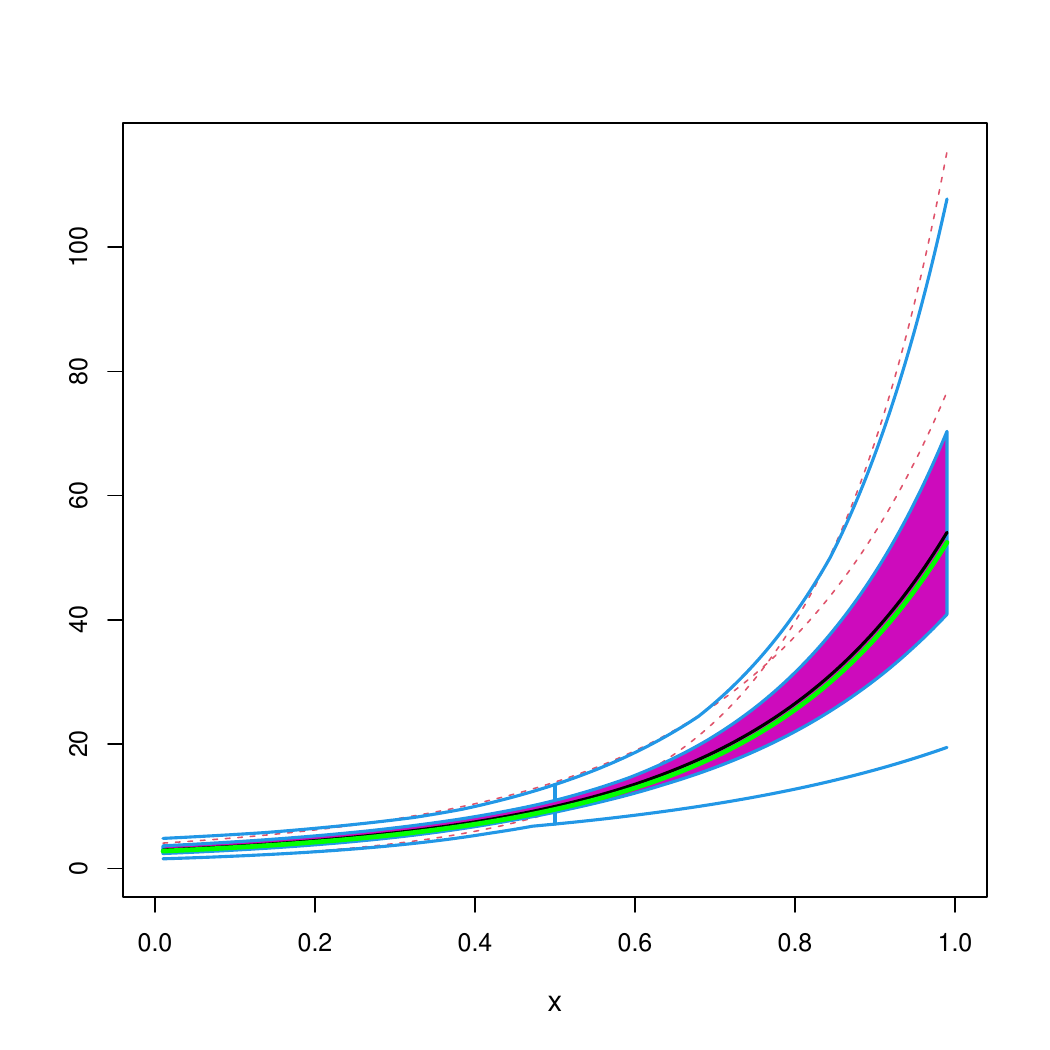}\\
			\multicolumn{3}{c}{$C_3$}\\	[-4ex]
			\includegraphics[scale=0.3]{homos-C7_3variance_LS_bis.pdf} &  
			\includegraphics[scale=0.3]{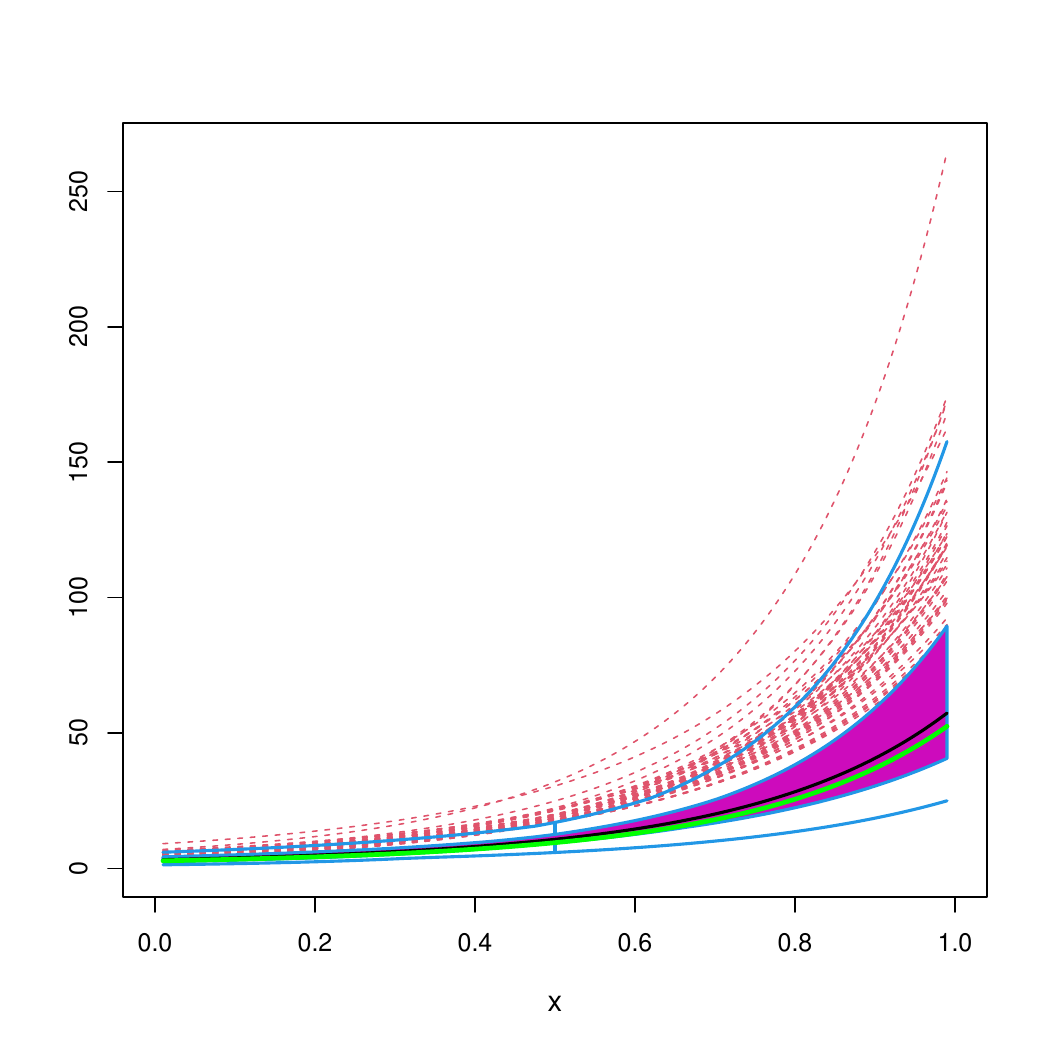}  &  
			\includegraphics[scale=0.3]{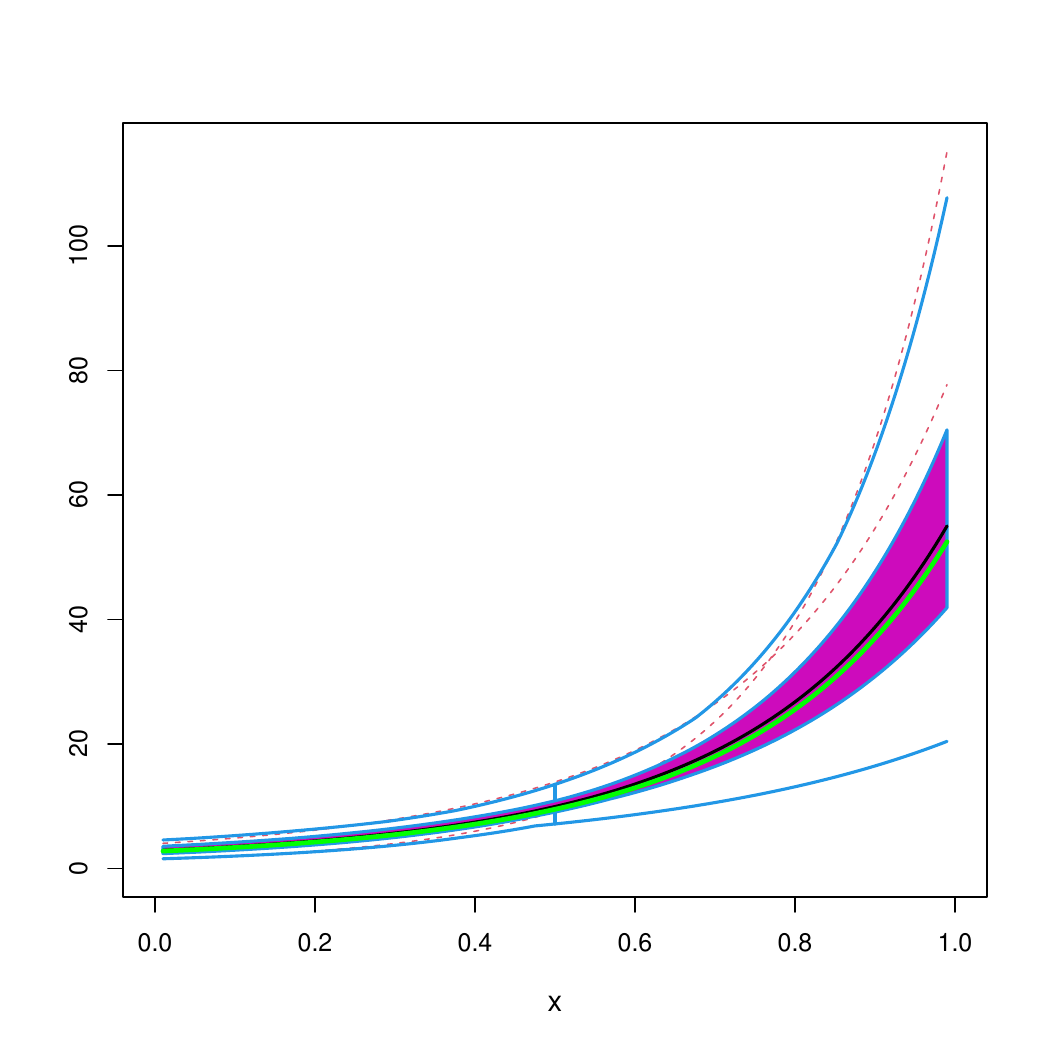}\\
			\end{tabular}
		\vskip-0.1in \caption{ \small \label{fig:variance_C123-N4} Functional boxplots of the estimated variance function, when considering the \LS ~ estimators and the robust ones defined through \textbf{Step N4},  for samples with vertical outliers. The green curve corresponds to the true variance function.}
	\end{center} 
\end{figure}

In order to visualize the influence of outliers in a whole picture, Figures \ref{fig:variance_C0} and \ref{fig:variance_C123} contain the functional boxplots of the estimated variance functions obtained with the classical estimators labelled \LS~ and with those obtained using \textbf{Step 4} to compute $\wlam$,  while Figure \ref{fig:variance_C123-N4} present the functional boxplots when computing the estimators through  \textbf{Step N4}. The functional boxplots related to $\wsigma(x)=\wsigma \upsilon(x, \wlam)$, when $ \wlam$ corresponds to the robust procedure defined in \textbf{Step 2} and \textbf{Step N2} are not presented here, since due to the bias observed in  Figure \ref{fig:lambda-bxp}, they lead to more biased curve estimates.  In these plots, the
area in magenta represents the central region containing the 50\% deepest curves; the dotted red lines correspond to outlying curves; the black line indicates the deepest curve, and the green line is the true variance function. Figure \ref{fig:variance_C0} shows that, except for a few outlying curves, all the methods accomplish the same goal when they  deal with clean data: they capture the essence of the variance function. The estimates obtained through \textbf{Step N4} seem to outperform those based on \textbf{Step 4}.

A different situation arises in the presence of atypical data. The impact of the introduced vertical outliers on the estimation of the variance function for the \LS ~ method is clear. In fact,  in Figure \ref{fig:variance_C123} it becomes evident that the estimated curves with \LS ~ are completely distorted. Indeed, under $C_1$ to $C_3$ the true variance function lies below the \LS ~ estimates. Conversely,  both robust estimators lead to very stable results. When considering  the estimators defined through \textbf{Step 4} the variance curves obtained with the robust \HMM ~and \HWMM ~methods reproduce approximately the true variance function, even when under $C_3$ the true function is below the limit of the central band containing the 50\% deepest curves, see Figure \ref{fig:variance_C123}. In contrast, as revealed in Figure  \ref{fig:variance_C123-N4} the estimators defined through \textbf{Step N4} not only capture the shape of the true variance function, but the deepest curve is close to the true one.

\clearpage
\subsection{High leverage outliers}{\label{sec:highlev}} 

%{OJO: estos estan calculados CON la modificacion de bis2. Corresponde a salidas en salidas_9_conbis2)

In this second experiment, we contaminate the data by replacing the last 5\% of observations of each sample by $(x_0,y_0)$ where $x_0=3.5+ u$, while $y_0$ takes the value $90$ in $D_1$  and $150$ in $D_2$, leading to two different severe contamination schemes with high leverage points. As in Section \ref{sec:vertical}, we added a  small noise $u$ to the value $3.5$ generated as $N(0,10^{-8})$ to avoid numerical instability.

\begin{figure}[ht!]
	\begin{center}
 	 \renewcommand{\arraystretch}{0.4}
 \newcolumntype{M}{>{\centering\arraybackslash}m{\dimexpr.1\linewidth-1\tabcolsep}}
   \newcolumntype{G}{>{\centering\arraybackslash}m{\dimexpr.35\linewidth-1\tabcolsep}}
\begin{tabular}{GGG}
			$C_0$  & $D_{1}$ & $D_{2}$\\[-3ex]
			\includegraphics[scale=0.3]{grafico_C0.pdf} &
			\includegraphics[scale=0.3]{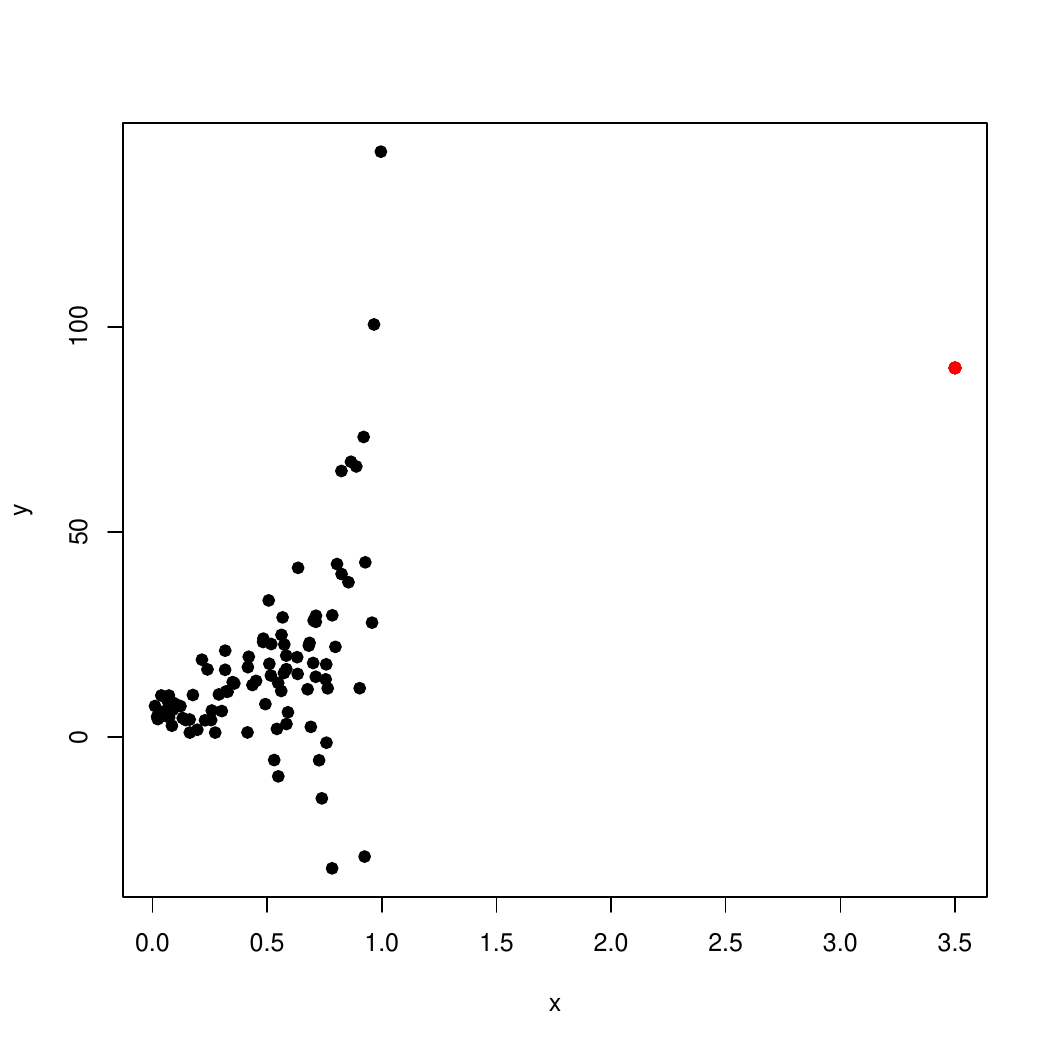} &
			\includegraphics[scale=0.3]{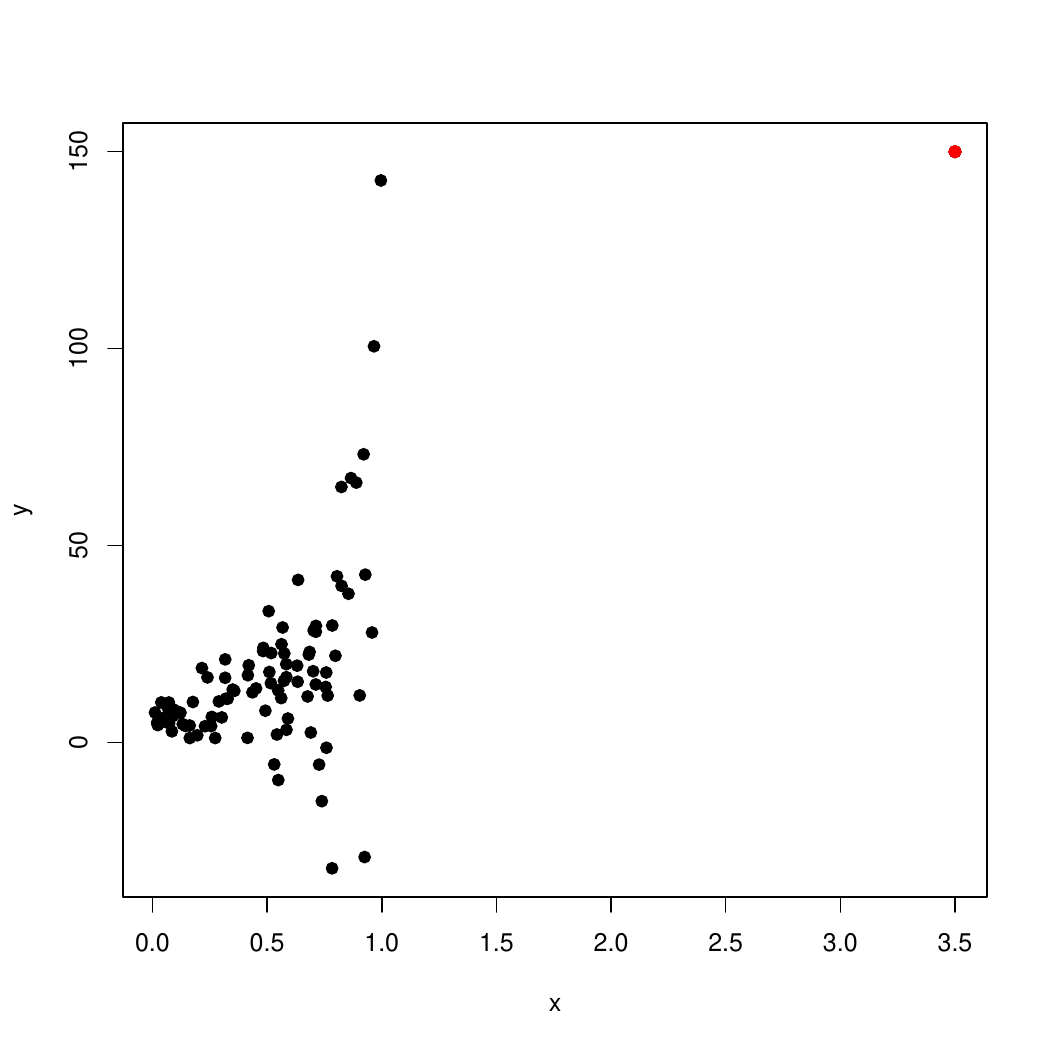}			
		\end{tabular}
		\vskip-0.1in \caption{\label{fig:datos_simulados2} \small Synthetic data obtained in one replication: the sample exhibited in the left panel was generated under scheme $C_0$, the   two other panels correspond to contaminations with high leverage points $D_{1}$ and $D_{2}$.}
	\end{center} 
\end{figure}

The central and right panels in Figure \ref{fig:datos_simulados2} illustrate the leverage and the size of the introduced outliers in one generated sample.

\begin{table}[ht!]
	\centering
	\begin{tabular}{|c|c|c|c|c|c|c|}
		\hline
		&\multicolumn{3}{c|}{MSE} &\multicolumn{3}{c|}{Bias}\\
		\cline{2-7}
		& $C_0$ & $D_1$ & $D_2$ & $C_0$ & $D_1$ & $D_2$\\ 
		\hline
        \LS & 1.953 & 7.266 & 6.363 & -0.012 & 6.991 & 6.072 \\ 
        \MM & 1.326 & 2.099 & 2.182 & 0.049 & 1.195 & 1.499 \\ 
        \WMM & 1.448 & 1.553 & 1.530 & 0.022 & 0.098 & 0.129 \\ 
        \hline
        \HLS & 0.637 & 3.646 & 3.218 & 0.043 & 1.786 & 1.655 \\ 
        \HMM & 0.646 & 1.748 & 1.857 & 0.036 & 0.580 & 0.511 \\ 
        \HWMM & 0.659 & 1.088 & 1.077 & 0.033 & 0.251 & 0.249 \\ 
\hline
  \HMM$_{\nuevo}$ & 0.648 & 0.728 & 0.725 & 0.039 & 0.117 & 0.118 \\ 
  \HWMM$_{\nuevo}$ & 0.656 & 0.758 & 0.760 & 0.037 & 0.040 & 0.040 \\ 
		\hline
	\end{tabular}
	\caption{\small \label{fig:betasD} MSE and bias of the estimators of $\beta_{01}$.} 
\end{table}

\begin{table}[ht!]
	\centering
	\begin{tabular}{|c|c|c|c|c|c|c|}
		\hline
       &\multicolumn{3}{c|}{MSE} &\multicolumn{3}{c|}{Bias}\\
       \cline{2-7}
       & $C_0$ & $D_1$ & $D_2$ & $C_0$ & $D_1$ & $D_2$\\ 
       \hline
\LS & 0.978 & 1.414 & 1.250 & 0.113 & -1.414 & -1.249 \\ 
\MM & 0.654 & 0.989 & 0.964 & 0.005 & -0.518 & -0.618 \\ 
\WMM & 0.703 & 0.756 & 0.743 & 0.029 & -0.011 & -0.028 \\ 
\hline
\HLS & 0.294 & 0.705 & 0.666 & -0.021 & -0.385 & -0.373 \\ 
\HMM & 0.300 & 0.771 & 0.830 & -0.017 & -0.212 & -0.170 \\ 
\HWMM & 0.305 & 0.516 & 0.511 & -0.014 & -0.111 & -0.110 \\
\hline
 \HMM$_{\nuevo}$  & 0.301 & 0.332 & 0.328 & -0.018 & -0.046 & -0.048 \\ 
 \HWMM$_{\nuevo}$  & 0.303 & 0.355 & 0.354 & -0.016 & -0.013 & -0.013 \\ 
  		\hline
	\end{tabular}
	\caption{\label{fig:alfasD}MSE and bias of the estimators of $\beta_{02}$.} 
\end{table}

\begin{figure}[ht!]
	\begin{center}
		 \renewcommand{\arraystretch}{0.4}
 \newcolumntype{M}{>{\centering\arraybackslash}m{\dimexpr.1\linewidth-1\tabcolsep}}
   \newcolumntype{G}{>{\centering\arraybackslash}m{\dimexpr.45\linewidth-1\tabcolsep}}
%\begin{tabular}{MGG}
\begin{tabular}{M GG} 
			& \small $\wbeta_{01}$  & \small $\wbeta_{02}$ \\[-4ex]
			$C_0$ & 
			\includegraphics[scale=0.33]{bxp-betas-C0_STEPN1toN4_2.pdf} & 
			\includegraphics[scale=0.33]{bxp-alfas-C0_STEPN1toN4_2.pdf} \\[-6ex]
			$D_1$ &  
			\includegraphics[scale=0.33]{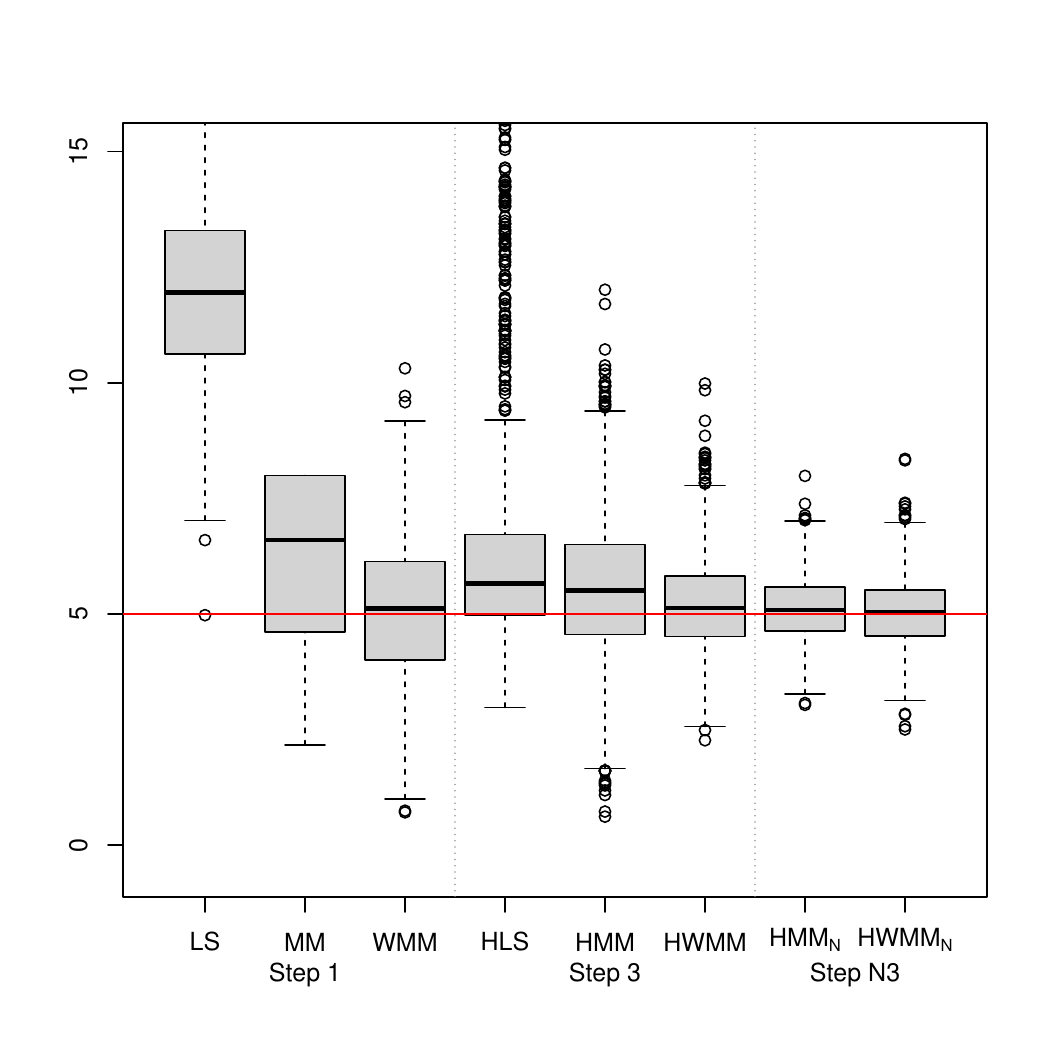} &  
			\includegraphics[scale=0.33]{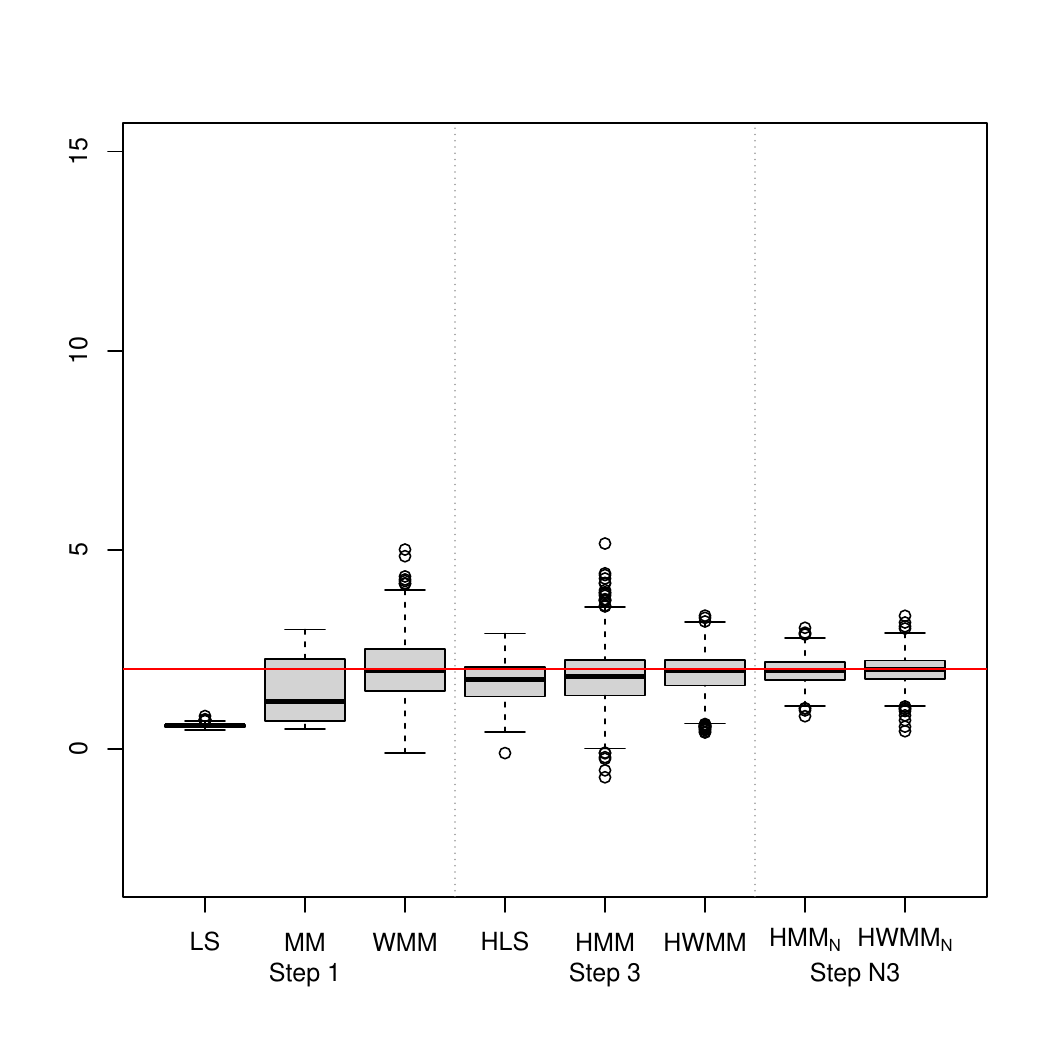} \\[-6ex]
			$D_2$ & 
			\includegraphics[scale=0.33]{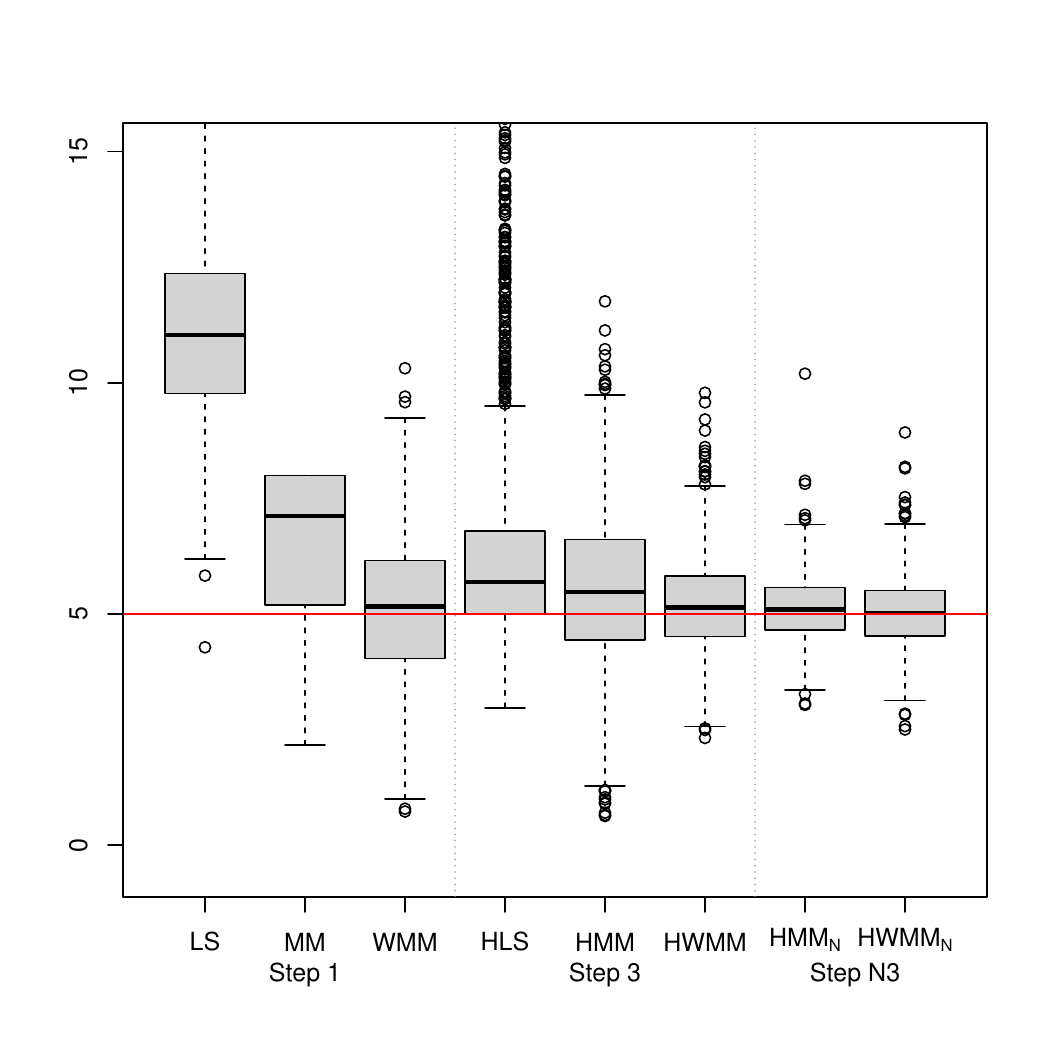} &  
			\includegraphics[scale=0.33]{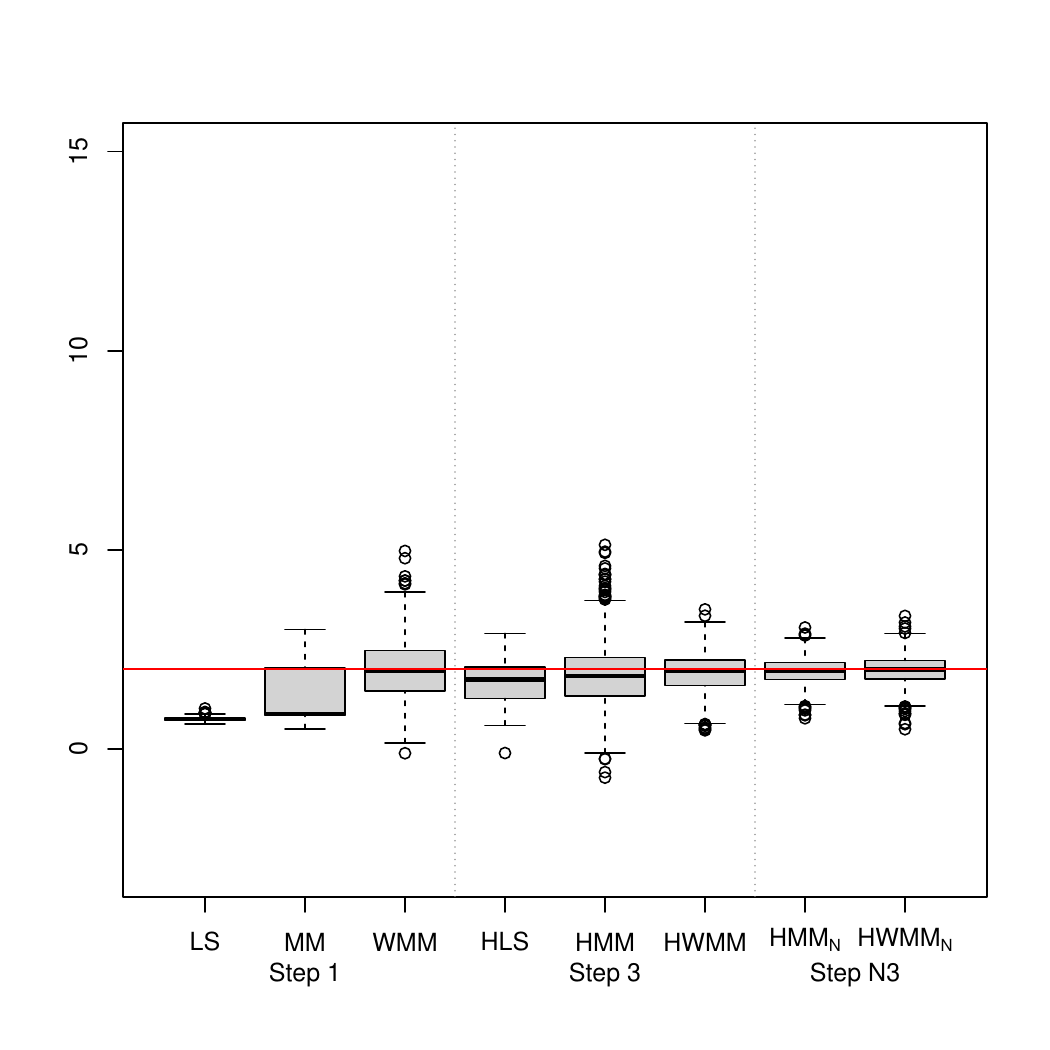}  
		\end{tabular}			
			\vskip-0.1in \caption{\small \label{fig:betaH0-C91-C92}  Boxplots of the estimates of the model parameters, under $C_{0}$ and contaminations  $D_{1}$ and $D_2$.}
	\end{center} 
\end{figure}

\begin{figure}[ht!]
	\begin{center}
		 \renewcommand{\arraystretch}{0.4}
 \newcolumntype{M}{>{\centering\arraybackslash}m{\dimexpr.1\linewidth-1\tabcolsep}}
   \newcolumntype{G}{>{\centering\arraybackslash}m{\dimexpr.45\linewidth-1\tabcolsep}}
%\begin{tabular}{MGG}
\begin{tabular}{M GG} 
			& \small $\wbeta_{01}$  & \small $\wbeta_{02}$ \\[-4ex]
			$C_0$ & 
			\includegraphics[scale=0.33]{bxp-betas-C0_STEPN1toN4_sinLS.pdf} & 
			\includegraphics[scale=0.33]{bxp-alfas-C0_STEPN1toN4_sinLS.pdf} \\[-6ex]
			$D_1$ &  
			\includegraphics[scale=0.33]{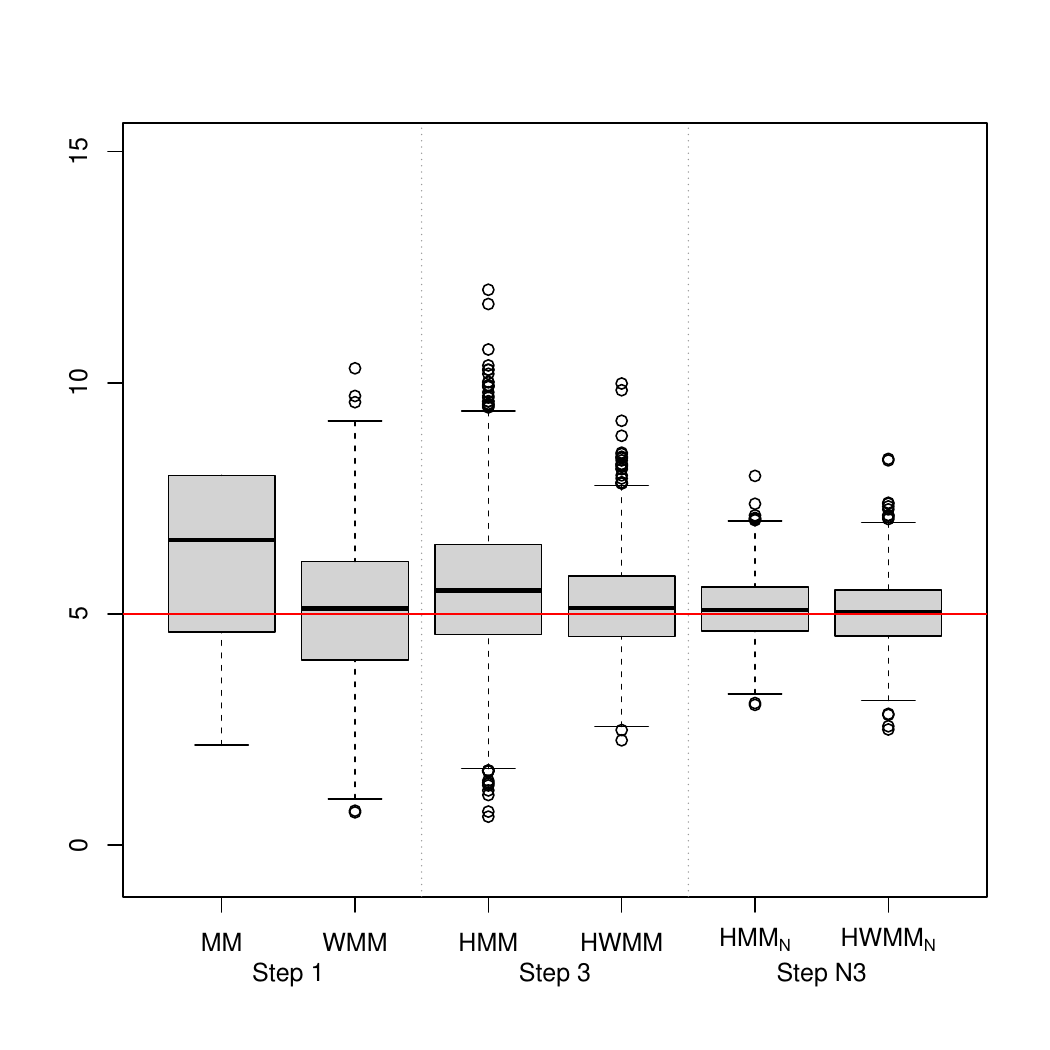} &  
			\includegraphics[scale=0.33]{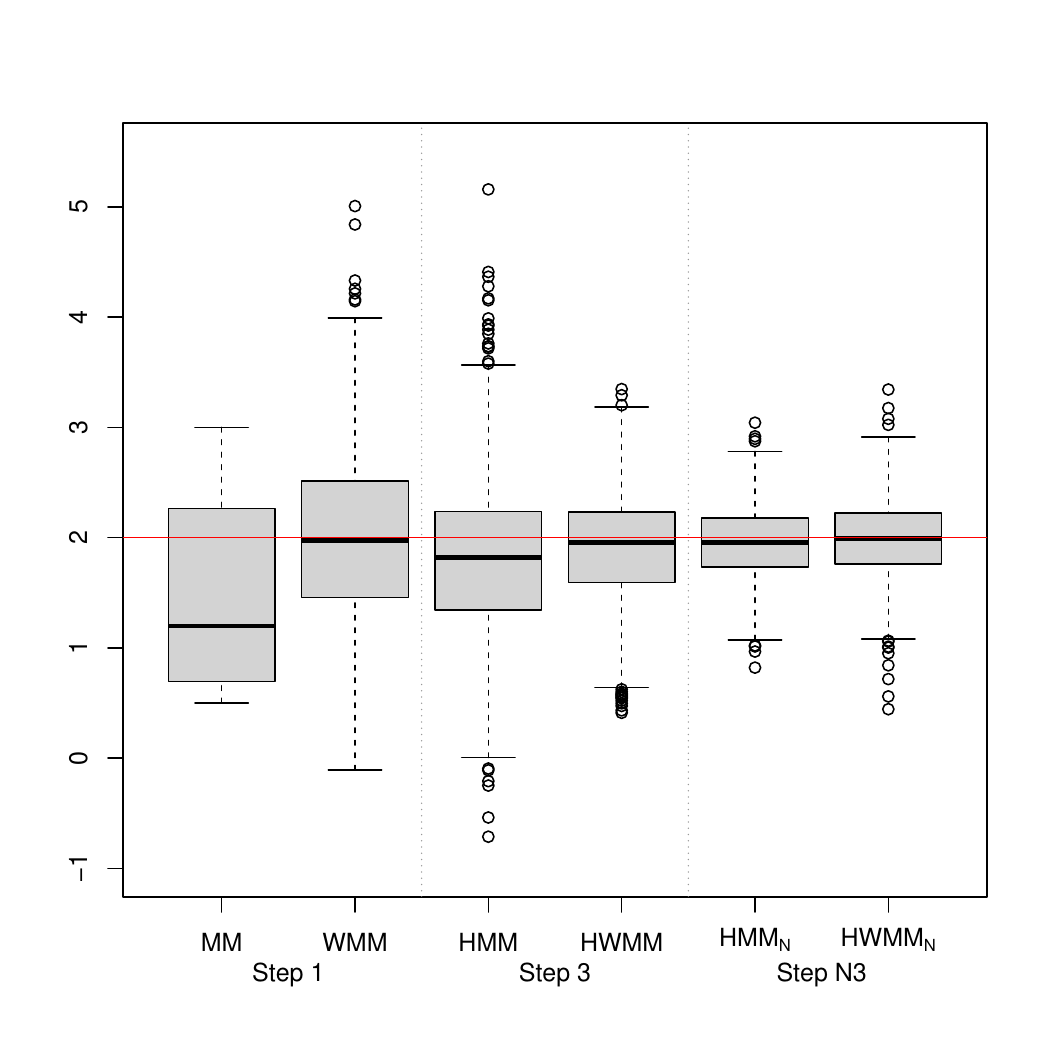} \\[-6ex]
			$D_2$ & 
			\includegraphics[scale=0.33]{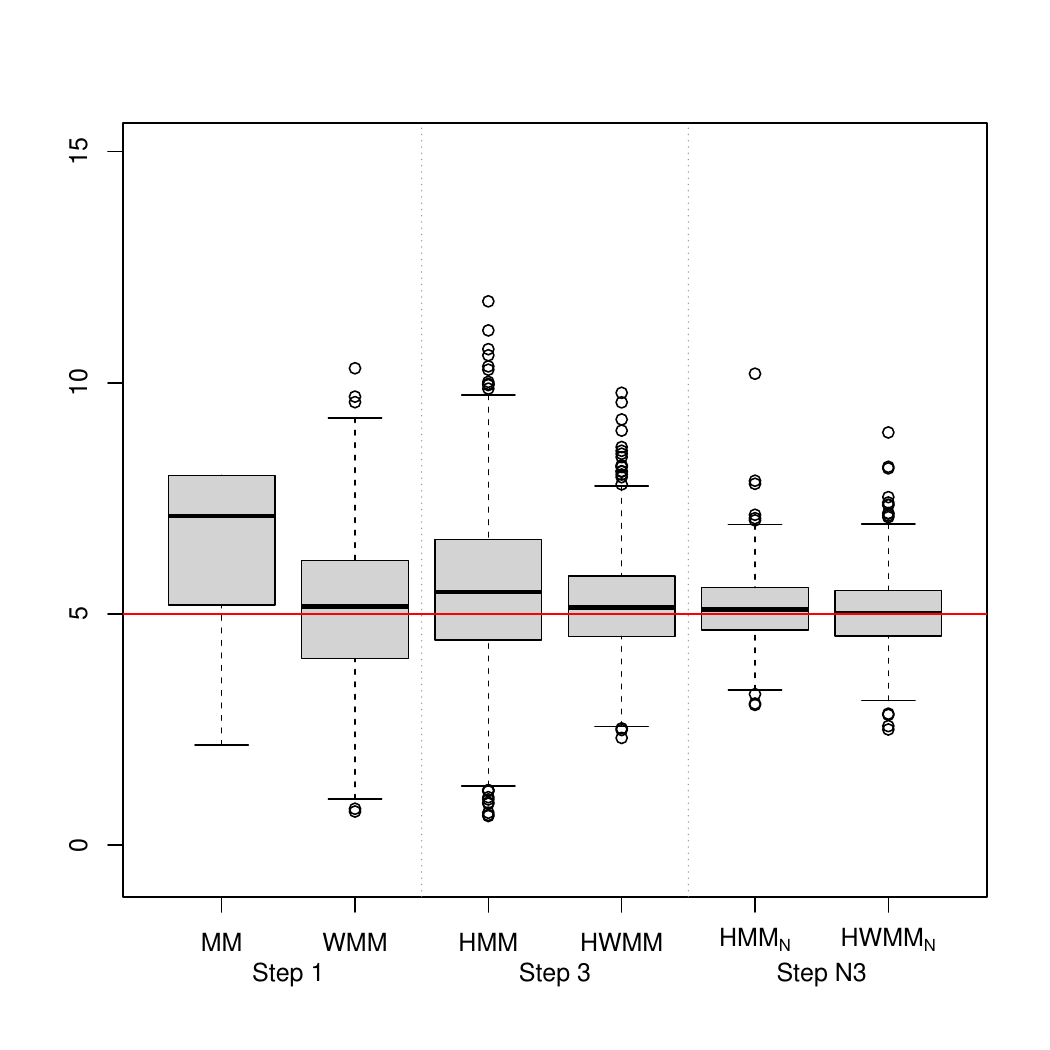} &  
			\includegraphics[scale=0.33]{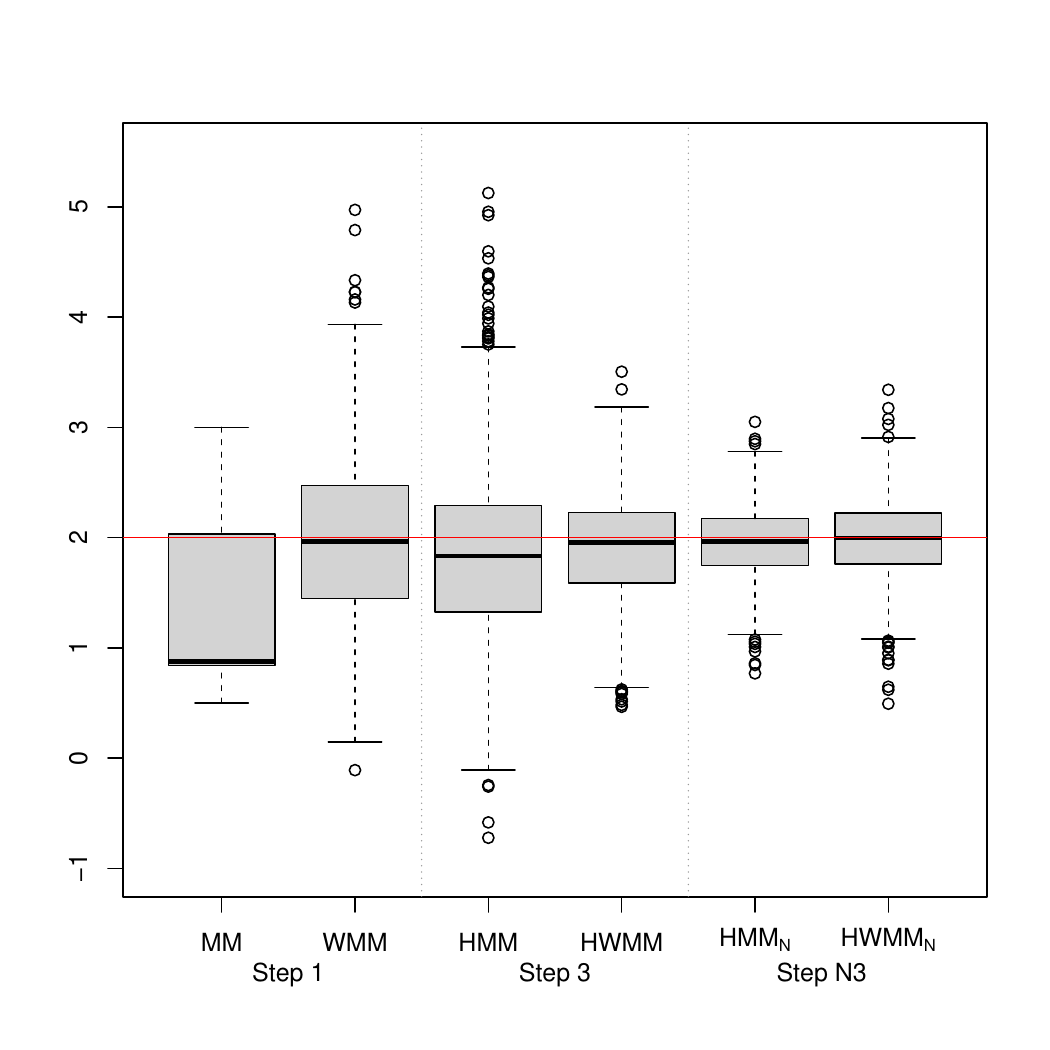}  
		\end{tabular}			
			\vskip-0.1in \caption{\small \label{fig:betaH0-C91-C92-sinLS}  Boxplots of the robust estimates of the model parameters, under $C_{0}$ and contaminations  $D_{1}$ and $D_2$.}
	\end{center} 
\end{figure}

On the one hand, Tables \ref{fig:betasD} and \ref{fig:alfasD} summarize the results of the numerical experiment in terms of the empirical mean squared error and bias of the estimators. Once again it becomes evident that the  estimators computed as if the model was homocedastic present larger MSE than their counterparts which take into account the heteroscedasticity,  in all scenarios. The estimators  \LS ~ and \MM ~ of both parameters are highly biased under $D_1$ and $D_2$, while the weighted $MM-$estimator \WMM~ presents an important decrease in the bias of  $\beta_{01}$ and $\beta_{02}$. Even when in all cases, the performance in terms of MSE  is improved when taking into account the heteroscedasticity in the estimation procedure, the \HWMM$_{\nuevo}$ estimators of both coefficients achieve the lowest  empirical mean squared errors and the smallest bias (in absolute value) in the contamination schemes $D_1$ and $D_2$. The reduction in mean square error and bias of  \HWMM$_{\nuevo}$ with respect to the other competitors is remarkable, leading to similar results than under $C_0$.
The  boxplots of the estimates displayed in Figures  \ref{fig:betaH0-C91-C92} and \ref{fig:betaH0-C91-C92-sinLS}  support these conclusions.

\begin{figure}[ht!]
  	 \renewcommand{\arraystretch}{0.4}
 \newcolumntype{M}{>{\centering\arraybackslash}m{\dimexpr.1\linewidth-1\tabcolsep}}
   \newcolumntype{G}{>{\centering\arraybackslash}m{\dimexpr.35\linewidth-1\tabcolsep}}
%\begin{tabular}{MGG} 
\begin{tabular}{G G G}
		$C_0$	& $D_1$  & $D_2$ \\[-2ex]
			\includegraphics[scale=0.33]{bxp-lambdas-C0_STEPN1toN4_2.pdf}   
			&
			\includegraphics[scale=0.33]{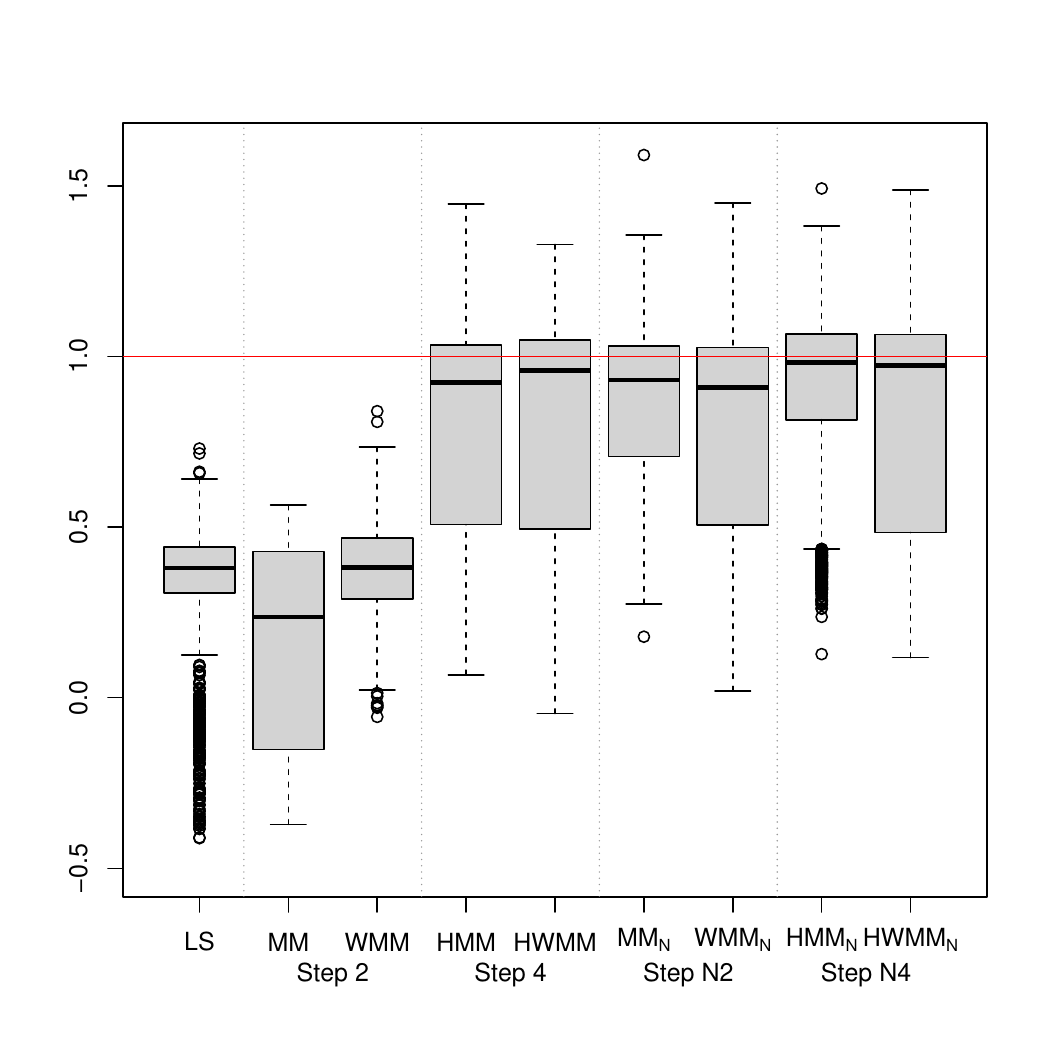}   
			&
			\includegraphics[scale=0.33]{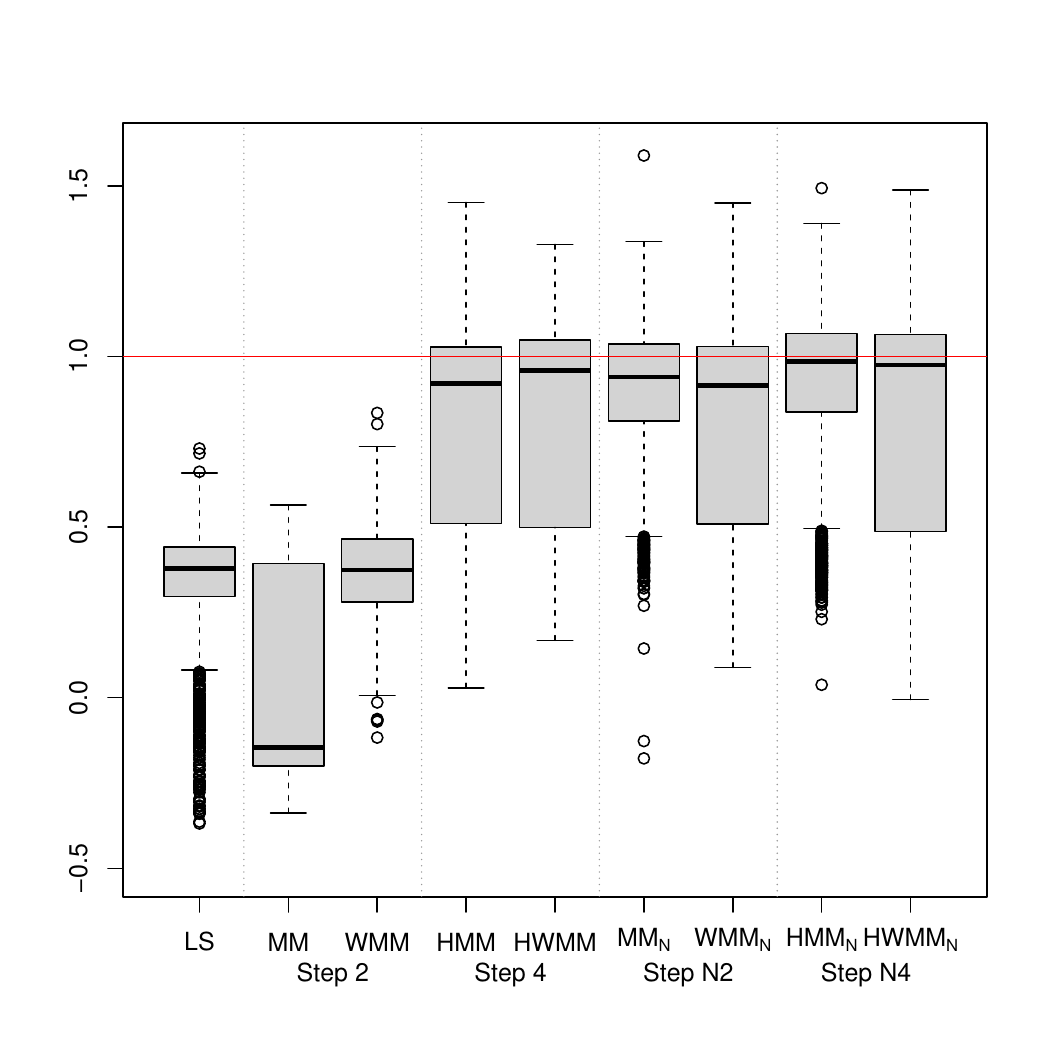}   
		\end{tabular}
		\vskip-0.1in 
		\caption{\small  \label{fig:lambda-bxp-C9}  Boxplots of the estimates of $\lambda_0$, under $C_{0}$,  $D_1$ and $D_2$.}
%	\end{center} 
\end{figure}

To study the behaviour of the estimates of the parameter $\lambda_0$ under these contaminations,  Figure \ref{fig:lambda-bxp-C9} presents the boxplots for the robust estimates obtained in \textbf{Step 2}, \textbf{Step 4}, \textbf{Step N2} and \textbf{Step N4}  as well as for the classical estimators. As above, we labelled  \MM~  and   \WMM ~ the boxplots of the estimators computed in \textbf{Step 2} when the initial estimator $\wbbe_{\ini}$ corresponds to an $MM-$ or a weighted $MM-$estimator assuming an homocedastic model. In contrast, the results for the estimators computed   in \textbf{Step 4} are indicated as \HMM ~ or  \HWMM~ when using the estimators of $\bbe$ defined in \textbf{Step 3} with weights equal to 1 or with the weights   $w(x)$ defined in \eqref{eq:pesos}, respectively.  To facilitate the comparisons all plots have the same vertical axis.

As in Figure \ref{fig:lambda-bxp}, the   effect of the outliers with high leverage on the classical estimator of ${\lambda_0}$ is devastating. As with vertical outliers, the   robust estimators of $\lambda_0$ computed through \textbf{Step 4}   show their advantage over those computed in   \textbf{Step 2}. The same conclusion is valid when comparing the results for the estimators obtained through  \textbf{Step N4} and  \textbf{Step N2}.  It is worth mentioning that the estimators obtained either in \textbf{Steps N2} or \textbf{N4} using  as regression estimator the \MM~ or the \HMM~ estimators have a lower dispersion than when considering their weighted versions.
 
\begin{figure}[ht!]
  	 \renewcommand{\arraystretch}{0.4}
 \newcolumntype{M}{>{\centering\arraybackslash}m{\dimexpr.1\linewidth-1\tabcolsep}}
   \newcolumntype{G}{>{\centering\arraybackslash}m{\dimexpr.35\linewidth-1\tabcolsep}}
%\begin{tabular}{MGG} 
\begin{tabular}{G G G}
		$C_0$	& $D_1$  & $D_2$ \\[-2ex]
			\includegraphics[scale=0.33]{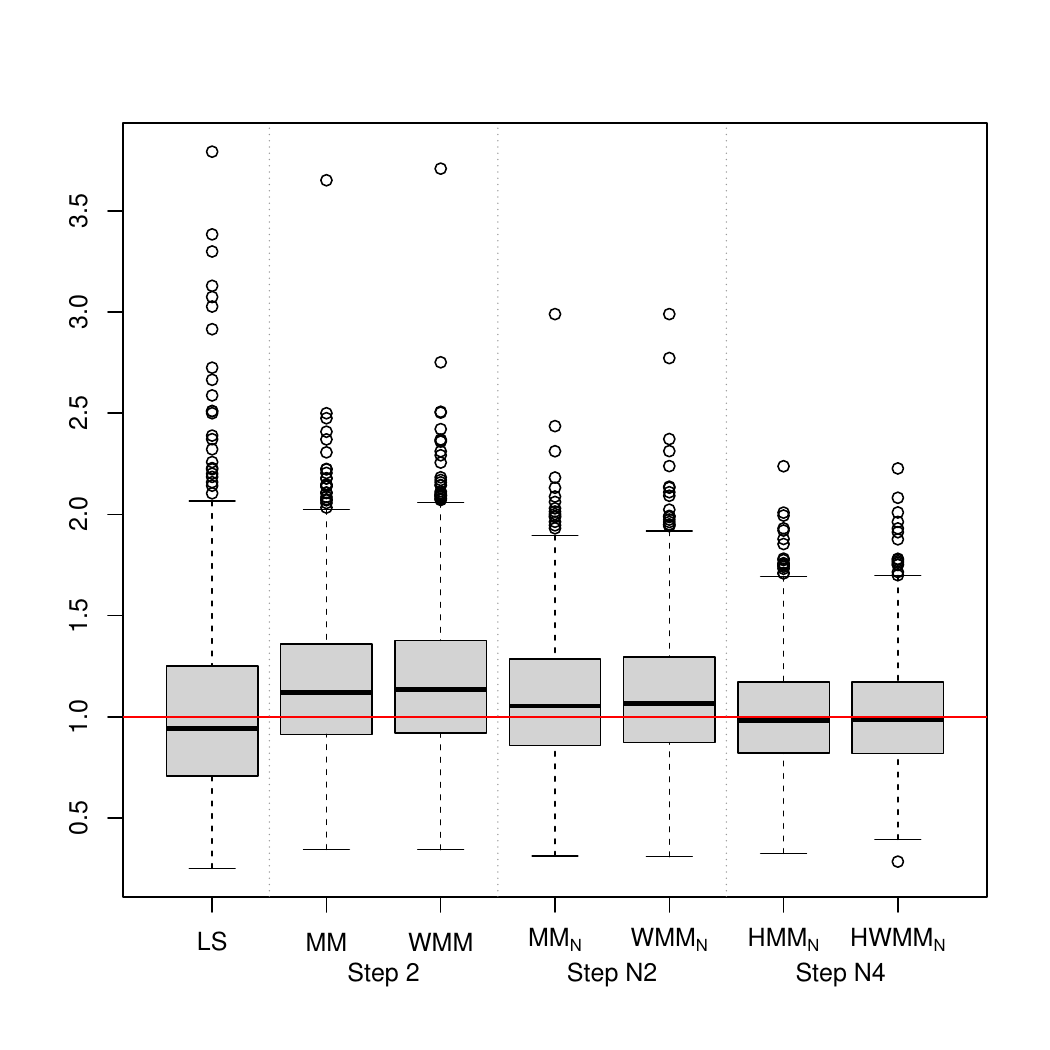}   
			&
			\includegraphics[scale=0.33]{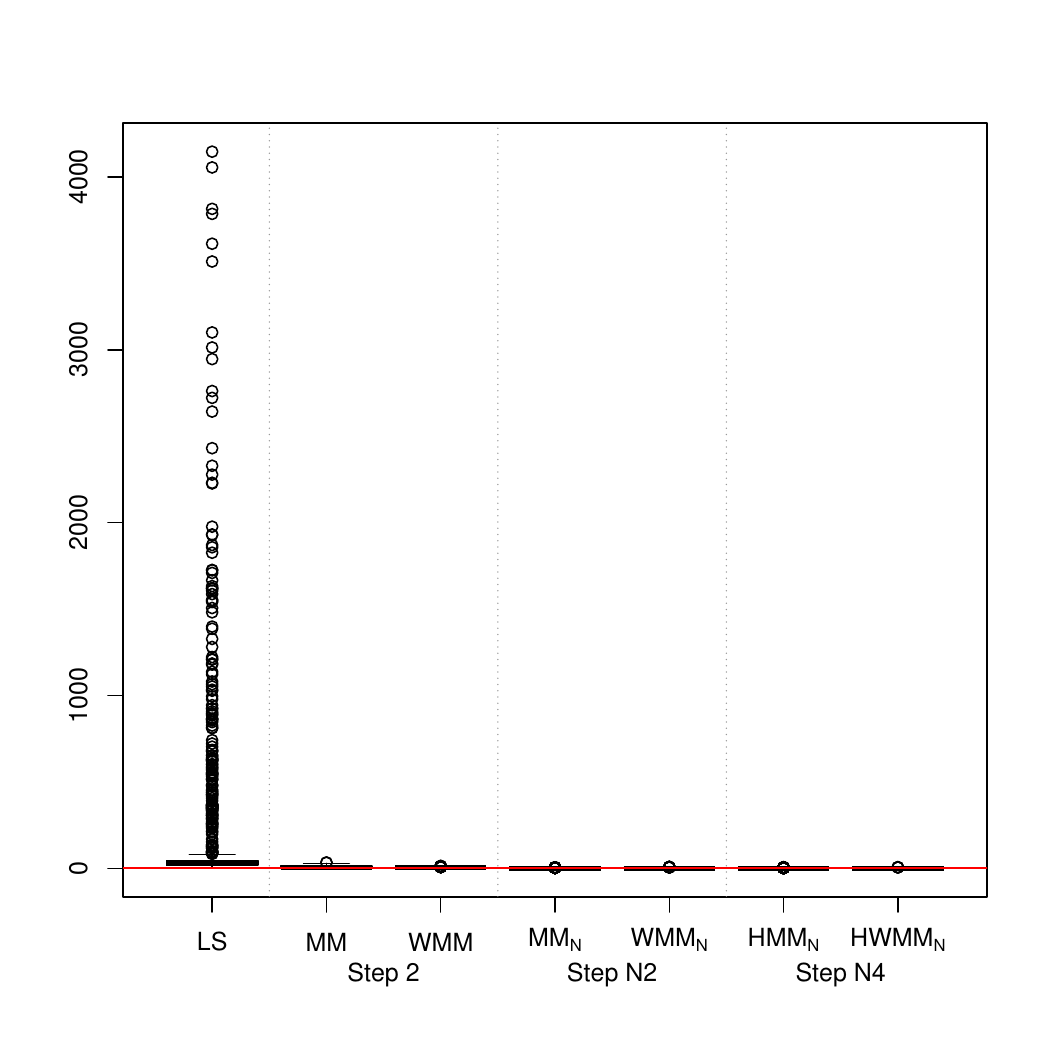}   
			&
			\includegraphics[scale=0.33]{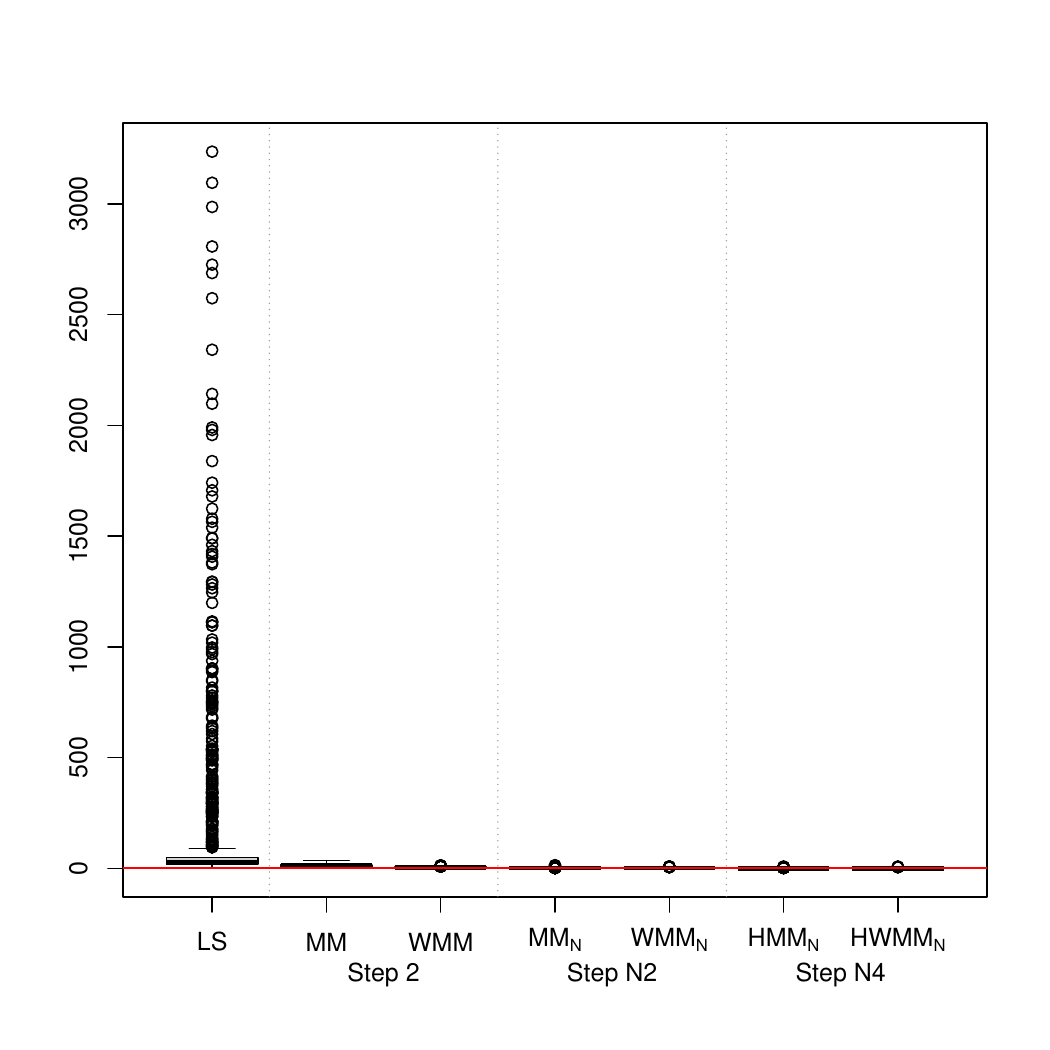}\\  
				$C_0$	& $D_1$  & $D_2$ \\[-2ex]
			\includegraphics[scale=0.33]{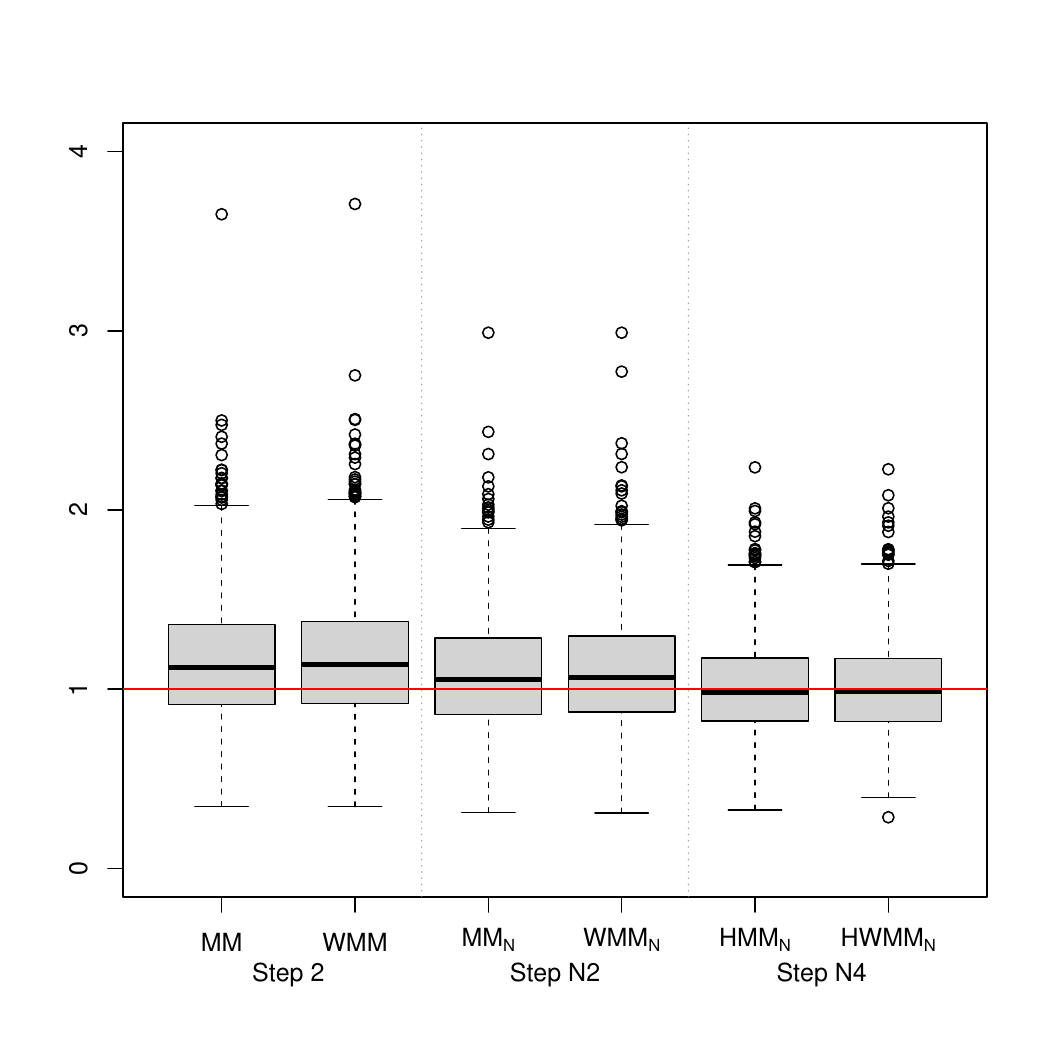}   
			&
			\includegraphics[scale=0.33]{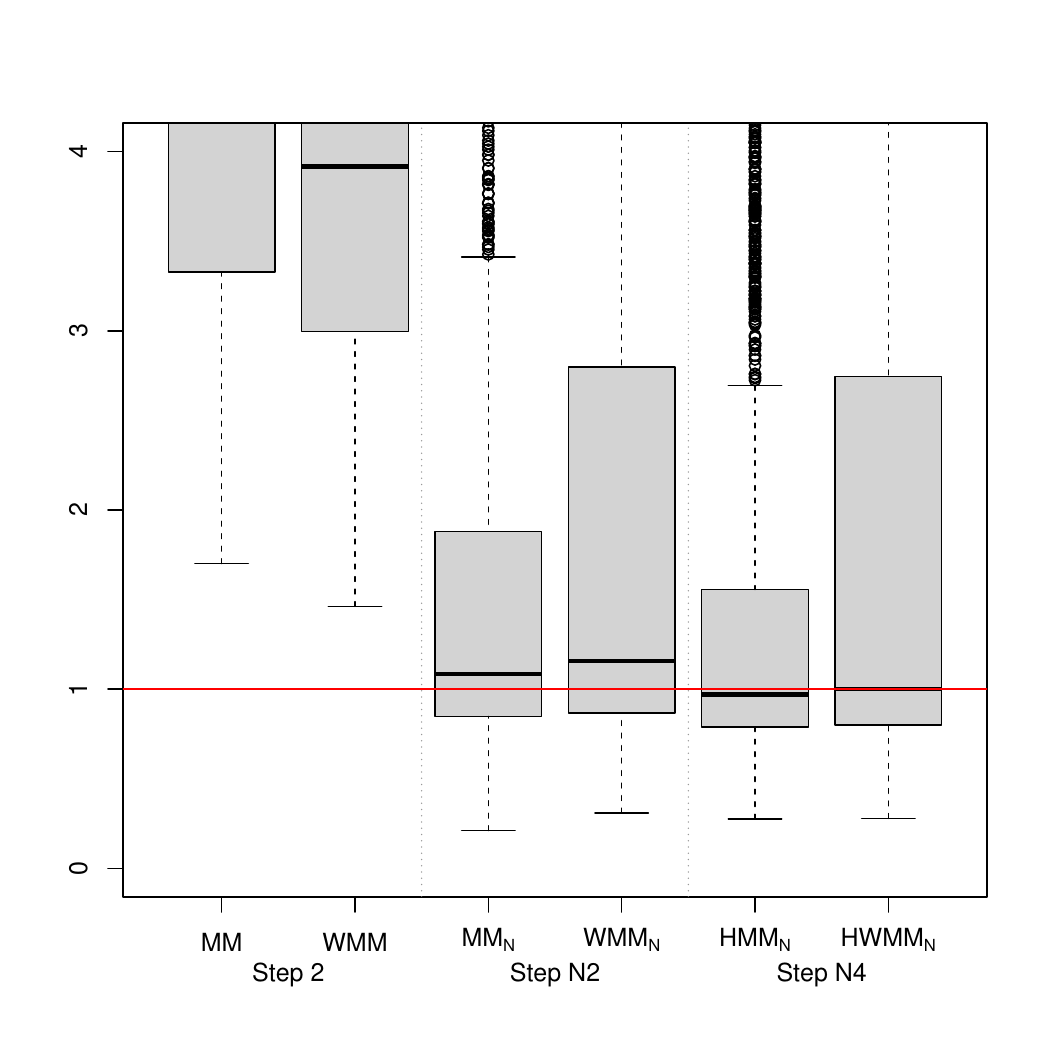}   
			&
			\includegraphics[scale=0.33]{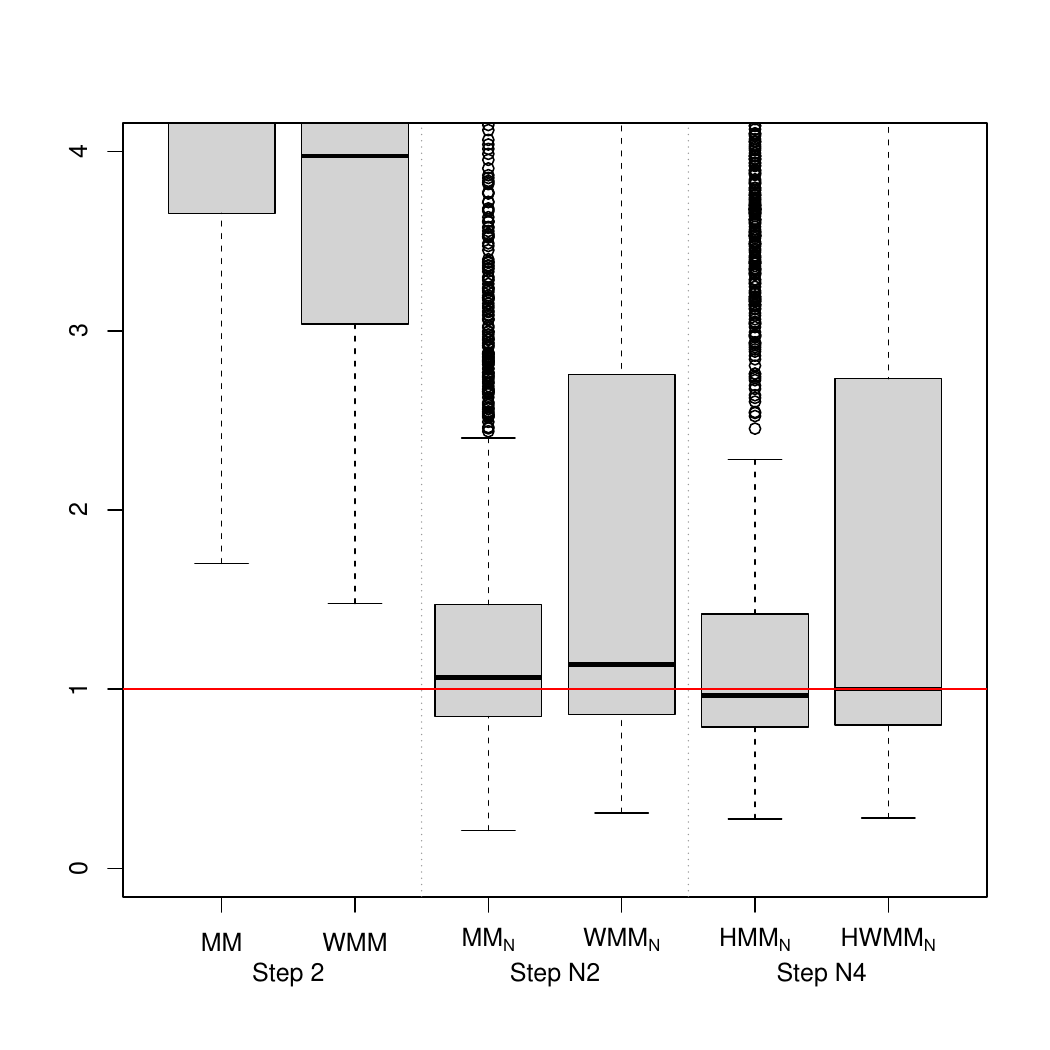}   
		\end{tabular}
		\vskip-0.1in 
		\caption{\small  \label{fig:sigma-bxp-C9}  Boxplots of the estimates of $\sigma$, under $C_{0}$,  $D_1$ and $D_2$. The lower panels present the boxplots of the robust estimators in the range $[0,4]$.}
%	\end{center} 
\end{figure}

\begin{figure}[ht!]
	\begin{center}
		\begin{tabular}{ccc}
			\HLS & \HMM & \HWMM\\		
			\multicolumn{3}{c}{$C_0$}\\	[-3ex]
			\includegraphics[scale=0.3]{homos-C0variance_LS_bis.pdf} &  
			\includegraphics[scale=0.3]{homos-C0variance_HMM_bis.pdf}  &  
			\includegraphics[scale=0.3]{homos-C0variance_HWMM_bis.pdf}\\
			\multicolumn{3}{c}{$D_{1}$}\\[-3ex]	
            \includegraphics[scale=0.3]{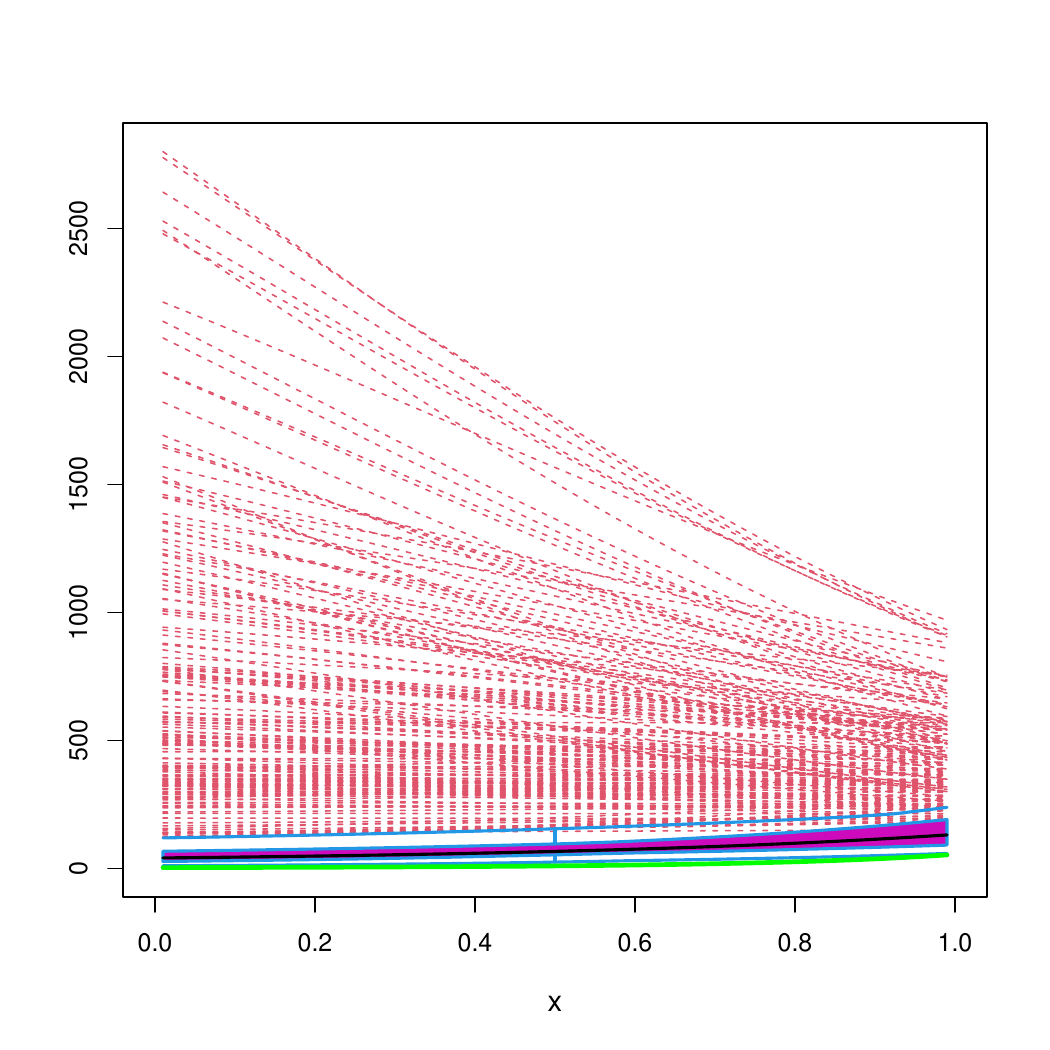} &  
            \includegraphics[scale=0.3]{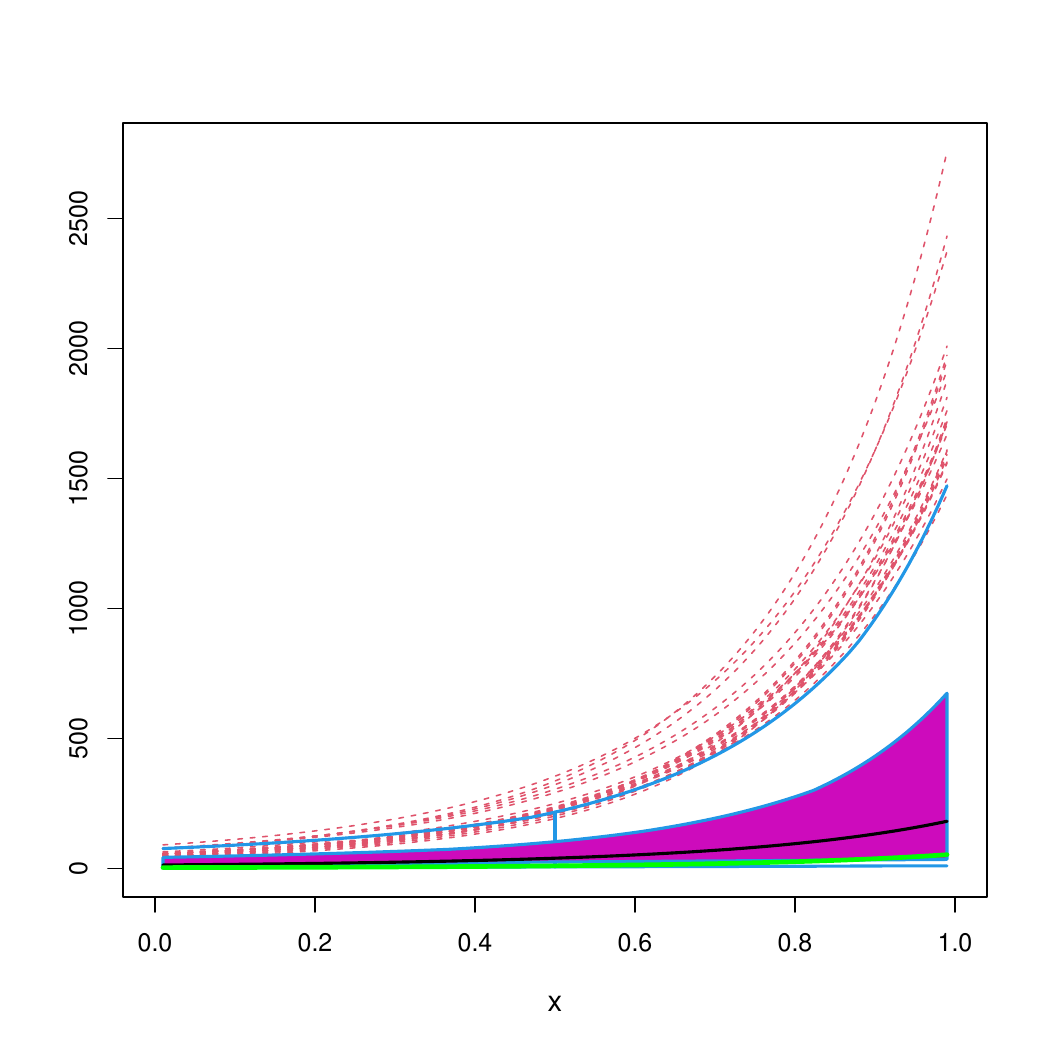}  &  
            \includegraphics[scale=0.3]{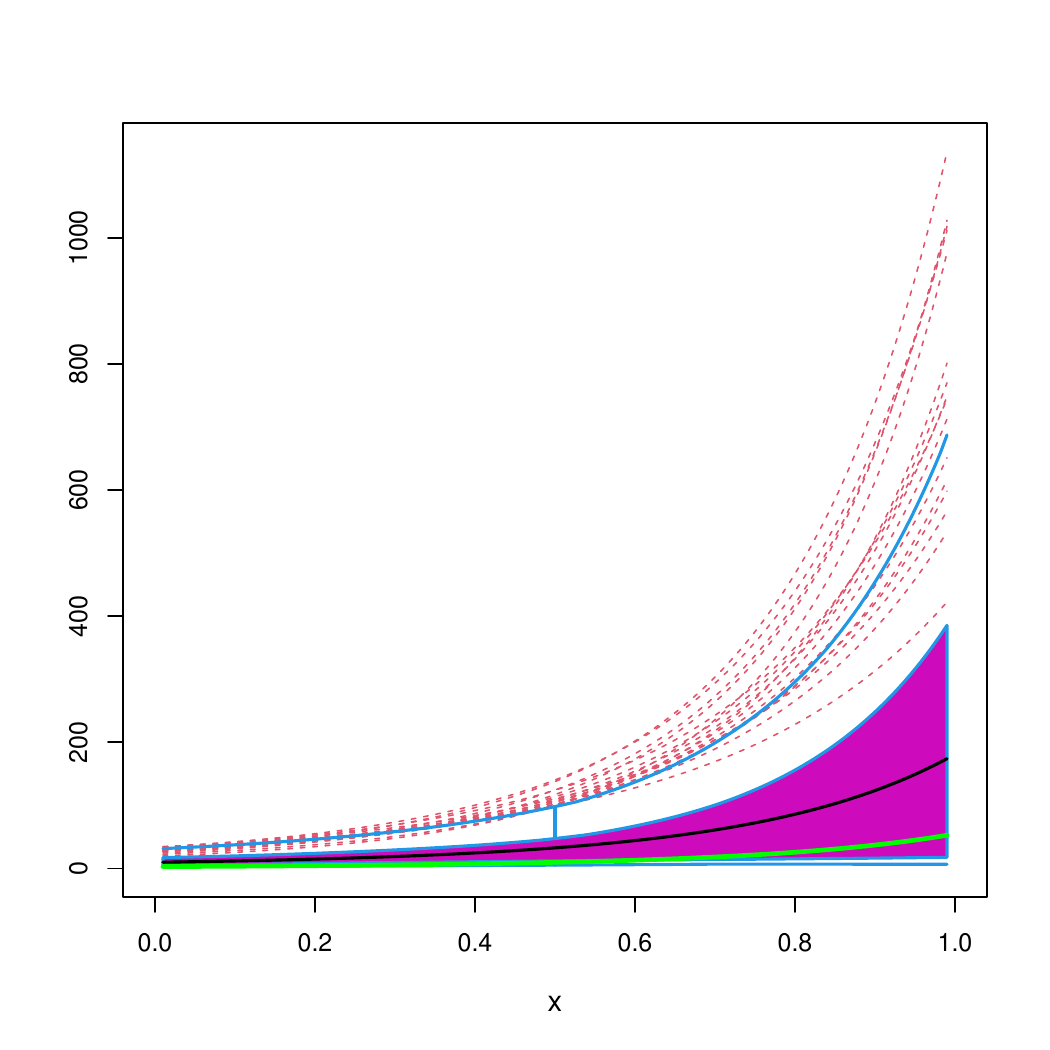}\\
            \multicolumn{3}{c}{$D_{2}$}\\[-3ex]
            \includegraphics[scale=0.3]{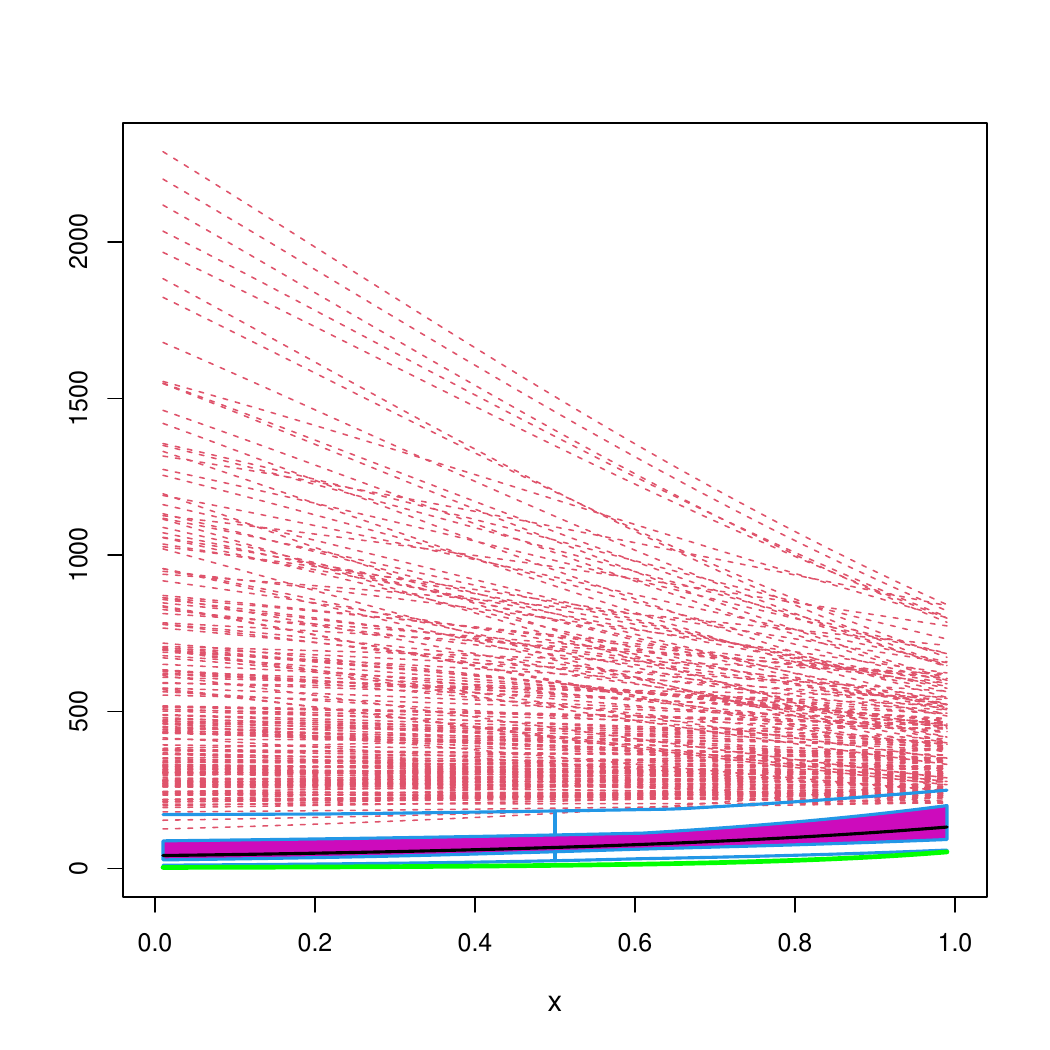} &  
            \includegraphics[scale=0.3]{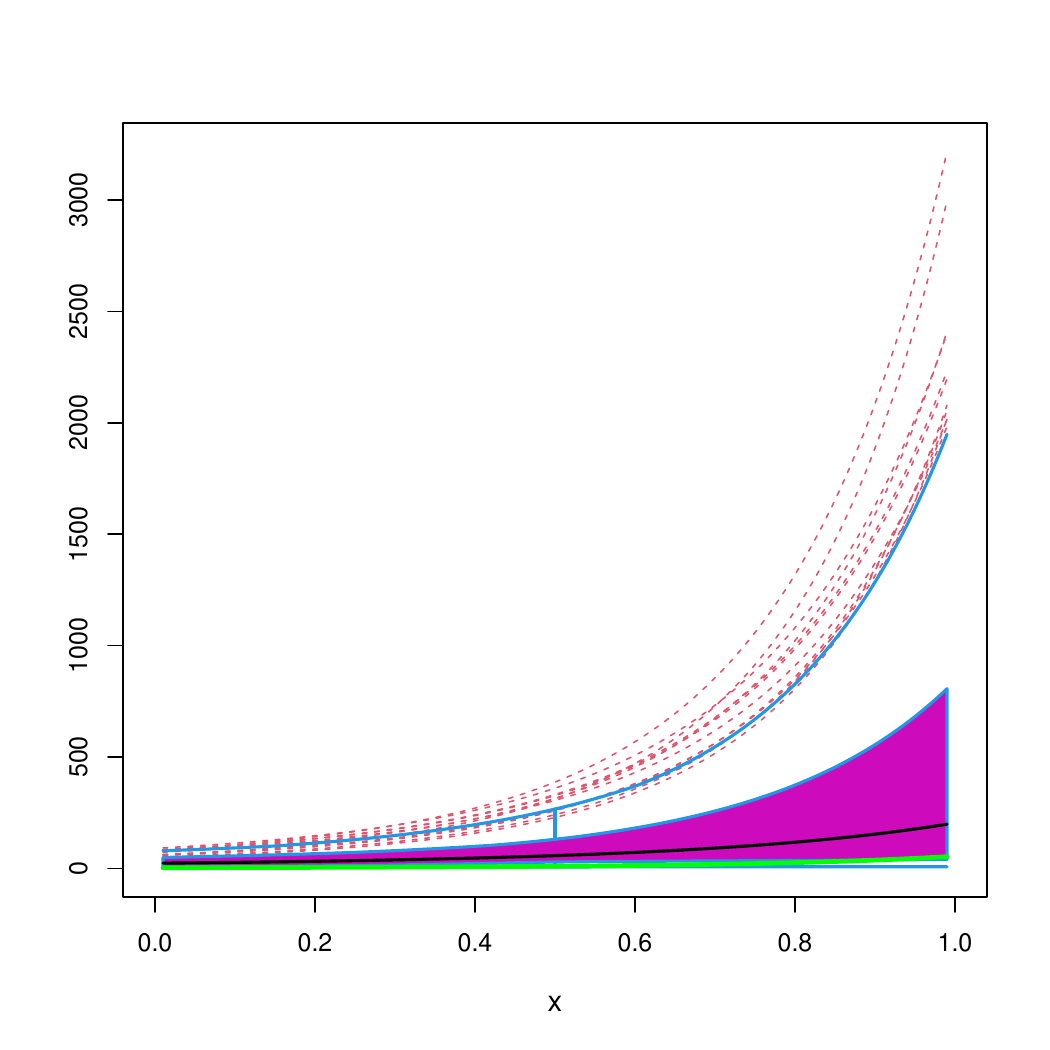}  &  
            \includegraphics[scale=0.3]{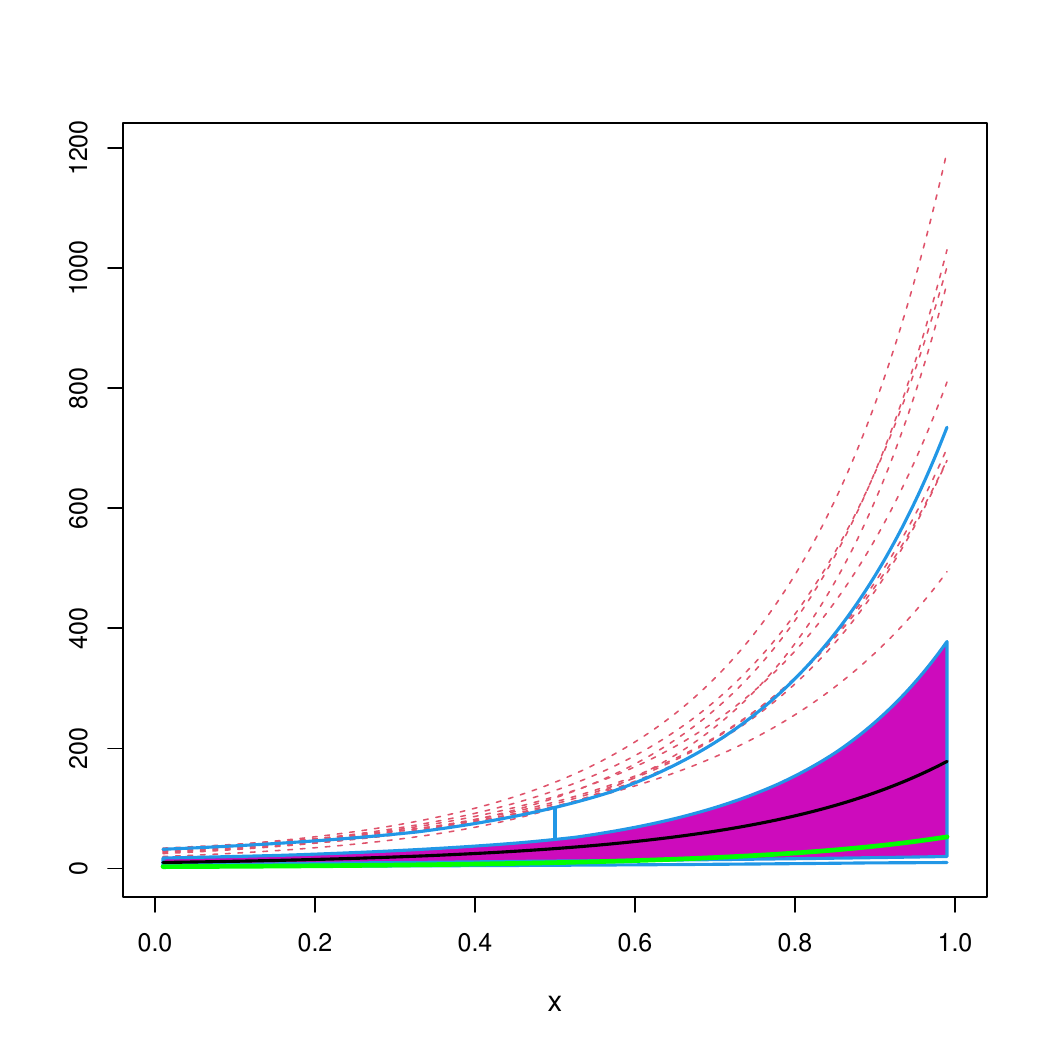}            
		\end{tabular}
		\vskip-0.1in \caption{ \small \label{fig:variance}  Functional boxplots of the estimated variance function   when considering the \LS ~ estimators and the robust ones defined through \textbf{Step 4}. The green curve corresponds to the true variance function.}
	\end{center} 
\end{figure}

\begin{figure}[ht!]
	\begin{center}
		\begin{tabular}{ccc}
			\HLS & \HMM$_{\nuevo}$ & \HWMM$_{\nuevo}$\\		
			\multicolumn{3}{c}{$C_0$}\\	[-3ex]
			\includegraphics[scale=0.3]{homos-C0variance_LS_bis.pdf} &  
			\includegraphics[scale=0.3]{fbplot-variance-C0_HMM_stepN4.pdf}  &  
			\includegraphics[scale=0.3]{fbplot-variance-C0_HWMM_stepN4.pdf}\\ 
			\multicolumn{3}{c}{$D_{1}$}\\[-3ex]	
            \includegraphics[scale=0.3]{homos-C9_1variance_LS_bis.pdf} &  
			\includegraphics[scale=0.3]{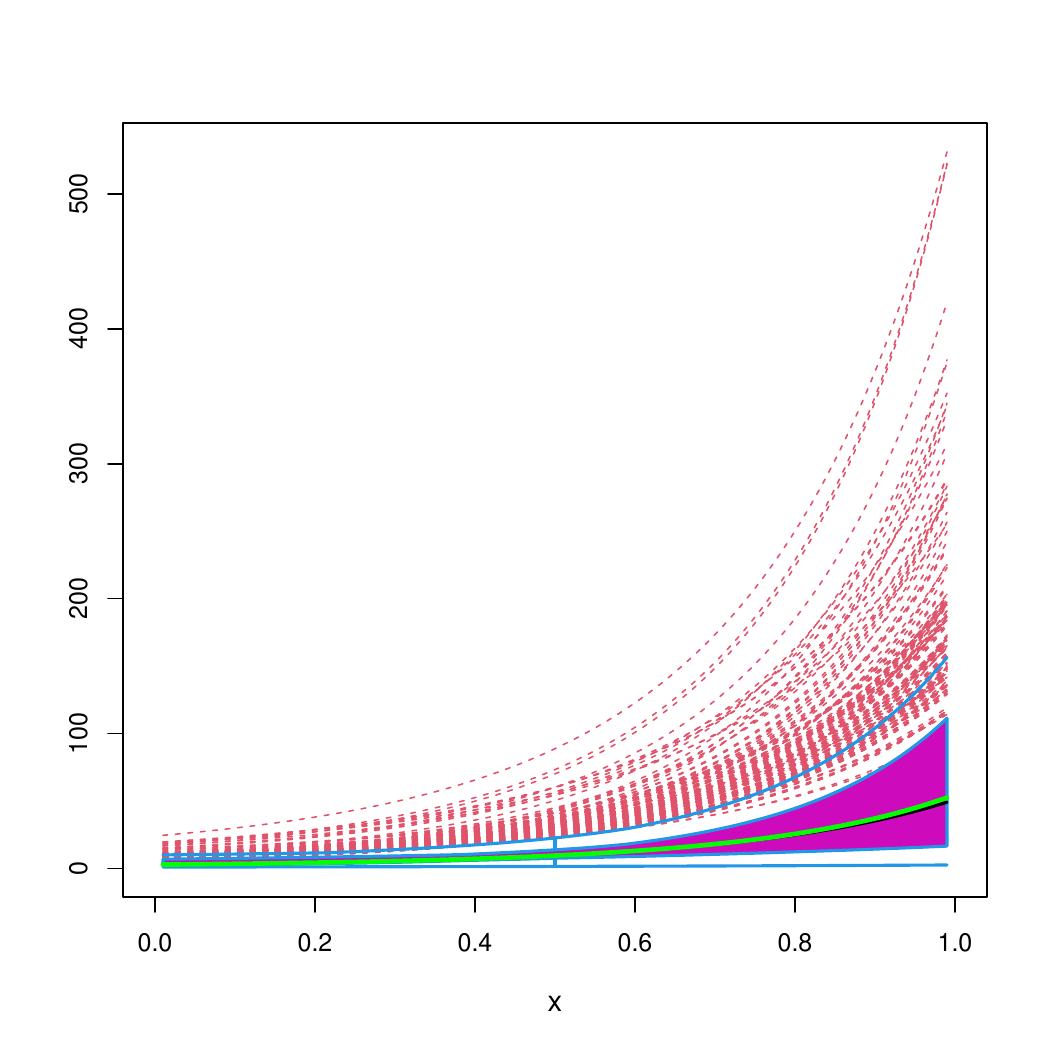}  &  
			\includegraphics[scale=0.3]{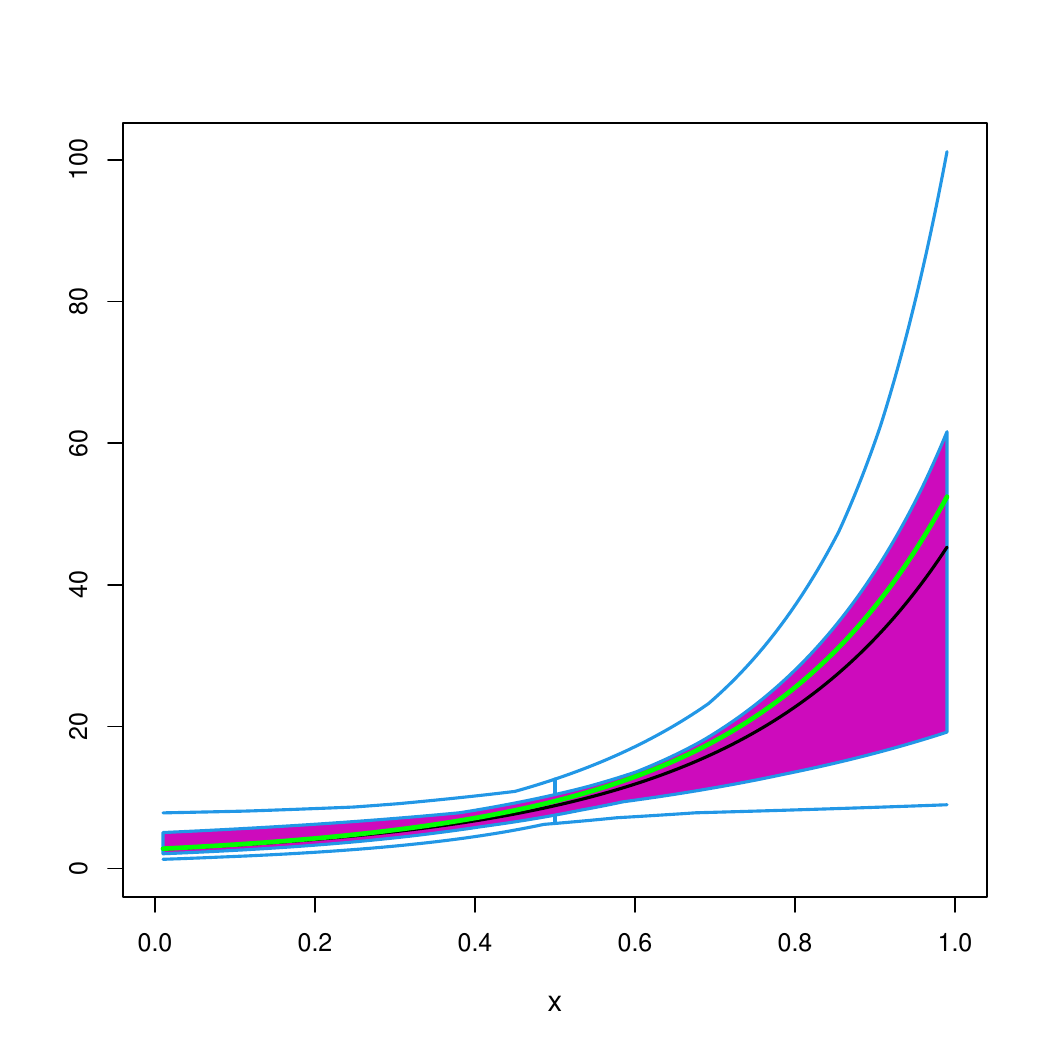}\\ 
			\multicolumn{3}{c}{$D_{2}$}\\[-3ex]
            \includegraphics[scale=0.3]{homos-C9_2variance_LS_bis.pdf} &  
			\includegraphics[scale=0.3]{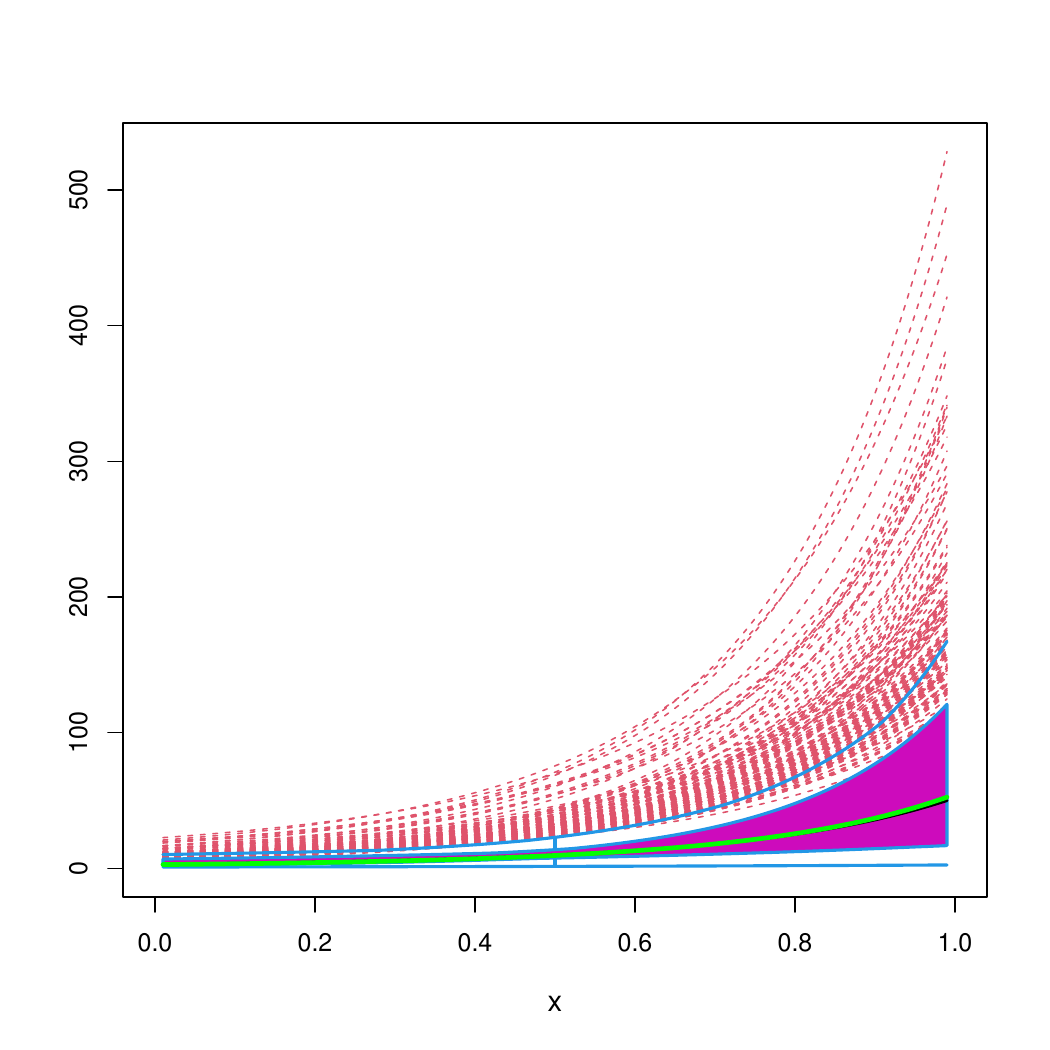}  &  
			\includegraphics[scale=0.3]{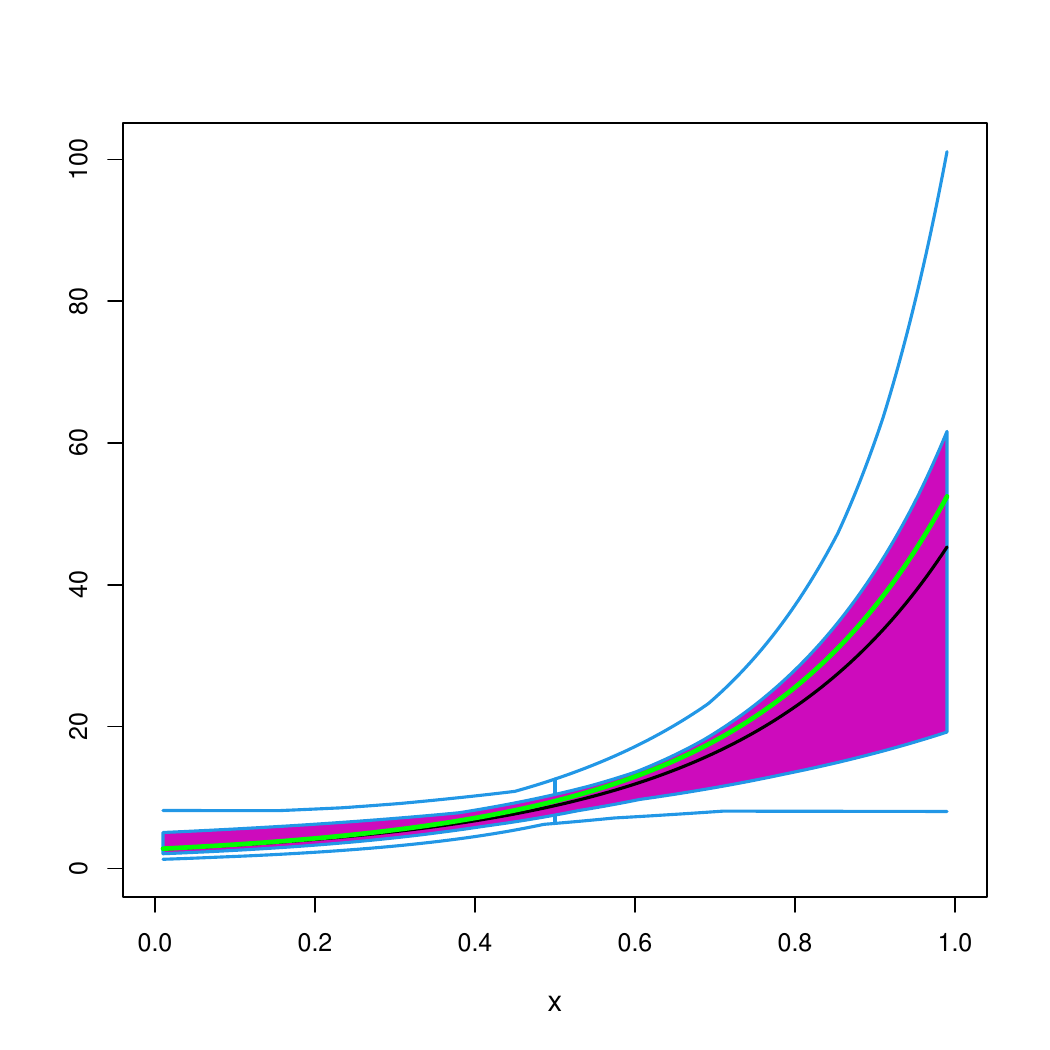}\\          
		\end{tabular}
		\vskip-0.1in \caption{ \small \label{fig:variance-stepN4}  Functional boxplots of the estimated variance function when considering the \LS ~ estimators and the robust ones defined through \textbf{Step N4}. The green curve corresponds to the true variance function.}
	\end{center} 
\end{figure}

The harmful impact of the introduced outliers on the classical estimation of the variance function becomes evident in Figure \ref{fig:variance},  where we display the functional boxplots of the estimated variance function  $\wsigma(x)=\wsigma \upsilon(x, \wlam)$, when $ \wlam$ corresponds to the classical estimators or to the robust procedure defined in \textbf{Step 4}. Figure \ref{fig:variance-stepN4} displays the functional boxplots when considering the classical procedure and the robust method given  in \textbf{Step N4}.  In both figures, the first row corresponds to $C_0$ and as mentioned above, except for some outlying curves, all the methods perform quite well for clean data. On the contrary, plots in the second and third rows of the same figure  reveal that the effect of outliers on the classical estimator is damaging, since most of the  estimated curves lie far from the true one and are identified as outliers. The situation is better with the robust estimators  \HMM~ and \HWMM, indeed, the results are much more stable. The estimators of the variance function obtained with the \HWMM ~estimator follow the pattern of the true curve, but for large values of $x$,  most of the estimates computed from the \HMM ~and \HWMM ~ methods lie above the true curve. This fact may be explained by the large variability of the estimates of both $\lambda_0$ and $\sigma_0$ observed in Figures \ref{fig:lambda-bxp-C9} and  \ref{fig:sigma-bxp-C9} under $D_1$ and $D_2$, respectively. Even when the boxplot is almost centered at the true value $\lambda_0=1$, the interquartile range is considerably enlarged with respect to that obtained for clean samples,  suggesting a certain  instability of the variance function estimators, under these two contaminations. In contrast, as revealed in Figure \ref{fig:variance-stepN4}, the procedure obtained through \textbf{Step N4} outperforms the \HMM ~ and  \HWMM ~ methods and provides remarkably stable estimators.
 
\section{Final comments}{\label{sec:comentarios}}
Non--linear regression models are widely used in applications, among them chemometrics. In some situations, non--homogenous variance errors arise, posing the challenging problem of distinguishing atypical observations from regular ones. In this paper, we address the problem of robust estimation under a non--linear heteroscedastic model. We focus not only on estimating the regression  function but also the variance function, as well. We provide   two robust stepwise procedures that  combine weighted $MM-$estimators of the regression parameters  and two different  methods to estimate the variance function.  Our simulation study illustrates the stability of our proposals  under different contamination scenarios. Furthermore, it highlights the benefit of our second stepwise approach, specifically the one characterized by \textbf{Steps N1} to \textbf{N4} in conjunction with weights for controlling leverage. This second procedure improves, in particular, the performance of the variance function estimators, ensuring reliable results, which may be useful when hypothesis testing and confidence interval problems are of interest.  

\noi\textbf{\small Acknowledgements.} {\small  This research was partially supported by  grants 20020220200037\textsc{ba}    from the Universidad de Buenos Aires  and  \textsc{pict} 2021-I-A-00260 from \textsc{anpcyt} at  Argentina (Ana M. Bianco and Graciela Boente), the Spanish Project {MTM2016-76969P} from the Ministry of Economy, Industry and Competitiveness, Spain (MINECO/AEI/FEDER, UE)  (Ana M. Bianco and Graciela Boente) and by National Funds through FCT and CEMAT/UL, projects UIDB/04621/2020 and UIDP/04621/2020 (Concei\c{c}\~{a}o Amado and Isabel M. Rodrigues). 
This work began while Ana M. Bianco and Graciela Boente were visiting the Departamento de Matem\'atica at Instituto Superior T\'ecnico and received support by National Funds through FCT and CEMAT/UL.}
%\clearpage 
\small

\bibliographystyle{apalike}
\bibliography{heterw2023}

\end{document}